\pdfoutput=1
\documentclass[oneside,fleqn,twocolumn]{article}

\usepackage[%
    paperwidth=210mm,
    paperheight=297mm,
    top={25mm},
    headheight={12pt},
    headsep={5.15mm},
    text={160mm,245mm},
    marginparsep=5mm,
    marginparwidth=12mm,
    footskip=12mm,
    hcentering,
    twocolumn]{geometry}
\makeatletter
\renewcommand\normalsize{%
   \@setfontsize\normalsize{10bp}{12bp}%
   \abovedisplayskip 12\p@ \@plus2\p@ \@minus1\p@
   \abovedisplayshortskip \z@ \@plus3\p@
   \belowdisplayshortskip 3\p@ \@plus3\p@ \@minus3\p@
   \belowdisplayskip \abovedisplayskip
   \let\@listi\@listI}
\renewcommand\small{%
   \@setfontsize\small\@ixpt{11}%
   \abovedisplayskip 5\p@ \@plus3\p@ \@minus4\p@
   \abovedisplayshortskip \z@ \@plus2\p@
   \belowdisplayshortskip 3\p@ \@plus2\p@ \@minus2\p@
   \def\@listi{\leftmargin\leftmargini
               \topsep 4\p@ \@plus2\p@ \@minus2\p@
               \parsep 2\p@ \@plus\p@ \@minus\p@
               \itemsep \parsep}%
   \belowdisplayskip \abovedisplayskip}
\renewcommand\footnotesize{%
   \@setfontsize\footnotesize{7}{8}%
   \abovedisplayskip 5\p@ \@plus2\p@ \@minus4\p@
   \abovedisplayshortskip \z@ \@plus\p@
   \belowdisplayshortskip 3\p@ \@plus\p@ \@minus2\p@
   \def\@listi{\leftmargin\leftmargini
               \topsep 3\p@ \@plus\p@ \@minus\p@
               \parsep 2\p@ \@plus\p@ \@minus\p@
               \itemsep \parsep}%
   \belowdisplayskip \abovedisplayskip}
\renewcommand\scriptsize{\@setfontsize\scriptsize\@ixpt\@ixpt}%
\renewcommand\tiny{\@setfontsize\tiny\@vpt\@vipt}%
\renewcommand\large{\@setfontsize\large{12}{14}}%
\renewcommand\Large{\@setfontsize\Large{16}{18}}%
\renewcommand\LARGE{\@setfontsize\LARGE\@xviipt{22}}%
\renewcommand\huge{\@setfontsize\huge\@xxpt{25}}%
\renewcommand\Huge{\@setfontsize\Huge\@xxvpt{30}}%
\DeclareMathSizes{\@ixpt}{\@ixpt}{7}{5}%
\DeclareMathSizes{\@xpt}{\@xpt}{7}{5}%
\DeclareMathSizes{\@xipt}{\@xipt}{7}{5}%
\makeatother
\normalsize

\makeatletter
\def\sectionfont{\reset@font\fontfamily{\rmdefault}\fontsize{14bp}{16bp}\bfseries\selectfont\raggedright\boldmath}%
\def\subsectionfont{\reset@font\fontfamily{\rmdefault}\fontsize{12bp}{14bp}\bfseries\selectfont\raggedright\boldmath}%
\def\subsubsectionfont{\reset@font\fontsize{11bp}{13bp}\bfseries\selectfont\raggedright\boldmath}%
\def\paragraphfont{\reset@font\fontsize{10bp}{12bp}\bfseries\itshape\selectfont\raggedright}%
\renewcommand\section{\@startsection{section}{1}{\z@}%
                                    {-12pt \@plus -4pt \@minus -2pt}%
                                    {9pt}{\sectionfont}}
\renewcommand\subsection{\@startsection{subsection}{2}{\z@}%
                                       {-12pt \@plus -4pt \@minus -2pt}%
                                       {6pt}{\subsectionfont}}
\renewcommand\subsubsection{\@startsection{subsubsection}{3}{\z@}%
                                          {-12pt \@plus -4pt \@minus -2pt}%
                                          {6pt}{\subsubsectionfont}}
\renewcommand\paragraph{\@startsection{paragraph}{4}{\z@}%
                                      {-12pt \@plus -4pt \@minus-2pt}%
                                      {3pt}{\paragraphfont}}
\makeatother

\usepackage{hyperref}
\hypersetup{
    colorlinks,
    citecolor=blue,
    linkcolor=blue,
    urlcolor=blue,
    breaklinks=true,
    pdftitle={Volcanite: commodity-hardware segmentation volume visualization
              for connectomics and beyond},
    pdfauthor={Max Piochowiak, Reiner Dolp, Julian Herold, Eric Behle,
               Carsten Dachsbacher, Alexander Schug},
    pdfsubject={Interactive voxel-precise visualization of terascale
                segmentation volumes on commodity hardware},
    pdfkeywords={Segmentation, Rendering, Compression, Voxel Data, Volumes,
                 Visualization, Connectomics},
}

\usepackage[numbers,sort&compress]{natbib}
\makeatletter
\def\bibfont{\reset@font\fontfamily{\rmdefault}\small\selectfont}
\makeatother

\makeatletter
\newif\ifdoublecol
\@ifclasswith{article}{twocolumn}{\doublecoltrue}{\doublecolfalse}
\makeatother

\usepackage{graphicx}%
\usepackage{multirow}%
\usepackage{amsmath,amssymb,amsfonts}%
\usepackage{amsthm}%
\usepackage{mathrsfs}%
\usepackage[title]{appendix}%
\usepackage{xcolor}%
\usepackage{xstring}%
\usepackage{textcomp}%
\usepackage{manyfoot}%
\usepackage{booktabs}%
\usepackage{algorithm}%
\usepackage{algorithmicx}%
\usepackage[noend]{algpseudocode}%
\usepackage{listings}%
\definecolor{ccomment}{rgb}{0,0.4,0.05}
\definecolor{cnumber}{rgb}{0.7,0.7,0.7}
\definecolor{ckeyword}{rgb}{0,0.1,0.72}
\definecolor{cmauve}{rgb}{0.58,0,0.82}
\lstdefinestyle{GLSL}{
    language=C,
    backgroundcolor=\color{white},
    commentstyle=\color{ccomment},
    keywordstyle=\color{ckeyword},
    numberstyle=\tiny\color{cnumber},
    stringstyle=\color{cmauve},
    basicstyle=\footnotesize,
    morekeywords={vec3, ivec3, max, fract, floor, argmin, inout},
    breakatwhitespace=false,
    breaklines=true,
    breakautoindent=true,
    captionpos=b,
    keepspaces=true,
    numbers=left,
    numbersep=2pt,
    showspaces=false,
    showstringspaces=false,
    showtabs=false,
    tabsize=1,
    lineskip=1.4pt,
    literate={.}{{.}}1,
}

\usepackage{caption}
\usepackage{wrapfig}
\usepackage{subcaption}
\usepackage[shortcuts]{glossaries}
\glsdisablehyper
\setacronymstyle{long-short}
\usepackage{ifthen}
\usepackage{booktabs}
\usepackage{multirow}
\usepackage{numprint}
\usepackage{enumitem}  %
\setlist{topsep=0pt, partopsep=0pt, parsep=0pt, itemsep=0pt, leftmargin=*}
\usepackage{import}    %
\usepackage{tikz}
\usepackage{placeins} %
\usepackage[normalem]{ulem} %
\usepackage{array} %
\usepackage[activate={true,nocompatibility},%
            final,%
            tracking=true,%
            kerning=true,%
            spacing=true,%
            stretch=10,%
            shrink=30]{microtype}
\microtypecontext{spacing=nonfrench}
\SetTracking{encoding={*}, shape=sc}{0}
\usepackage{cleveref}
\crefname{table}{Tab.}{Tabs.}
\Crefname{table}{Tab.}{Tabs.}
\crefname{section}{sec.}{secs.}
\Crefname{section}{Sec.}{Secs.}
\crefname{figure}{Fig.}{Figs.}
\Crefname{figure}{Fig.}{Figs.}

\usepackage{pgfplotstable}
\pgfplotsset{compat=1.18,
    select coords between index/.style 2 args={
        x filter/.code={
            \ifnum\coordindex<#1\def\pgfmathresult{}\fi
            \ifnum\coordindex>#2\def\pgfmathresult{}\fi
        }
    }
}
\usetikzlibrary{pgfplots.statistics}
\usetikzlibrary{external}
\IfFileExists{REBUILD_FIGURES}{%
    \typeout{VOLCANITE: rebuilding externalized figures (shell-escape required).}%
}{%
    \tikzset{external/mode=only graphics}%
}
\usepgfplotslibrary{colorbrewer}
\pgfplotsset{colormap/BuGn-7}
\pgfplotsset{
    cycle list/Set3-7,
    cycle multiindex* list={
       mark list*\nextlist
       Set3-7\nextlist
    },
}

\usepackage{ext/comptools}
\definecolor{spylenscolor}{HTML}{000000} %
\definecolor{spylensleftcolor}{HTML}{000000}
\definecolor{spylensrightcolor}{HTML}{000000}

\newcommand{\textcode}[1]{{\small \texttt{#1}}}

\graphicspath{{./paper/volcanite}{./}}

\newacronym{csgv}{CSGV}{Fast Compressed Segmentation Volumes}
\newacronym{csgvr}{CSGV-R}{Random Access Compressed Segmentation Volumes}
\newacronym[plural=SVDAGs,longplural=Sparse Voxel Directed Acyclic Graphs]{svdag}{SVDAG}{Sparse Voxel Directed Acyclic Graph}
\newacronym{rle}{RLE}{run length encoding}
\newacronym{ans}{ANS}{asymmetric numerical systems}
\newacronym[plural=$\overline{LOD}$s,longplural=inverse levels-of-detail]{ilod}{$\overline{LOD}$}{inverse level-of-detail}
\newacronym{chc}{CHC}{canonical Huffman-codes}
\newacronym[plural=LSB,longplural=least significant bits]{lsb}{LSB}{least significant bit}
\newacronym[plural=MSB,longplural=most significant bits]{msb}{MSB}{most significant bit}
\newacronym{dda}{DDA}{digital differential analyzer}
\newacronym[plural=LODs,longplural=levels-of-detail]{lod}{LOD}{level-of-detail}
\newacronym[plural=BRDFs,longplural=bidirectional distribution functions]{brdf}{BRDF}{bidirectional distribution function}
\newacronym[plural=TTFFs,longplural=times to first frame]{ttff}{TTFF}{time to first frame}
\newacronym{vram}{VRAM}{video memory}
\newacronym{gpu}{GPU}{graphics processing unit}
\newacronym{gpc}{GPC}{graphics processing cluster}
\newacronym{sm}{SM}{streaming multiprocessor}
\newacronym{simd}{SIMD}{Single Instruction Multiple Data}
\newacronym{simt}{SIMT}{Single Instruction Multiple Threads}
\newacronym{cpu}{CPU}{central processing unit}
\newacronym{gui}{GUI}{graphical user interface}
\newacronym{cli}{CLI}{command line interface}
\newacronym{em}{EM}{electron microscopy}

\newcommand*{\inlineimg}[2][no alt]{%
    \raisebox{-.3\baselineskip}{%
        \includegraphics[
        alt={#1},
        height=0.9\baselineskip,
        width=0.9\baselineskip,
        keepaspectratio,
        ]{#2}%
    }%
    {}
}
\newcommand{\opparent}{\inlineimg[parent]{fig/op/op_parent}}
\newcommand{\opneighbor}{\inlineimg[neighbor]{fig/op/op_neighbor}}
\newcommand{\opxneighbor}{\inlineimg[X neighbor]{fig/op/op_neighbor_x}}
\newcommand{\opyneighbor}{\inlineimg[Y neighbor]{fig/op/op_neighbor_y}}
\newcommand{\opzneighbor}{\inlineimg[Z neighbor]{fig/op/op_neighbor_z}}
\newcommand{\oppalette}{\smash{\inlineimg[palette advance]{fig/op/op_palette}}}
\newcommand{\oplast}{\inlineimg[palette last]{fig/op/op_palette_last}}
\newcommand{\opdelta}{\inlineimg[palette delta]{fig/op/op_palette_delta}}
\newcommand{\opdeltaold}{\inlineimg[palette delta]{fig/op/op_palette_delta}~\kern-1em~$\star$}

\newcommand{\polyamid}{\textsc{Polyamid}}
\newcommand{\ara}{\textsc{MouseBA}}
\newcommand{\griesser}{\textsc{Fabric}}
\newcommand{\griesserval}{\textsc{Fabric$_{\textsc{val}}$}}
\newcommand{\wolny}{\textsc{Plant}}
\newcommand{\xtm}{\textsc{Battery}}
\newcommand{\azba}{\textsc{AZBA}}
\newcommand{\hone}{\textsc{H01}}
\newcommand{\honebv}{\textsc{H01$_{\textsc{BV}}$}}
\newcommand{\honewm}{\textsc{H01$_{\textsc{WM}}$}}
\newcommand{\liconn}{\textsc{Liconn}}
\newcommand{\cortexsmall}{\textsc{MouseL4$_\textsc{S}$}}
\newcommand{\cortex}{\textsc{MouseL4}}
\newcommand{\cells}{\textsc{Cells}}
\newcommand{\fiber}{\textsc{Fiber}}

\newcommand{\dataNameFromCSV}[1]{%
    \IfStrEqCase{#1}{%
    {pa66}{\polyamid}
    {Ara2016}{\ara}
    {Griesser2022-sample}{\griesser}
    {Griesser2022-validation}{\griesserval}
    {Wolny2020}{\wolny}
    {xtm-battery}{\xtm}
    {azba}{\azba}
    {H01-bloodvessel}{\honebv}
    {H01-wm}{\honewm}
    {liconn}{\liconn}
    {Motta2019-small}{\cortexsmall}
    {Motta2019}{\cortex}
    {cells}{\cells}
    {fiber}{\fiber}
    }[UNKNOWN DATA SET]
}

\newcommand{\dataCitationFromCSV}[1]{%
    \IfStrEqCase{#1}{%
    {pa66}{\cite{Bertoldo:2021:pa66}}
    {Ara2016}{\cite{Allen:2017:ara2016}}
    {Griesser2022-sample}{\cite{Griesser:2022:sample}}
    {Griesser2022-validation}{\cite{Griesser:2022:validation}}
    {Wolny2020}{\cite{Wolny:2020:plant}}
    {xtm-battery}{\cite{Mueller:2021:xtm}}
    {azba}{\cite{Kenney:2021:azba}}
    {H01-bloodvessel}{\cite{Shapson-Coe:2024:h01}}
    {H01-wm}{\cite{Shapson-Coe:2024:h01}}
    {liconn}{\cite{Tavakoli:2025:liconn}}
    {Motta2019-small}{\cite{Motta:2019:DCR}}
    {Motta2019}{\cite{Motta:2019:DCR}}
    {cells}{\cite{Rosenbauer:2020:ETD}}
    {fiber}{\cite{Maurer:2022:Fiber}}
    }[UNKNOWN DATA SET #1]
}

\newcommand{\domainFromCSV}[1]{%
    \IfStrEqCase{#1}{%
    {pa66}{Fiber-reinf. Polyamide}
    {Ara2016}{Mouse Brain Atlas}
    {Griesser2022-sample}{Nonwoven Material}
    {Griesser2022-validation}{Generated Nonwoven}
    {Wolny2020}{Plant Cells}
    {xtm-battery}{Composite Battery}
    {azba}{Zebrafish Brain Atlas}
    {H01-bloodvessel}{Blood Vessels (EM)}
    {H01-wm}{Human (EM CONN)}
    {liconn}{Mammal (LICONN)}
    {Motta2019-small}{Mouse (EM CONN)}
    {Motta2019}{Mouse (EM CONN)}
    {cells}{CPM Tumor Simulation}
    {fiber}{Fiber-reinf. Polymer}
    }[UNKNOWN DATA SET #1]
}

\npdecimalsign{.}
\npthousandsep{ }
\nprounddigits{3}

\newcolumntype{T}{>{\raggedleft}p{0.6cm}}

\newcommand{\subfiglabel}[2][0pt]{%
  {\vspace{#1}\textbf{\textsf{#2}}\vspace{-#1}}%
}

\newcommand{\sectionref}[2]{see '\hyperref[#1]{#2}'}
\newcommand{\Sectionref}[2]{See '\hyperref[#1]{#2}'}
\newcommand{\sectionrefmanual}[2]{#2}

\pgfplotsset{colormap={tab20}{
rgb(0pt)=(0.12156862745098039,0.4666666666666667,0.7058823529411765);
rgb(1pt)=(0.6823529411764706,0.7803921568627451,0.9098039215686274);
rgb(2pt)=(1.0,0.4980392156862745,0.054901960784313725);
rgb(3pt)=(1.0,0.7333333333333333,0.47058823529411764);
rgb(4pt)=(0.17254901960784313,0.6274509803921569,0.17254901960784313);
rgb(5pt)=(0.596078431372549,0.8745098039215686,0.5411764705882353);
rgb(6pt)=(0.8392156862745098,0.15294117647058825,0.1568627450980392);
rgb(7pt)=(1.0,0.596078431372549,0.5882352941176471);
rgb(8pt)=(0.5803921568627451,0.403921568627451,0.7411764705882353);
rgb(9pt)=(0.7725490196078432,0.6901960784313725,0.8352941176470589);
rgb(10pt)=(0.5490196078431373,0.33725490196078434,0.29411764705882354);
rgb(11pt)=(0.7686274509803922,0.611764705882353,0.5803921568627451);
rgb(12pt)=(0.8901960784313725,0.4666666666666667,0.7607843137254902);
rgb(13pt)=(0.9686274509803922,0.7137254901960784,0.8235294117647058);
rgb(14pt)=(0.4980392156862745,0.4980392156862745,0.4980392156862745);
rgb(15pt)=(0.7803921568627451,0.7803921568627451,0.7803921568627451);
rgb(16pt)=(0.7372549019607844,0.7411764705882353,0.13333333333333333);
rgb(17pt)=(0.8588235294117647,0.8588235294117647,0.5529411764705883);
rgb(18pt)=(0.09019607843137255,0.7450980392156863,0.8117647058823529);
rgb(19pt)=(0.6196078431372549,0.8549019607843137,0.8980392156862745);
}}

\definecolor{volcanitecol1}{RGB}{141,211,199} %
\definecolor{volcanitecol2}{RGB}{255,255,179} %
\definecolor{volcanitecol3}{RGB}{190,186,218} %
\definecolor{volcanitecol4}{RGB}{251,128,114} %
\definecolor{volcanitecol5}{RGB}{128,177,211} %
\definecolor{volcanitecol6}{RGB}{253,180,98}  %
\definecolor{volcanitecol7}{RGB}{179,222,105} %

\pgfplotscreateplotcyclelist{volcanitecolors}{
  {color=volcanitecol5!60!black, fill=volcanitecol5, no markers},   %
  {color=volcanitecol2!60!black, fill=volcanitecol2, no markers},   %
  {color=volcanitecol4!60!black, fill=volcanitecol4, no markers},   %
  {color=volcanitecol7!60!black, fill=volcanitecol7, no markers},   %
  {color=volcanitecol3!60!black, fill=volcanitecol3, no markers},   %
  {color=volcanitecol1!60!black, fill=volcanitecol1, no markers},   %
  {color=volcanitecol6!60!black, fill=volcanitecol6, no markers},   %
}
\pgfplotsset{cycle list name=volcanitecolors}

\newcommand{\pltvideoframes}[2]{%
    \centering
    \setlength{\fboxrule}{0.25pt}
    \setlength{\fboxsep}{0pt}

    \hfill
    \begin{minipage}[t]{0.23\textwidth}
    \centering
    \fbox{\includegraphics[width=\textwidth]{paper/volcanite/results/video/#1_#2/#1_#2_0020.jpg}}
    {Frame 20}
    \end{minipage}%
    \hfill
    \begin{minipage}[t]{0.23\textwidth}
    \centering
    \fbox{\includegraphics[width=\textwidth]{paper/volcanite/results/video/#1_#2/#1_#2_0200.jpg}}
    {Frame 200}
    \end{minipage}%
    \hfill
    \begin{minipage}[t]{0.23\textwidth}
    \centering
    \fbox{\includegraphics[width=\textwidth]{paper/volcanite/results/video/#1_#2/#1_#2_0400.jpg}}
    {Frame 400}
    \end{minipage}%
    \hfill
    \begin{minipage}[t]{0.23\textwidth}
    \centering
    \fbox{\includegraphics[width=\textwidth]{paper/volcanite/results/video/#1_#2/#1_#2_0580.jpg}}
    {Frame 580}
    \end{minipage}
    \hfill

    \vspace{0.2cm}
}

\newcommand{\pltimageframe}[2]{%
    \begin{minipage}[t]{0.24\textwidth}\vspace{0pt}%
        \centering
        \fbox{\includegraphics[width=\textwidth]{paper/volcanite/results/image/#1_#2.jpg}}

        \vspace{0.1cm}
        {\raggedleft Final image frame \newline \dataNameFromCSV{#1} (#2)}
    \end{minipage}
}

\NewDocumentCommand{\pltimagetimings}{s m m}{%
    \pgfplotstableread[col sep=comma,header=true]{paper/volcanite/results/image/#2_#3_timing.csv}\datatable
        
    \begin{tikzpicture}
    \begin{axis}[
        ybar stacked,
        width=\textwidth,
        height=4cm,
        bar width=6pt,
        enlarge x limits=0.03,
        xtick={0,1,...,20},
        ytick={0,10,...,60},
        ymajorgrids=true,
        major grid style={line width=0.1pt,draw=black!20},
        ylabel={$\varnothing$ frame [ms]},
        xlabel={Image Frame},
        legend style={at={(0.98,0.9)}, anchor=north east,legend columns=-1,align=left,font=\raggedright},
        cycle list name=volcanitecolors,
    ]
    
    \addplot+[fill, select coords between index={0}{20}] table[x expr={\coordindex}, y=Cache] {\datatable};
    \addplot+[fill, select coords between index={0}{20}] table[x expr={\coordindex}, y=Decompress] {\datatable};
    \addplot+[fill, select coords between index={0}{20}] table[x expr={\coordindex}, y=Render] {\datatable};
    \addplot+[fill, select coords between index={0}{20}] table[x expr={\coordindex}, y={Post-Process}] {\datatable};
    \addplot+[fill, select coords between index={0}{20},stack plots=false] table[x expr=\coordindex, y=Total] {\datatable};

    \IfBooleanTF{#1}{}{\legend{Cache,Decompress,Render,Post-Process, Other}}
    \end{axis}
    \end{tikzpicture}
}

\NewDocumentCommand{\pltimagetimingsandframe}{s m m}{%

    \begin{minipage}[t]{0.73\textwidth}\vspace{0pt}%
        \vspace{-0.2cm}
        \includegraphics[width=\columnwidth]{results/plots/gpu-image-timings_#2_#3.pdf}
    \end{minipage}%
    \begin{minipage}[t]{0.24\textwidth}\vspace{0pt}%
        \centering
        \fbox{\includegraphics[width=\textwidth]{paper/volcanite/results/image/#2_#3.jpg}}
        \caption*{\raggedleft Final image frame \newline \dataNameFromCSV{#2} (#3)}
    \end{minipage}
}

\newcommand{\supvideooffline}{Video A1}
\newcommand{\supvideocells}{Video A2}

\usepackage[linecolor=red,backgroundcolor=red!10!white,bordercolor=red!75!black,textcolor=red,size=footnotesize]{todonotes}
\let\old@todo\@todo
\RenewDocumentCommand{\todo}{O{} m}{%
  \IfValueTF{#1}{
    \IfStrEq{#1}{inline}{%
      \textcolor{red}{TODO: #2}%
    }{%
      \tikzexternaldisable\old@todo[#1]{TODO: #2}\tikzexternalenable%
    }%
  }{
    \tikzexternaldisable\old@todo{TODO: #2}\tikzexternalenable%
  }%
}

\usepackage{tcolorbox}
\AtBeginEnvironment{tcolorbox}{\footnotesize}
\newtcolorbox{todobox}{colback=red!10!white,colframe=red!75!black,coltext=red}

\title{\bfseries Volcanite: commodity-hardware segmentation volume\\ visualization for connectomics and beyond \vspace{0.5cm}}

\author{%
  Max Piochowiak$^{1,2}$ \and
  Reiner Dolp$^{1,2}$ \and
  Julian Herold$^{2,4}$ \and
  Eric Behle$^{3}$ \and
  Carsten Dachsbacher$^{1}$ \and
  Alexander Schug$^{3,4*}$
}
\date{}

\newcommand{\affiliations}{%
  \begingroup\setlength{\parskip}{0pt}\setlength{\parindent}{0pt}
  {\small
  $^{1}$Institute for Visualization and Data Analysis, Karlsruhe Institute of Technology, Karlsruhe, Germany\\
  $^{2}$HIDSS4Health -- Helmholtz Information and Data Science School for Health, Karlsruhe / Heidelberg, Germany\\
  $^{3}$J\"ulich Supercomputing Center, Research Center J\"ulich, J\"ulich, Germany\\
  $^{4}$Scientific Center for Computing, Karlsruhe Institute of Technology, Karlsruhe, Germany%
  }
  \par\medskip
  \texttt{max.piochowiak@kit.edu}; \texttt{reiner.dolp@kit.edu}; \texttt{julian.herold@kit.edu}\\
  \texttt{eric.behle@hhu.de}; \texttt{dachsbacher@kit.edu}; \texttt{schug@kit.edu};
  \endgroup
}

\begin{document}

\makeatletter
\twocolumn[
  \begin{@twocolumnfalse}
    \maketitle
    \begin{center}
      \affiliations
    \end{center}
    \vspace{1.5em}
    \begin{abstract}
      \noindent
      Modern imaging produces terabyte-scale segmentation volumes, assigning each voxel an object label. These categorical, boundary-sensitive and label-rich data underpin connectomics and other imaging-driven fields, yet their scale often forces interpretation through slices, approximate meshes or distributed workflows that obscure spatial context and voxel-level defects. Here we show that such volumes can be explored directly on commodity hardware with \emph{Volcanite}, an open-source framework for dense-segmentation rendering. Combining compression-aware data handling, a Vulkan GPU backend and segmentation-specific rendering, \emph{Volcanite} enables low-latency exploration without meshing or distributed infrastructure, including for unpublished, sensitive or proprietary data. Across multi-domain datasets, it preserves voxel labels, adds shadows and global illumination, and renders up to half a trillion voxels at 100 frames per second. By replacing lengthy preprocessing with direct inspection, \emph{Volcanite} decouples hypothesis generation from data preparation and turns teravoxel label fields into interactive evidence for discovery, validation and cross-domain spatial analysis.

      \medskip
      \noindent
      \textbf{Keywords:} Segmentation, Rendering, Compression, Voxel Data, Volumes, Visualization, Connectomics
    \end{abstract}
    \vspace{1.4cm}
  \end{@twocolumnfalse}
]
\makeatother

Many scientific breakthroughs begin with seeing structure clearly, but modern imaging increasingly outpaces our ability to interpret three-dimensional organization at scale.
In many imaging-driven domains, this information takes the form of segmentation volumes, in which each voxel is assigned to a labeled object or region.
Such volumes arise across domains ranging from connectomics and computational biology to medicine, materials science, industrial tomography, computer vision and paleontology~\cite{Shapson-Coe:2024:h01,Motta:2019:DCR,Behle:2026:mousebrain,Isensee:2021:nnu,Maurer:2022:Fiber,Gruber:2024:Me163,Tian:2023:occ3d,Ikegami:2025:fossil,Zhao:2025:medSegFoundation}.
In connectomics in particular, recent advances in imaging and segmentation have driven volume sizes to scales that challenge interactive three-dimensional analysis on conventional systems~\cite{Shapson-Coe:2024:h01,Dorkenwald:2025:CAVE}.

Yet for many scientific questions, central challenges towards interpretation and understanding encompass not only storage and processing of these data, but also the interactive exploration of their three-dimensional spatial organization.
In practice, this often relies on coarse slice-based inspection for dense segmentation data or on mesh-based views that require intense preprocessing and substantial backend infrastructure, rather than on direct voxel-precise rendering of large labeled volumes~\cite{neuroglancer, Berger:2018:VAST, Fedorov:2012:using3Dslicer, Boergens:2017:webknossos, Hulbert:2021:PainteraProposal, Zhang:2024:MarchingWindows}.
The problem is compounded by the fact that densely labeled segmentation volumes impose requirements that differ fundamentally from those of conventional scalar volume rendering, including categorical rather than interpolable values, large label spaces and the need for precise boundary-preserving visualization~\cite{paraview, Beyer:2015:STARLS, Ljung:2016:TF, Shapson-Coe:2024:h01, Kenney:2021:azba}.
As a result, interactive visualization of large segmentation volumes at full voxel precision remains difficult on widely accessible single-system hardware~\cite{Weissenboeck:2014:FiberScout, Lesar:2022:VC, Beyer:2013:ETC, Ayachit:2015:ParaviewGuide, Moreland:2011:IceT}.

To address this gap, we present \emph{Volcanite}, an interactive visualization system for large segmentation volumes designed for voxel-precise rendering of terascale labeled data on commodity hardware.
Volcanite combines compression-aware data handling~\cite{Piochowiak:2024:csgv}, a modern Vulkan-based GPU backend~\cite{Vulkan:2024:Spec} and rendering methods tailored specifically to dense segmentation volumes.
This design supports high-quality interactive exploration without requiring multi-system streaming architectures or specialized HPC installations~\cite{Beyer:2013:ETC, Ayachit:2015:ParaviewGuide, Moreland:2011:IceT}. In our evaluation, Volcanite renders up to half a trillion voxels at 100 frames per second on a regular desktop system.
We release \emph{Volcanite} as open-source software to provide a practical resource for domain scientists and a reproducible basis for future work on segmentation volume visualization.

In summary, our key contributions are:
\begin{itemize}
    \item addressing dense segmentation volumes as a distinct volume visualization problem that is not well served by existing scalar-volume or mesh-based workflows, %
    \item introducing an integrated rendering framework for voxel-precise interactive visualization of terascale segmentation volumes on commodity hardware through compression-aware data handling and segmentation-specific rendering,
    \item enabling accessible exploratory three-dimensional analysis of large segmentation volumes without slow meshing pipelines or specialized distributed infrastructure, and
    \item providing open-source software and a curated multi-domain benchmark dataset for reproducible evaluation of large-scale segmentation volume visualization.
\end{itemize}

\begin{figure*}[ht]
    \begin{subfigure}{0pt}
        \phantomcaption\label{fig:vcnt:visual-abstract:gui}
    \end{subfigure}
    \begin{subfigure}{0pt}
        \phantomcaption\label{fig:vcnt:visual-abstract:rendering}
    \end{subfigure}
    \begin{subfigure}{0pt}
        \phantomcaption\label{fig:vcnt:visual-abstract:image-timings}
    \end{subfigure}
    \begin{subfigure}{0pt}
        \phantomcaption\label{fig:vcnt:visual-abstract:h01wm}
    \end{subfigure}
\begin{subfigure}{0pt}
        \phantomcaption\label{fig:vcnt:visual-abstract:shading}
    \end{subfigure}
    
    \centering
    \includegraphics[width=\linewidth]{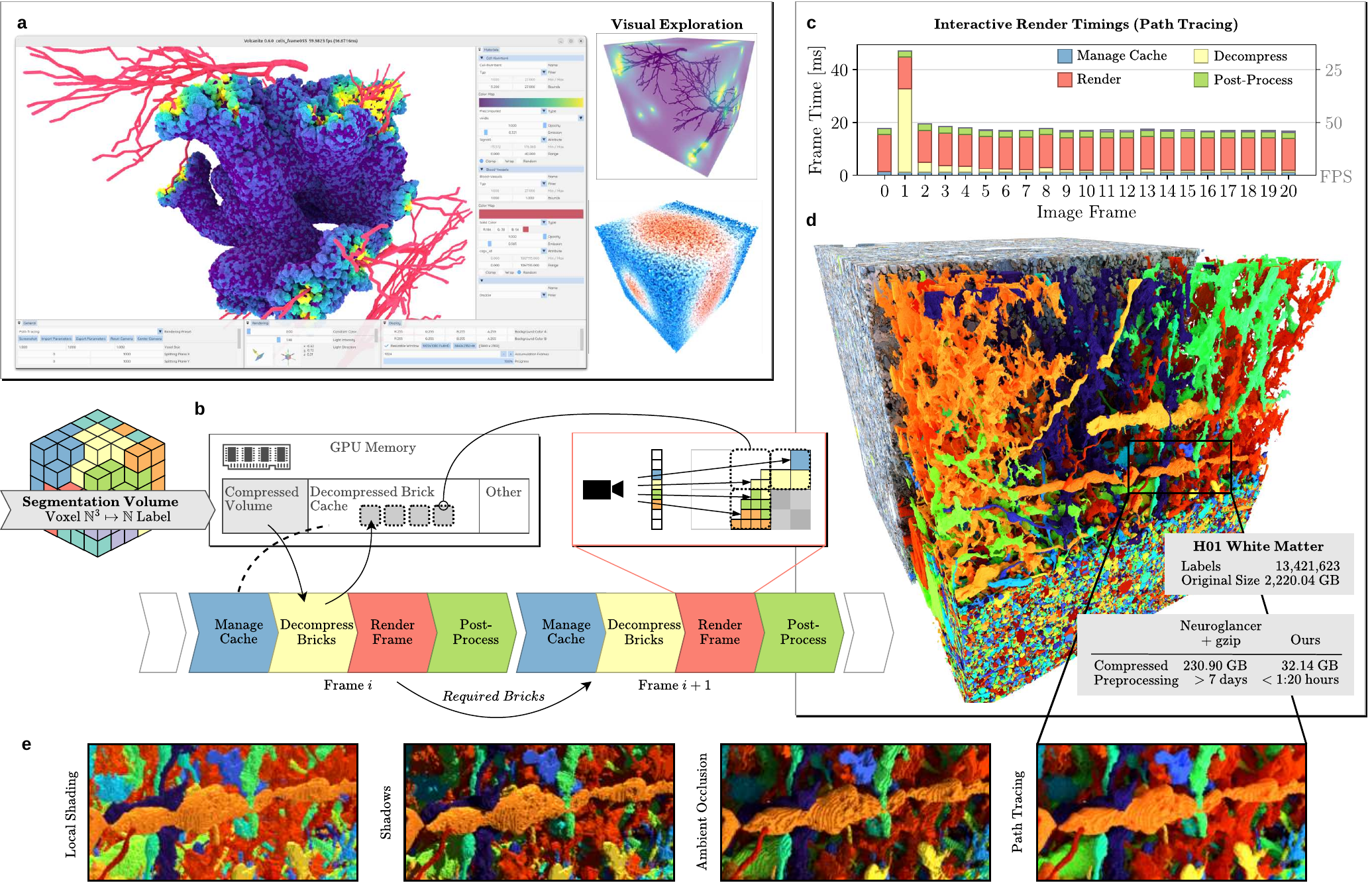}

    \vspace{1em}

    \caption{Visual Abstract. \subref{fig:vcnt:visual-abstract:gui} Our interactive application Volcanite allows visual exploration of large segmentation volumes on commodity hardware leveraging state-of-the-art lossless compression.
    \subref{fig:vcnt:visual-abstract:rendering} Our voxel-precise \gls{gpu}-based rendering pipeline decompresses visible volume regions into a cache for fast access.
    This pipeline enables interactive exploration of different rendering configurations down to examining single voxels (right).
    \subref{fig:vcnt:visual-abstract:image-timings} Rendering is interactive at milliseconds per rendered frame. First frames have higher decompression loads that quickly vanish.
    \subref{fig:vcnt:visual-abstract:h01wm} With Volcanite, we are able to interactively explore a 2.2 TB section of the H01 human cortex white matter segmentation volume with Full-HD path-traced illumination at 66 frames per second on a commodity hardware system at a significantly faster time to first frame compared to other methods.
    Generating imprecise meshes for visualization in Neuroglancer takes over a week, while our preprocessing creates significantly smaller and lossless encodings in under two hours on our system.
    \subref{fig:vcnt:visual-abstract:shading} Shading modes in Volcanite can be changed from fast local shading and high-quality path tracing with immediate feedback.
    }   
    \label{fig:vcnt:visual-abstract}

    \vspace{2em}
\end{figure*}

\section*{Results}
Segmentation volumes are 3D grids that store an object or instance label per element (voxel) as $V: \mathbb{N}^3 \to \mathbb{N}$.
The (usually contiguous) label regions segment the space, for example into individual neurons in a connectome.
We present our Volcanite segmentation volume renderer in the following.

\subsection*{Novel integrated segmentation volume rendering framework}
\Cref{fig:vcnt:visual-abstract} shows a segmentation volume visualization workflow with Volcanite.
First, Volcanite compresses a segmentation volume into our \gls{csgv} format (previously published in~\cite{Piochowiak:2024:csgv}), the current strongest lossless compressor for segmentation volumes.
Compression times mostly depend on reading input files from disk (\cref{tab:vcnt:tools-preprocess}) with an encoding throughput at Gigabytes per second.
Apart from visualizations, Volcanite enables compact volume storage. %
The rendering pipeline to create all visualizations is, except for special cases (\sectionref{sec:vcnt:renderer:detail-streaming}{detail streaming}), fully GPU-based for high performance (\cref{fig:vcnt:visual-abstract:rendering}).
\gls{csgv} encoded volumes are stored fully in \gls{gpu} \gls{vram} with on-demand decompression of visible areas into a cache.

Employing this pipeline, Volcanite's graphics application (\cref{fig:vcnt:visual-abstract:gui}) enables interactive exploration of the data at milliseconds per frame (\cref{fig:vcnt:visual-abstract:image-timings}). %
Interactive exploration is essential as desired visualization configurations are not inherently clear for unknown volumes.
Other tools usually only render large volumes as 2D slices, and thus at reduced spatial clarity~\cite{neuroglancer,Boergens:2017:webknossos,Berger:2018:VAST}.
If 3D visualizations are provided, these are limited to subsets or meshed 
approximations of the label regions~\cite{neuroglancer,Plaza:2022:neuprint,Boergens:2017:webknossos,Pieper:2004:3dslicer}.
Volcanite renders the data in 3D as it is, with full, lossless access to each single voxel.

High quality rendering can be performed interactively or through Volcanite's \gls{cli}.
Existing workflows usually require switching data representations and tools for advanced rendering.
For example, after exploration in ParaView or Neuroglancer without global illumination support, data is manually transferred to Blender~\cite{Velicky:2023:useBlender, Blender:2018}. %
Volcanite's \gls{cli} also enables remote execution or batch processing.
\supvideooffline{} shows the \cortex{} volume rendered offline through Volcanite's \gls{cli},
\supvideocells{} an interactive exploration of \cells{}.

\subsection*{Enabling scientific insights through segmentation volume rendering}

\begin{figure*}[pt]

    \begin{minipage}[t]{0.33\textwidth}
        \begin{subfigure}[t]{\textwidth}
            \caption{}
            \label{fig:vcnt:toolsrender:vtk-renderings}

            \vspace{-0.5cm}
            
            \begin{center}        
                \includegraphics[width=\linewidth,trim={7cm 2cm 13cm 2cm},clip]{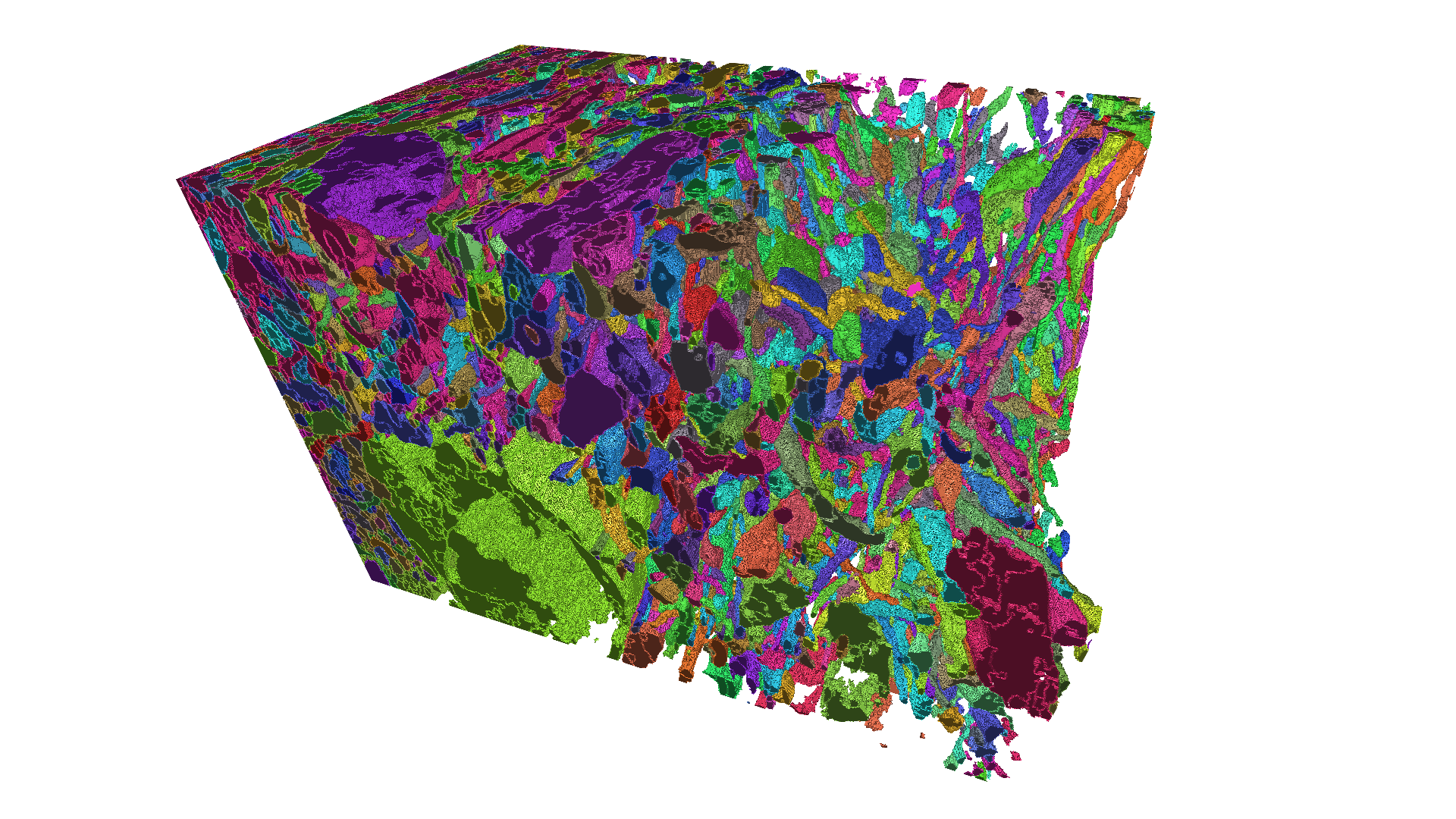}%
            
                VTK

                \vspace{0.5cm}
                
                \includegraphics[width=\linewidth,trim={7cm 2cm 13cm 2cm},clip]{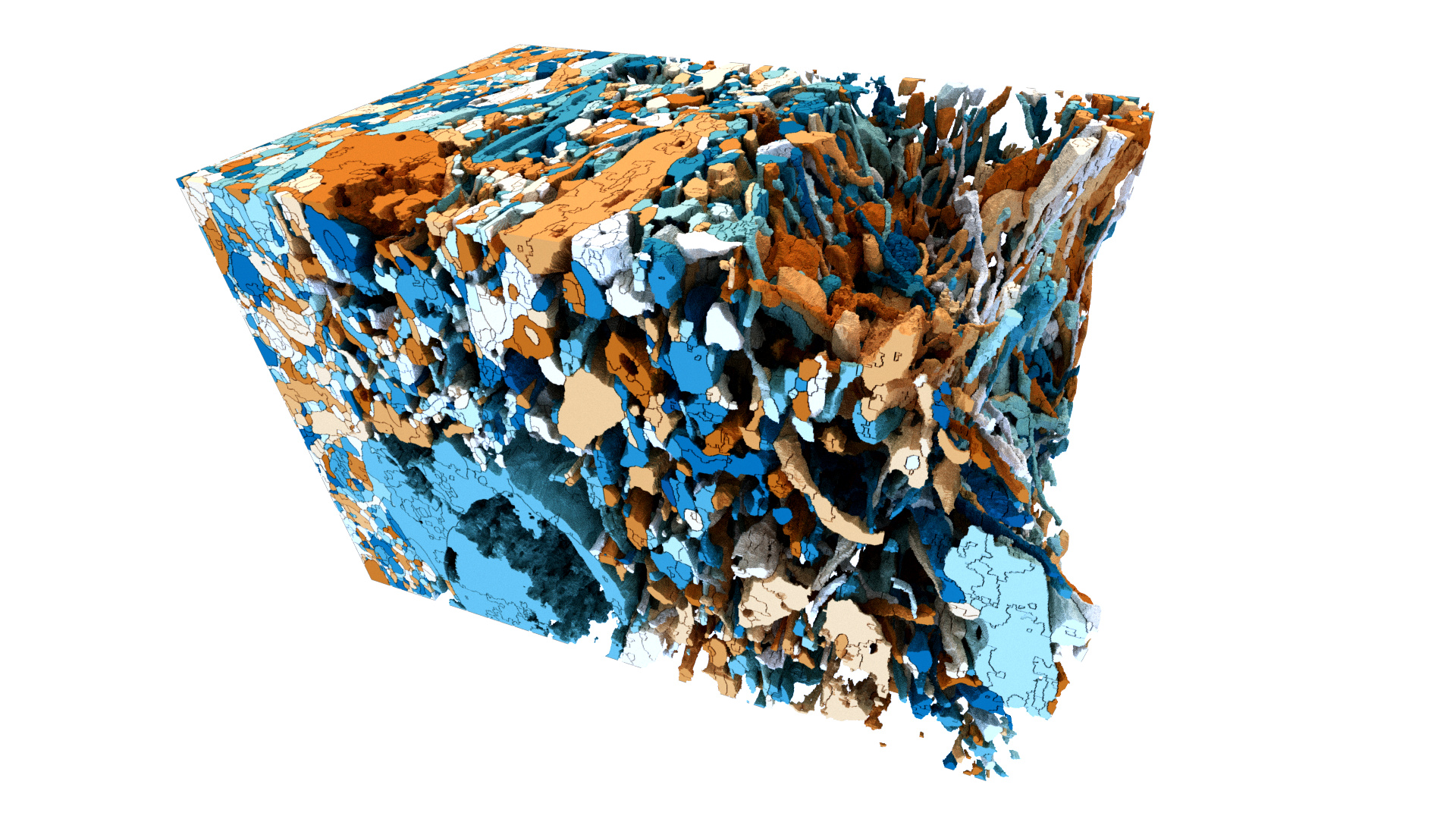}
                
                Volcanite (pt)
            \end{center}

        \end{subfigure}%
    \end{minipage}
    \begin{minipage}[t]{0.66\textwidth}

        \begin{subfigure}[t]{\linewidth}
            \caption{}
            \label{fig:vcnt:toolsrender:inaccuracy}     

            \vspace{-0.5cm}

            \begin{minipage}[t]{0.5\linewidth}
                \centering
                \includegraphics[width=\linewidth,trim={15cm 3cm 15cm 3cm},clip]{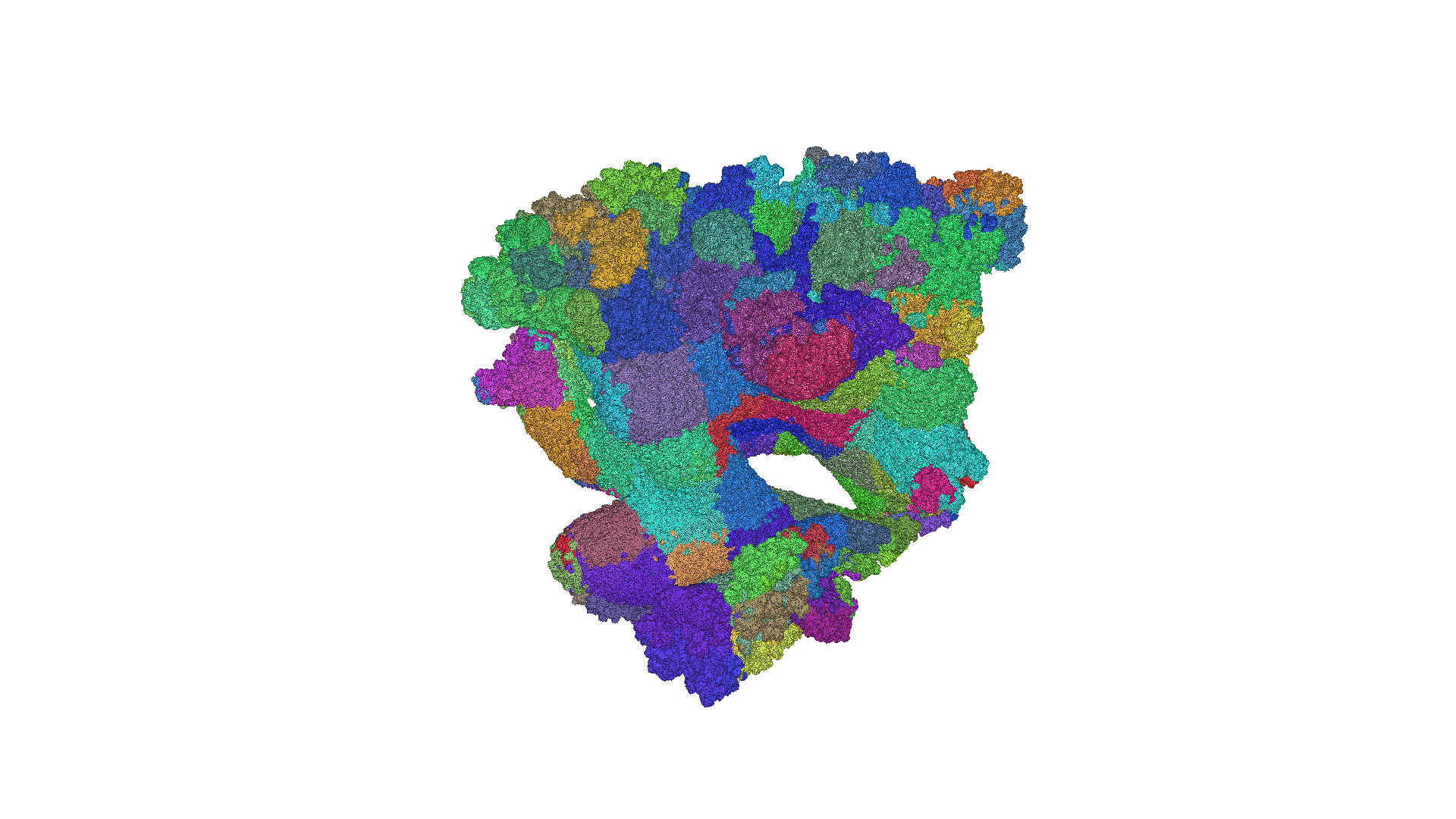}%

                VTK
            \end{minipage}
            \begin{minipage}[t]{0.5\linewidth}
                \centering
                \includegraphics[width=\linewidth,trim={15cm 3cm 15cm 3cm},clip]{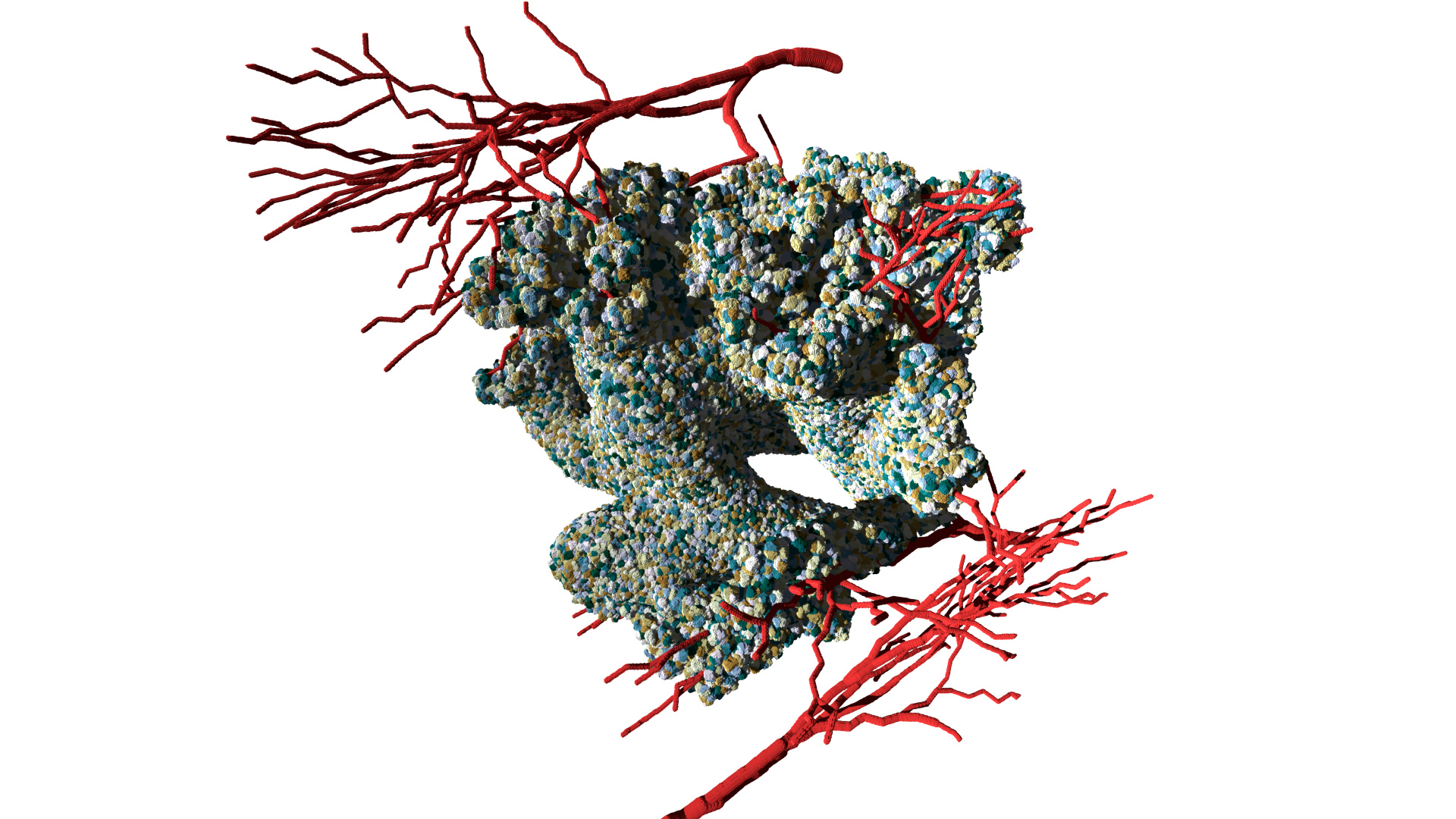}

                Volcanite (shadows)
            \end{minipage}

        \end{subfigure}

        \begin{subfigure}[t]{\linewidth}
            \vspace{-0.4cm}
            \caption{}
            \label{fig:vcnt:toolsrender:closeup-vtk}

            \hfill
            \begin{minipage}[t]{0.49\linewidth}               
                \centering 
                
                \magnifyBottom{
                    \includegraphics[width=0.99\linewidth,trim={0 10cm 0 0},clip]{results/vtk/Wolny2020-closeup.png}
                }{8}{0.2}{0.6}{0.65}{0.7}{}{} %

                \vspace{-0.3cm}

                VTK
            \end{minipage}%
            \hfill
            \begin{minipage}[t]{0.49\linewidth}
                \centering
                
                \magnifyBottom{
                    \includegraphics[width=0.99\linewidth,trim={0 10cm 0 0},clip]{results/vtk/Wolny2020-closeup-volcanite-pt.jpg}
                }{8}{0.2}{0.6}{0.65}{0.7}{}{} %

                \vspace{-0.3cm}

                Volcanite (path tracing)
            \end{minipage}

        \end{subfigure}
        
    \end{minipage}

    \vspace{-0.5cm}
    
    \begin{subfigure}[t]{\textwidth}
        \caption{}
        \label{fig:vcnt:toolsrender:alignment-compare}
        \centering
        
        \hfill
        \begin{minipage}[t]{0.48\textwidth}
            \centering
            
            \colorlet{oldpylenscolor}{spylenscolor}
            \colorlet{oldspylensleftcolor}{spylensleftcolor}
            \colorlet{oldpylensrightcolor}{spylensrightcolor}

            \colorlet{spylenscolor}{white}
            \colorlet{spylensleftcolor}{white}
            \colorlet{spylensrightcolor}{white}
        
            \magnifyBottom{
                \includegraphics[width=0.99\textwidth,trim={10cm 0cm 10cm 2cm},clip]{paper/volcanite/fig/neuroglancer-cmp/435_mini_subs2_neuroglancer.jpg}
            }{2}{0.48}{0.6}{0.5}{0.8}{}{} %

            \colorlet{spylenscolor}{oldpylenscolor}
            \colorlet{spylensleftcolor}{oldspylensleftcolor}
            \colorlet{spylensrightcolor}{oldpylensrightcolor}

            \vspace{-0.3cm}

            Google Neuroglancer (Mesh)
        \end{minipage}%
        \hfill
        \begin{minipage}[t]{0.48\textwidth}
            \centering
            \magnifyBottom{
                \includegraphics[width=0.99\textwidth,trim={13cm 2cm 10cm 5cm},clip]{paper/volcanite/fig/neuroglancer-cmp/435_mini_subs2_volcanite.jpg}
            }{2}{0.43}{0.57}{0.45}{0.70}{}{} %

            \vspace{-0.3cm}
            
            Volcanite (path tracing)
        \end{minipage}
        \hfill

        \vspace{-6cm}
        {            
            \centering
            \setlength{\fboxsep}{0pt}
            \setlength{\fboxrule}{1pt}
            \fbox{%
            \includegraphics[width=0.2\linewidth,trim={270pt 0 300pt 0},clip]{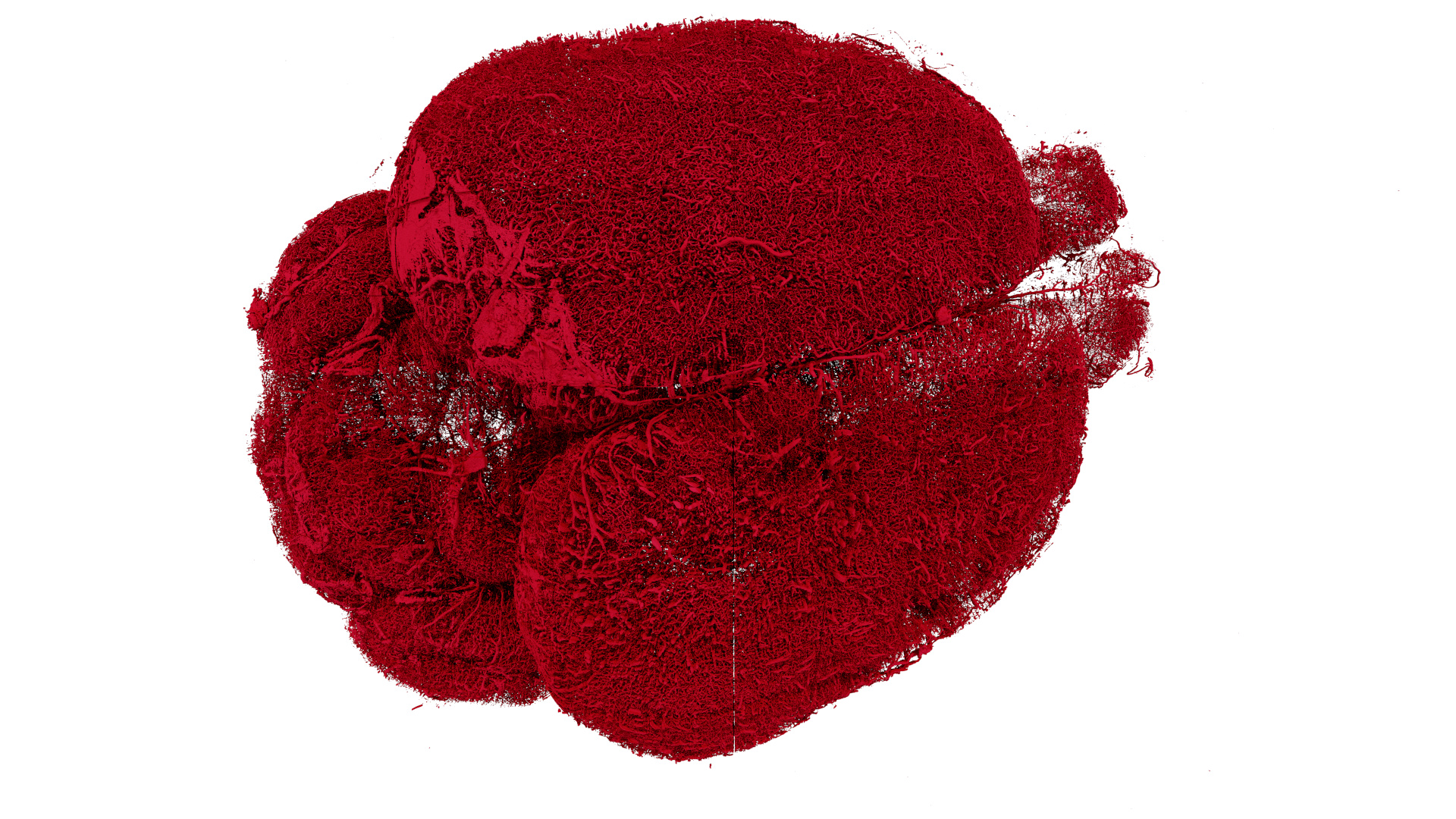}%
            }
        }
        \vspace{3.8cm}

    \end{subfigure}

    \caption{%
    3D rendering in VTK and Neuroglancer can exhibit inaccuracies and reduced spatial perception compared to our method (Volcanite).
    \subref{fig:vcnt:toolsrender:vtk-renderings} VTK does not support global illumination, limiting spatial perception. Volcanite supports indirect lighting and shadows with path tracing.
    \subref{fig:vcnt:toolsrender:inaccuracy} VTK's voxel transfer function is designed for quantitative (grayscale) volumes and inaccurately samples densely labeled segmentation data. The filtered undersampling misses the blood vessels in \cells{}.
    \subref{fig:vcnt:toolsrender:closeup-vtk} VTK's ray marching creates artifacts through inaccurate voxel sampling. Volcanite's improved DDA traversal does not.
    \subref{fig:vcnt:toolsrender:alignment-compare} Neuroglancer and Volcanite rendering of a faulty mouse brain vascular segmentation (inset) chunk alignment. Voxel errors (flat cuts) are less visible in Neuroglancer's 3D mesh approximations compared to our voxel-precise rendering. Again, missing global illumination (shadows) impedes spatial perception in Neuroglancer.
    }
    
    \label{fig:vcnt:toolsrender}
\end{figure*}

\begin{figure*}[pt]
    \begin{subfigure}[t]{\linewidth}
        \caption{}
        \label{fig:vcnt:tools:vtk-timings}

        \vspace{-0.2cm}

        \includegraphics[width=\linewidth]{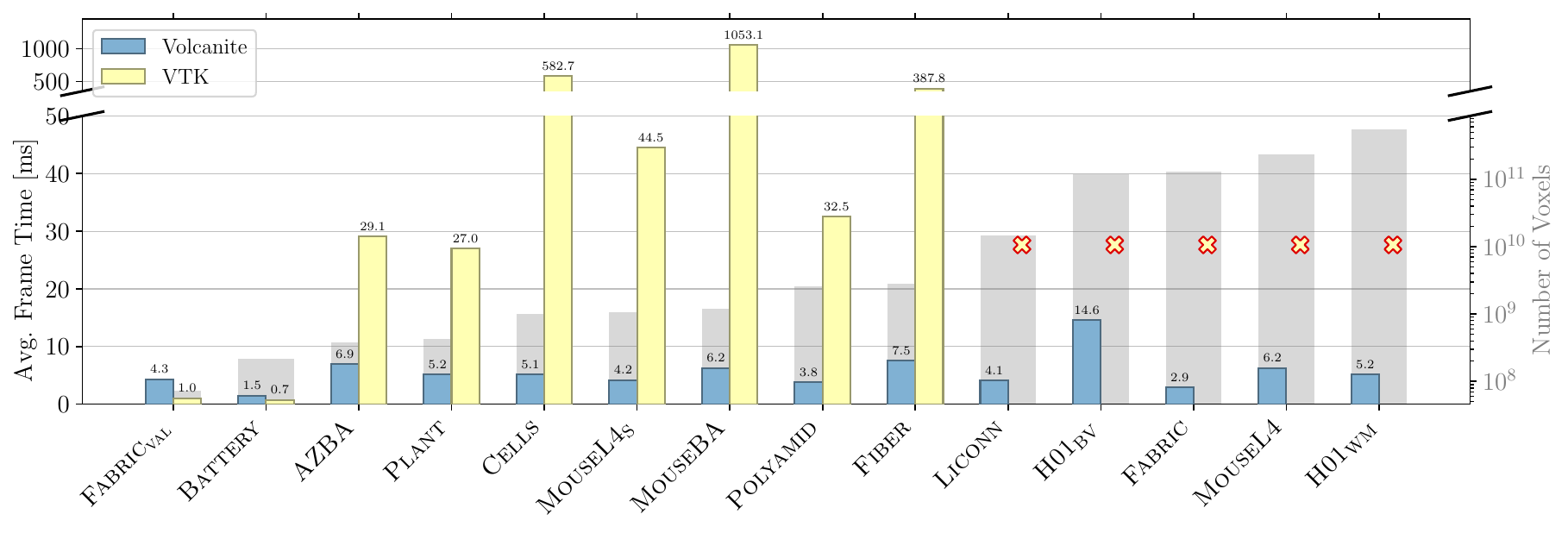}
    \end{subfigure}

    \vspace{-0.4cm}

    \begin{subfigure}[t]{\linewidth}
        \centering
        \caption{}
        \label{fig:vcnt:tools:azba-comparison}

        \begin{minipage}[t]{0.31\linewidth}
            \centering
            Volcanite
            \includegraphics[width=\linewidth,trim={2cm 0 10cm 0},clip]{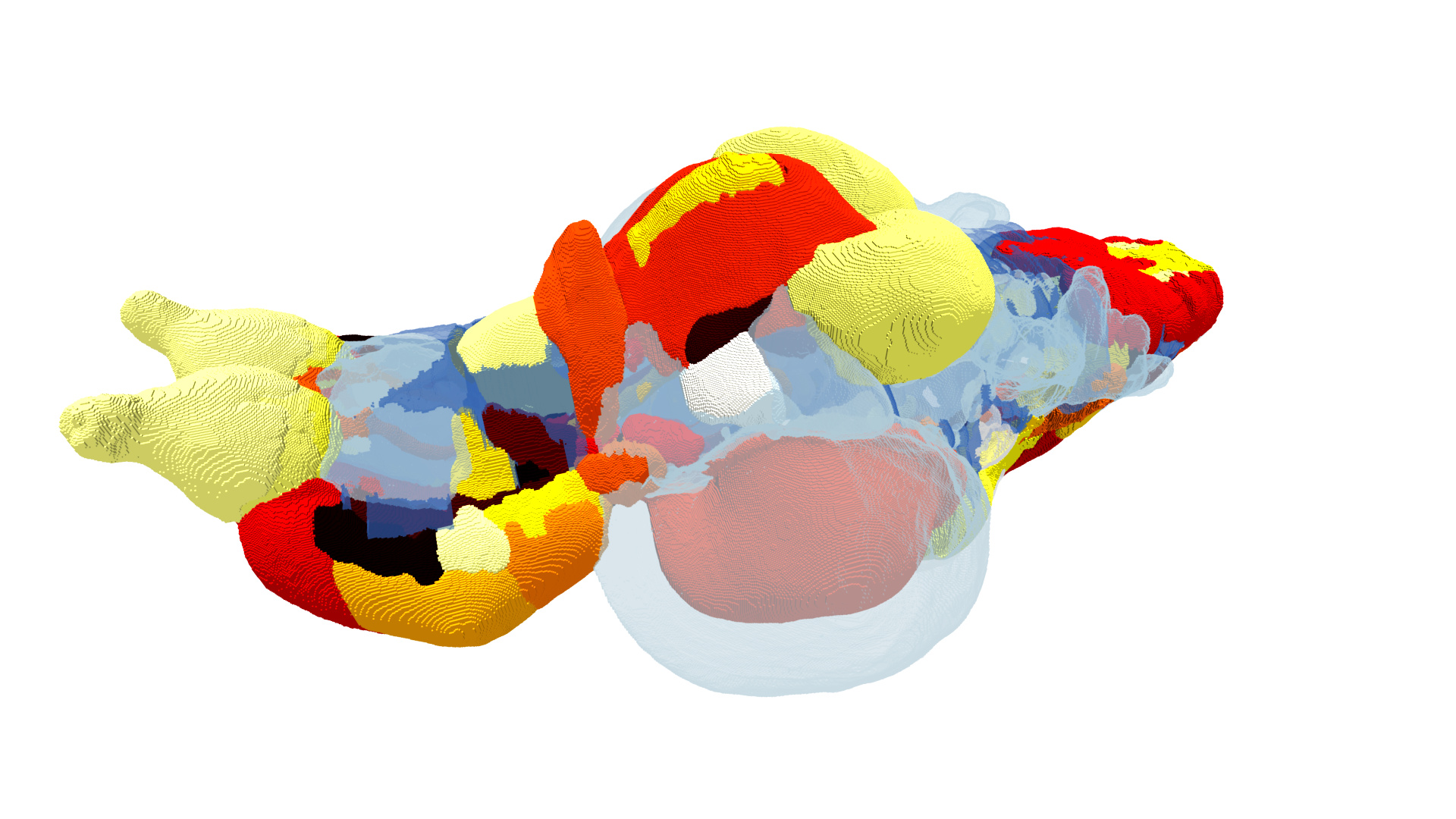}
        \end{minipage}
        \hfill
        \begin{minipage}[t]{0.31\linewidth}
            \centering
            VTK / ParaView
            
            \includegraphics[width=\linewidth,trim={2cm 0 10cm 0},clip]{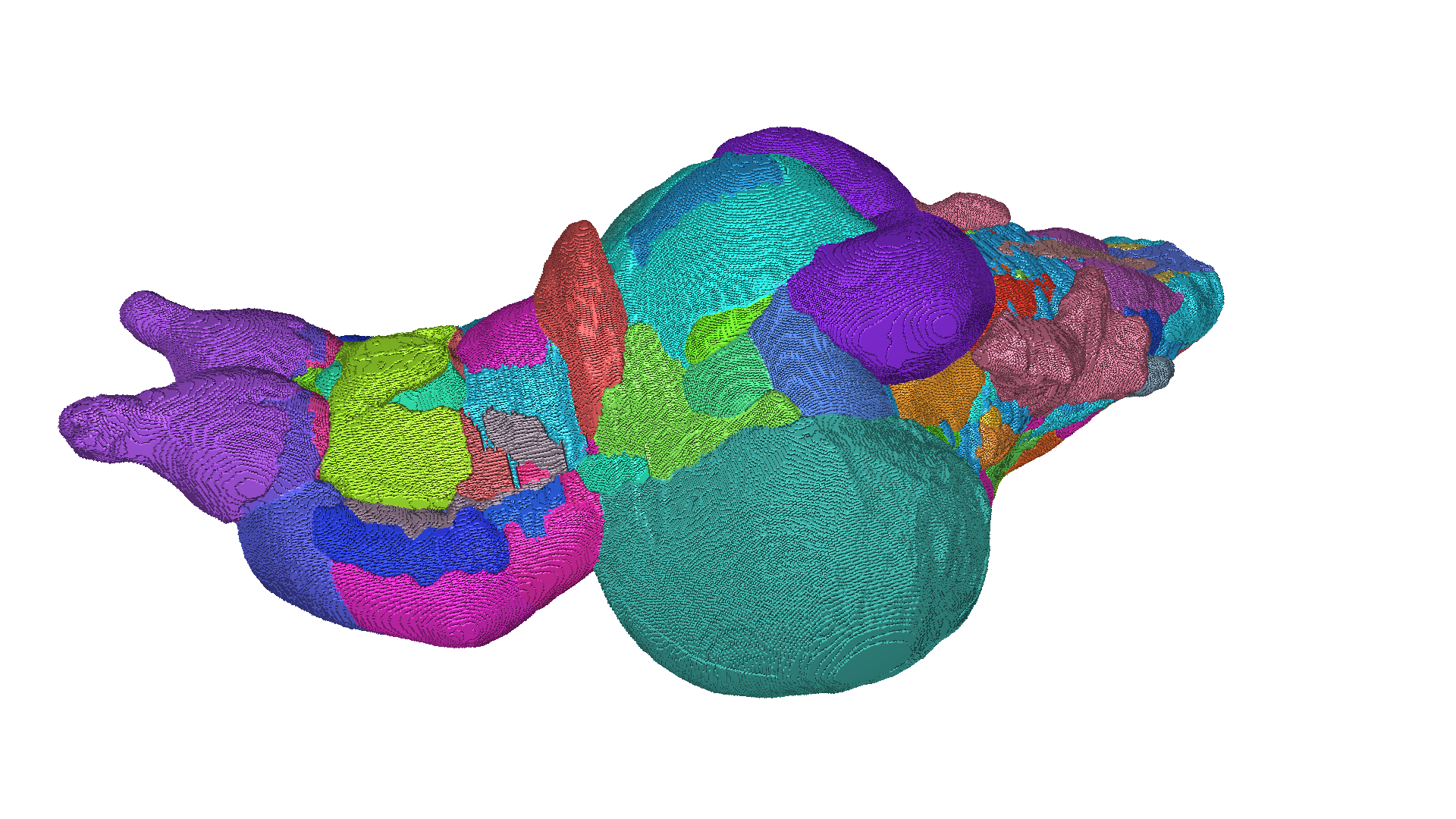}
        \end{minipage}
        \hfill
        \begin{minipage}[t]{0.31\linewidth}
            \centering
            Neuroglancer

            \vspace{0.3em}
            
            \includegraphics[width=\linewidth,trim={0 0 0 0cm},clip]{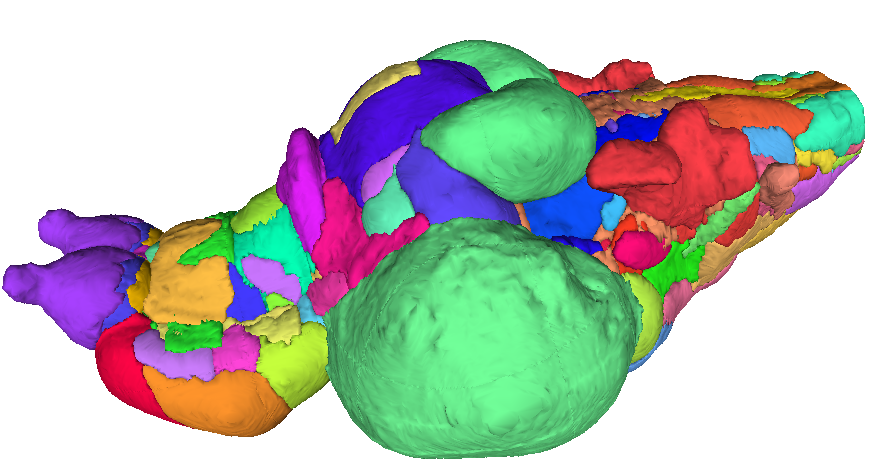}
        \end{minipage}
    \end{subfigure}

    \vspace{-0.7cm}

    \begin{subfigure}[t]{\textwidth}
        \caption{}
        \label{fig:vcnt:tools:preprocessing-times}

        \vspace{-0.3cm}
                
        \includegraphics[width=\columnwidth]{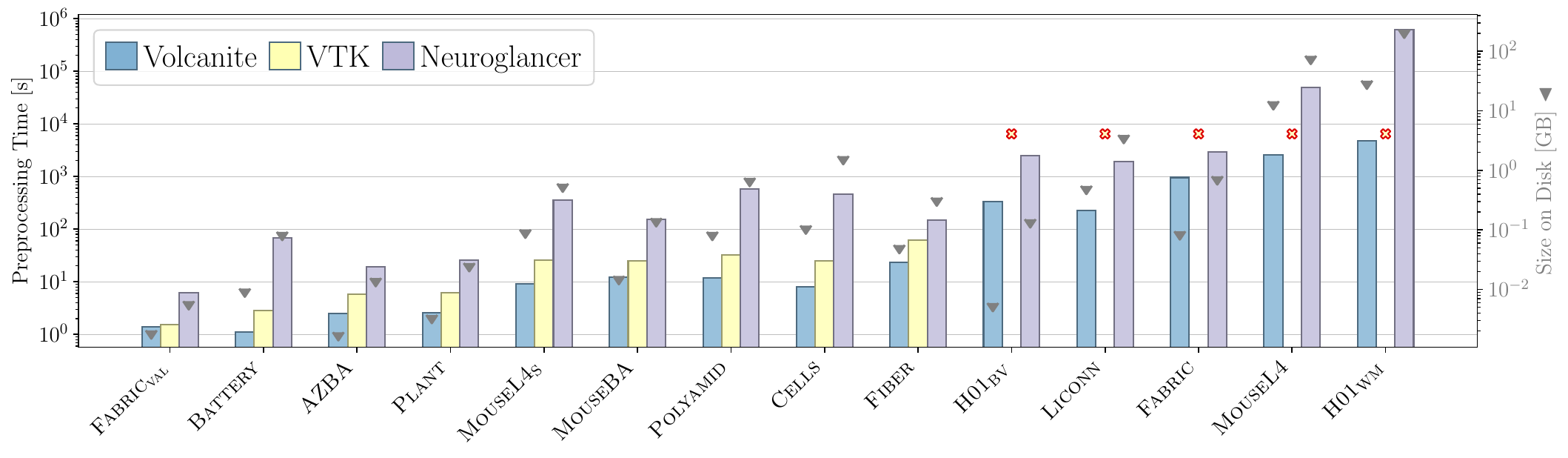}
    \end{subfigure}

    \caption{%
    \subref{fig:vcnt:tools:vtk-timings} Frame timings for voxel-based rendering with Volcanite (blue) and VTK (yellow) with local shading from a fixed camera (lower is better). Large data cannot be rendered due to memory limits. Frame timings were not available for Neuroglancer.
    \subref{fig:vcnt:tools:azba-comparison} \azba{} rendered with matching perspectives and simple local shading as in \subref{fig:vcnt:tools:vtk-timings} in Volcanite and VTK and in a similar configuration in Neuroglancer. Since VTK direct segmentation volume rendering and Neuroglancer only support local shading, we compare against Volcanite's local shading in performance measurements as well, even though Volcanite supports shadows and indirect lighting.
    \subref{fig:vcnt:tools:preprocessing-times} Even without secondary gzip compression as applied for Neuroglancer, Volcanite stores the smallest data and has the shortest preprocessing times overall.%
    }
    
    \label{fig:vcnt:tools}
\end{figure*}

\Cref{fig:vcnt:toolsrender:alignment-compare} presents a 3D rendering of a stitched blood vessel segmentation in a mouse brain~\cite{Behle:2026:mousebrain}, reconstructed from individual 2D microscopy image patches as described in \cite{DiGiovanna:2018:wholeBrainReconstruction}. This stitching process critically depends on precise alignment of the individual patches across all three spatial dimensions. During development of this pipeline, Volcanite was critical for detecting and diagnosing alignment errors in the reconstructed volume.

Using Volcanite’s voxel-precise rendering, misalignments can be identified immediately through clearly visible discontinuities in the blood vessels (center). In contrast, the same errors are significantly less apparent when visualizing the data as 3D meshes in Neuroglancer. This is due to the required mesh simplification (using Igneous default parameters~\cite{Silversmith:2022:igneous}), which smooths fine surface details of labeled regions. Such simplification is necessary in mesh-based rendering approaches to control file sizes, which would otherwise exceed those of compressed voxel representations by orders of magnitude~\cite{Zhang:2024:MarchingWindows,Hulbert:2021:PainteraProposal}. When simplification includes removing regions below a voxel-count threshold (an optional Igneous parameter), even single-voxel errors may be entirely eliminated from the visualization. Additionally, the absence of global illumination and shadowing further reduces spatial perception in the densely structured vascular data when using Neuroglancer.

Beyond visual fidelity, these differences have direct practical implications. Volcanite not only improves visualization quality but also accelerates the analysis workflow: its substantially reduced rendering times enable rapid iteration and interactive exploration, eliminating the need to wait hours for individual renderings. This responsiveness is critical during pipeline development and debugging. Furthermore, alternative voxel-based visualization tools, such as ParaView, are unable to handle datasets at the full data scale (up to 163 GB at 1 bit/voxel), whereas Volcanite remains capable of interactively rendering and inspecting such large volumes.

\subsection*{Effective segmentation volume material editor}
\begin{figure*}[pt]
    \centering
    
    \begin{subfigure}[t]{0.52\textwidth}
        \caption{}
        \label{fig:vcnt:renderer:transferfunction:mapping}
        \centering
        \includegraphics[width=0.9\textwidth]{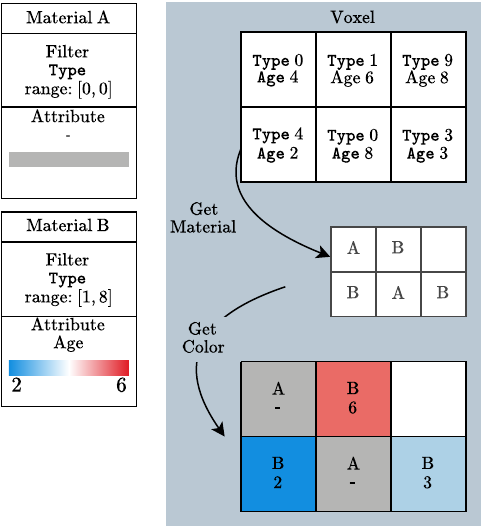}
    \end{subfigure}
    \hfill
    \begin{subfigure}[t]{0.47\textwidth}
        \caption{}
        \label{fig:vcnt:renderer:transferfunction:preprocessing}
        \centering
        \includegraphics[width=\textwidth,trim={0 0 0 0.9cm},clip]{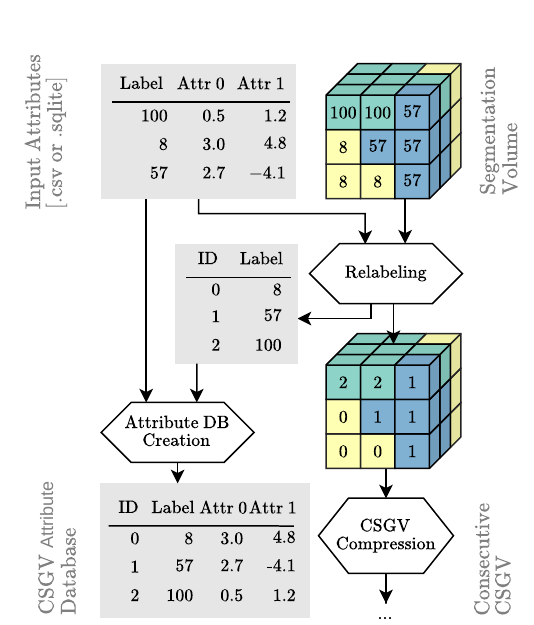}
    \end{subfigure}

    \begin{subfigure}[t]{0.52\textwidth}
        \caption{}
        \label{fig:vcnt:renderer:transferfunction:rendering}
        \begin{center}
            \includegraphics[width=\textwidth,trim={0 20cm 0cm 0},clip]{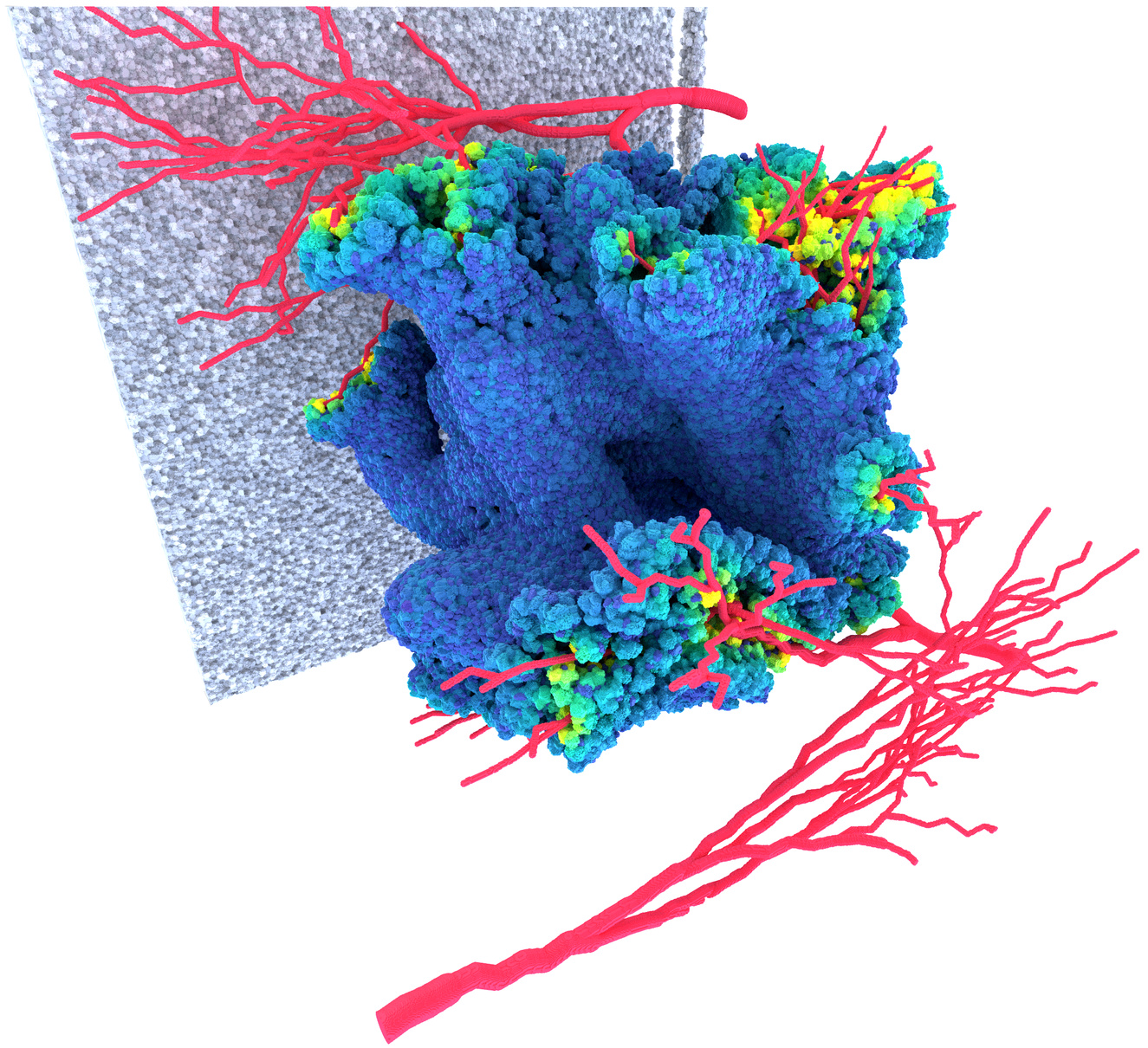}
        \end{center}
    \end{subfigure}
    \hfill
    \begin{subfigure}[t]{0.47\textwidth}
        \caption{}
        \label{fig:vcnt:renderer:transferfunction:editor}
        \includegraphics[width=\textwidth,trim={0 0cm 0 0},clip]{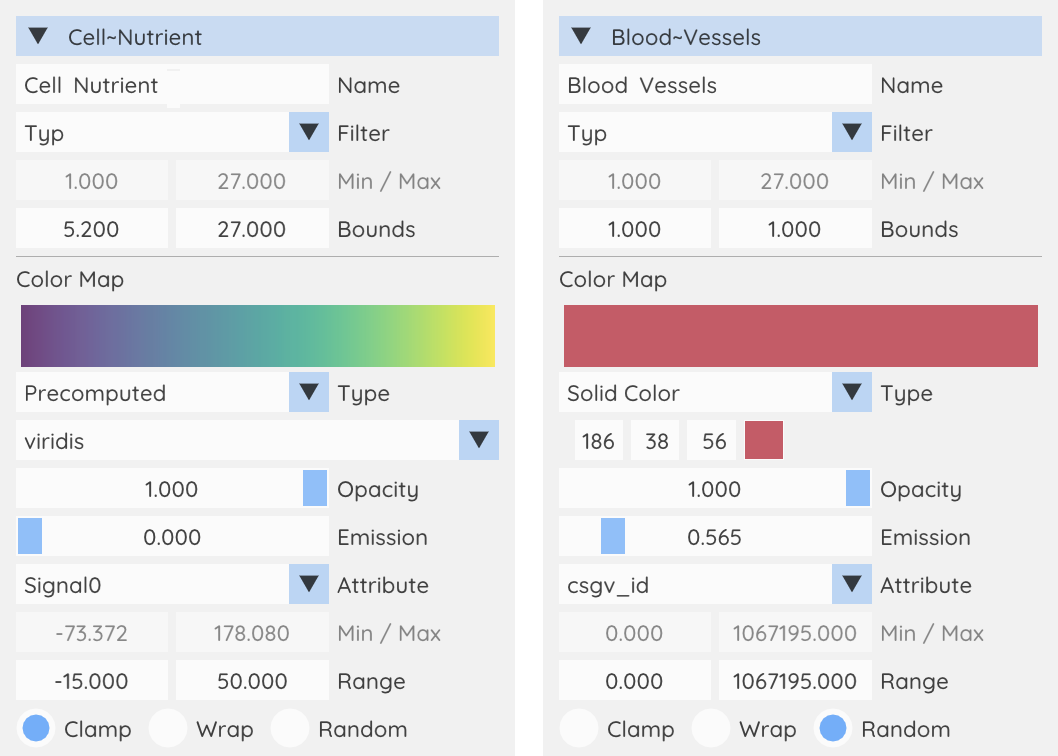}
    \end{subfigure}
    
    \caption{\subref{fig:vcnt:renderer:transferfunction:mapping} Example assignment of colors to voxels. The labels of the six voxels map to two attributes (age and type) each. The first material for which the filter attribute falls into a specified range, is assigned to a voxel. Another attribute range can be used to assign colors from a colormap.
    \subref{fig:vcnt:renderer:transferfunction:preprocessing} Volumes are preprocessed to store consecutive ID values for efficiently mapping voxel to attributes in memory.
    \subref{fig:vcnt:renderer:transferfunction:editor} The transfer function editor specifies shading materials based on label attribute intervals. For each material, an interval of the filter attribute specifies for which voxel labels it is relevant. For these labels, the values of the attribute property can be mapped to different colors.
    \subref{fig:vcnt:renderer:transferfunction:rendering} Rendering of \cells{} with three materials: blood vessels in red, nutrient density of tumor cells displayed with the viridis colormap, ECM cutout in gray.}
    \label{fig:vcnt:renderer:transferfunction}
\end{figure*}
Transfer functions in volume visualization map data values to optical attributes like color and opacity.
Existing interfaces mainly address grayscale volume rendering with quantitative value domains~\cite{Ljung:2016:TF}.
Additional segmentations are sometimes used to select different mappings for voxels $\mathbf{x}$~\cite{Hadwiger:2003:TwoLevelVolRen, Bruckner:2005:VolumeShop, Bruckner:2007:styleTF, Ljung:2016:TF}.
In segmentation volumes $V: \mathbf{x} \mapsto l$, labels $l \in \mathbb{N}$ are nominal but can map to \textit{secondary} attributes $A: l \mapsto (a_0, ..., a_{N-1}) \in \mathbb{R}^N$ (see \textit{nutrient density} of labeled cells in \cref{fig:vcnt:renderer:transferfunction:rendering}).
For segmentation volume rendering, existing transfer functions are usually limited to specific domains~\cite{Weissenboeck:2014:FiberScout} or apply to small label sets only~\cite{Pieper:2004:3dslicer, neuroglancer}.
For example, the ParaView~\cite{paraview} backend VTK, one of the leading volume visualization tools, supports 32 labels.
For more labels, one must rely on inaccurate sampling with non-categorical transfer functions which only supports local shading (\cref{fig:vcnt:tools:azba-comparison}) as opposed to our method.
Unlike ours, such VTK rendering creates sampling errors on segmentations with large label ranges (\cref{fig:vcnt:toolsrender:inaccuracy}).
Neuroglancer~\cite{neuroglancer} displays more labels as approximated 3D meshes, but its interface is limited and handles only several thousand labels.
Our multi-material transfer function for segmentation volumes is more flexible, supports dense segmentations with millions of labels by grouping intervals of visible labels as materials, and supports secondary attributes, either automatically computed or from .sqlite or .csv files.
\supvideocells{} shows visualizations of different attributes in \cells{}.

\subsection*{Segmentation volume collection for visualization evaluation}

Datasets for benchmarking segmentation volume visualization systems usually cover few properties or domains (e.g.\ connectomics~\cite{Beyer:2013:ETC} or material science~\cite{Weissenboeck:2014:FiberScout}).
Volume databases (BossDB, Dryad, or Zenodo) cover a wider range, but with varying quality and over-representation of certain properties.
We present a curated benchmarking dataset of publicly available (except \cells{} and \fiber{}) segmentation volumes (\Cref{tab:vcnt:results:datasets}).
The volumes cover a wide range of sizes, acquisition methods, and domains from plant analysis~\cite{Wolny:2020:plant}, cellular simulations~\cite{Rosenbauer:2020:ETD}, brain atlases~\cite{Kenney:2021:azba, Allen:2017:ara2016}, composite materials~\cite{Bertoldo:2021:pa66, Maurer:2022:Fiber}, battery design~\cite{Mueller:2021:xtm}, nonwoven fabrics~\cite{Griesser:2022:article}, or human and animal connectomics~\cite{Tavakoli:2025:liconn, Motta:2019:DCR, Shapson-Coe:2024:h01}.
Sizes reach from 72 MB, 8-bit voxels for \griesserval{} to over 2 TB, 32-bit voxels for a white matter subset of the H01 human cortex segmentation~\cite{Shapson-Coe:2024:h01}.

\subsection*{CSGV segmentation volume compression}
\label{sec:vcnt:results:csgv}

\begin{figure*}[ht]
\centering

\begin{subfigure}[t]{\textwidth}
\centering

\caption{}
\label{tab:vcnt:results:datasets}

\vspace{-0.5em}

\footnotesize

\pgfplotstableread[col sep=comma,header=true]{paper/volcanite/results/csgv/csgv-eval.csv}\datasetcsv
\pgfplotstablecreatecol[
  create col/assign/.code={
        \edef\temp{\thisrow{DimX}\space $\times$ \thisrow{DimY}\space $\times$ \thisrow{DimZ}}
        \pgfkeyslet{/pgfplots/table/create col/next content}\temp
  }
]{Dimension}{\datasetcsv}
\pgfplotstablecreatecol[
    create col/expr={\thisrow{DimX}*\thisrow{DimY}*\thisrow{DimZ}}
]{VoxelCount}{\datasetcsv}
\pgfplotstablecreatecol[
    create col/expr={
        \thisrow{Labels}/\thisrow{VoxelCount}*1000000
    }
]{LabelsPerMMVoxels}{\datasetcsv}
\pgfplotstablecreatecol[
    create col/expr={
        \thisrow{Orig bits/voxel}/8
    }
]{byte/voxel}{\datasetcsv}
\pgfplotstablecreatecol[
    create col/assign/.code={
        \edef\temp{\thisrow{Data Set}}
        \pgfkeyslet{/pgfplots/table/create col/next content}\temp
    }
]{Citation}{\datasetcsv}
\pgfplotstablecreatecol[
    create col/assign/.code={
        \edef\temp{\thisrow{Data Set}}
        \pgfkeyslet{/pgfplots/table/create col/next content}\temp
    }
]{Domain}{\datasetcsv}
\pgfplotstabletypeset[
    col sep=comma,
    header=true,
    every head row/.style={after row=\midrule},
    columns={{Data Set},{Citation},{Dimension},{VoxelCount},{Orig Size [GB]},{byte/voxel},{Labels},{LabelsPerMMVoxels}},
    columns/{Data Set}/.style={string type, column type={l}, postproc cell content/.code={
            \pgfkeyssetvalue{/pgfplots/table/@cell content}{\dataNameFromCSV{##1}}
        }},
    columns/{Citation}/.style={column name={Origin / Domain},column type={p{3.4cm}}, string type, postproc cell content/.code={
            \pgfkeyssetvalue{/pgfplots/table/@cell content}{\dataCitationFromCSV{##1} \domainFromCSV{##1}}
        }},
    columns/{byte/voxel}/.style={column name={B}, column type={r}},
    columns/{Dimension}/.style={string type, column type={r}, column name={Dimension [Voxels]}},
    columns/{Labels}/.style={column type={r}},
    columns/{VoxelCount}/.style={column name={Voxels}, column type={r}},
    columns/{LabelsPerMMVoxels}/.style={column name={/ $10^6$ Voxels}, fixed, precision=3, column type={r}},
    columns/{Orig Size [GB]}/.style={column name={Size [GB]}, column type={r}, fixed, precision=3, fixed zerofill},
    empty cells with={},
    sort,
    sort key={Orig Size [GB]},
]{\datasetcsv}

\end{subfigure}

\begin{subfigure}[t]{\textwidth}
\caption{}
\label{fig:vcnt:csgv:encoding}
\centering

\vspace{-0.5em}
\includegraphics[width=\textwidth]{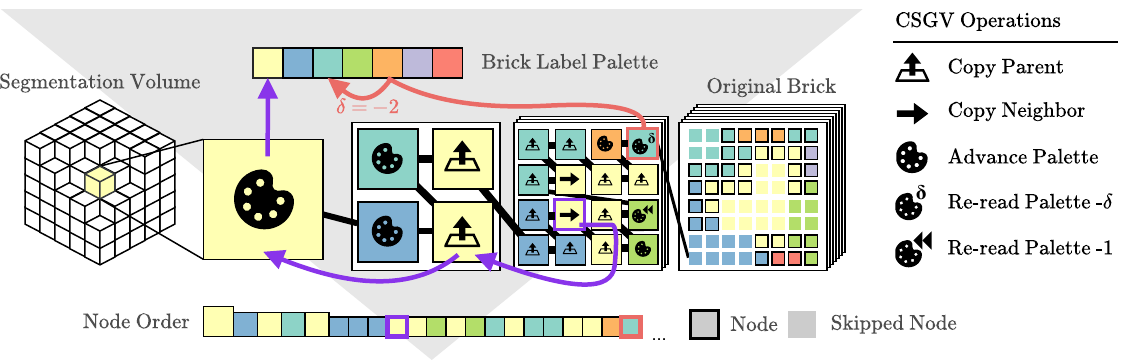}
\end{subfigure}

\vspace{-0.5em}

\begin{subfigure}[t]{\textwidth}
\caption{}
\label{fig:vcnt:csgv:results}
\centering
\pgfplotstableread[col sep=comma,header=true]{paper/volcanite/results/csgv/csgv-eval.csv}\csgvcsv
\pgfplotstablecreatecol[
  create col/assign/.code={
        \edef\temp{\thisrow{DimX}\space $\times$ \thisrow{DimY}\space $\times$ \thisrow{DimZ}}
        \pgfkeyslet{/pgfplots/table/create col/next content}\temp
  }
]{Dimension}{\csgvcsv}
\vspace{-1em}
\footnotesize
\pgfplotstabletypeset[
    col sep=comma,
    header=true,
    column type={r},
    every head row/.style={after row=\midrule},
    columns={{Data Set},{Labels},{Orig Size [GB]},{Orig bits/voxel},
             {CSGV Size [GB]},{Compression Rate [Pcnt]},{CSGV bits/voxel},{Detail Encoding [Pcnt]}
             },
    columns/{Data Set}/.style={string type, column type={l}, postproc cell content/.code={
            \pgfkeyssetvalue{/pgfplots/table/@cell content}{\dataNameFromCSV{##1}}
        }},
    columns/{Dimension}/.style={string type, column type={r}, column name={Dimension [Voxels]}},
    columns/{Orig Size [GB]}/.style={column name=Orig [GB], fixed, precision=3, fixed zerofill},
    columns/{Orig bits/voxel}/.style={column name=b/voxel},
    columns/{CSGV Size [GB]}/.style={column name=CSGV [GB], fixed, precision=3, fixed zerofill},
    columns/{Compression Rate [Pcnt]}/.style={column name={\%}, fixed, precision=3, fixed zerofill},
    columns/{CSGV bits/voxel}/.style={column name=b/voxel, fixed, precision=3, fixed zerofill},
    columns/{Detail Encoding [Pcnt]}/.style={column name={Detail \%}, fixed, precision=3, fixed zerofill},
    columns/{Palette Length avg}/.style={column name=$\varnothing |P|$, fixed, precision=1},
    columns/{Palette Duplicates avg}/.style={column name=$\varnothing |\mathrm{dupl}|$, fixed, precision=1},
    columns/{Palette Length max}/.style={column name=$\mathrm{max} |P|$, fixed, precision=1},
    columns/{Palette Duplicates max}/.style={column name=$\mathrm{max} \; |\mathrm{dupl}|$, fixed, precision=1},
    empty cells with={},
    sort,
    sort key={Orig Size [GB]},
]{\csgvcsv}

\vspace{0.5em}
\end{subfigure}

\caption{\subref{tab:vcnt:results:datasets} Our collection of segmentation volumes with different properties for evaluating visualizations.
While connectome (CONN) analysis is a main application, many domains utilize segmentation volumes.
\protect\subref{fig:vcnt:csgv:encoding} Our CSGV format compresses segmentation volumes as a set of brick encodings. Each brick is encoded by an independent stream of operations that reconstruct it in form of a multi-resolution grid. Small operation codes instruct nodes to copy their label from a neighbor \opneighbor or parent grid node \opparent, or from the brick's palette of 32-bit labels \oppalette.
\protect\subref{fig:vcnt:csgv:results} It significantly reduces memory consumption of the datasets requiring significantly less than 1 bit of storage per voxel on average (\textit{b/voxel}). Volcanite uses this to store TB sized volumes in \gls{gpu} \gls{vram} for a fully \gls{gpu}-based visualization pipeline. The encoding of the finest \gls{lod} can be separated in the \gls{csgv} to be streamed to the \gls{gpu} only on demand during rendering. The relative size of this detail \gls{lod} is given as \textit{Detail \%}.}
\label{fig:vcnt:csgv}

\end{figure*}
\glsreset{vram}
Volcanite's fully \gls{gpu}-based rendering pipeline operates on \gls{csgv}~\cite{Piochowiak:2024:csgv} encoded volumes. %
\Cref{fig:vcnt:csgv:results} lists compression results for all datasets (plots in \cref{fig:vcnt:csgvplots}).
Lossless compression allows processing terabyte volumes on a single commodity \gls{gpu} where \gls{vram} is limited to few GiB.
\gls{csgv} is largely independent of dataset label counts (\honewm{} has over 13 million labels) (\cref{fig:vcnt:csgvplots:label-count}) but rather depends on brick-local label densities which are usually low in segmentation volumes (\cref{fig:vcnt:csgvplots:label-density}).
This effectively results in worse compression rates for data with 8- or 16-bit raw voxels as those are packed more tightly compared to 32-bit data in their uncompressed representation to begin with (\cref{fig:vcnt:csgv:results}).
Even including this compression (once per dataset), our preprocessing is significantly faster than in VTK or Neuroglancer at much smaller data storage sizes (\cref{fig:vcnt:tools:preprocessing-times}) with the fastest \gls{ttff} (\cref{tab:vcnt:tools-preprocess}).
For 3D visualizations, Neuroglancer only stores vertex mesh approximations to conserve space (\cref{fig:vcnt:tools:azba-comparison}) while our encoding is lossless.

\gls{csgv} encodes multi-\gls{lod} volume bricks as serialized streams of label-copy operations (\cref{fig:vcnt:csgv:encoding}).
The finest \gls{lod} (45\% - 68\% total size) can be streamed to the \gls{gpu} on demand to reduce \gls{vram} consumption.
Each grid node stores a stop bit to mark all-constant children (no outline) to be omitted in further \glspl{lod}.
An operation per node determines to copy its label from another node if possible.
Otherwise, the label is read from the brick's label palette. %
By default, previous palette entries cannot be reused, requiring duplicate entries.
\gls{csgv} originally solves this with operations to re-read entries from $\delta$ indices ago (\cref{fig:vcnt:csgv:encoding}).
The back reference operation stores $1 < \delta \leq 16$ within the operation stream.

\cref{fig:vcnt:csgv-ablation} presents an ablation study of operations.
Palette back references reduce palette lengths (\cref{fig:vcnt:vram:cache-palette-result}) but rarely improve and sometimes worsen compression rates.
This is due to $\delta$ entries in the data stream skewing symbol frequencies unfavorably for the applied secondary ANS frequency coding~\cite{Duda:2015:ANS}.
Nevertheless, shorter palettes streamline processing, e.g.\ testing bricks for visible labels and \gls{vram} packing (\sectionref{sec:vcnt:renderer:memory}{memory management}).
We therefore replace the old back reference operation with a new variant that supports unlimited $\delta$.
It creates optimal palette lengths without duplicates (\cref{fig:vcnt:vram:cache-palette-result}).

\subsection*{Image and video rendering performance}
\begin{figure*}[pt]
    \begin{subfigure}[t]{\textwidth}
        \caption{}
        \label{fig:vcnt:gpu-rendering:pipeline}
        \vspace{-0.8cm}
        \begin{center}
            \includegraphics[width=\textwidth]{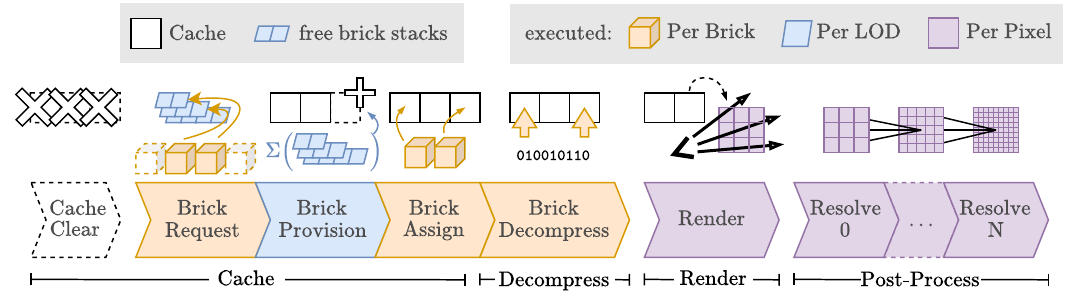}
        \end{center}
    \end{subfigure}
    
    \begin{subfigure}[t]{\textwidth}        
        \caption{}
        \label{fig:vcnt:gpu-rendering:timing-image}
        
        \includegraphics[width=\textwidth]{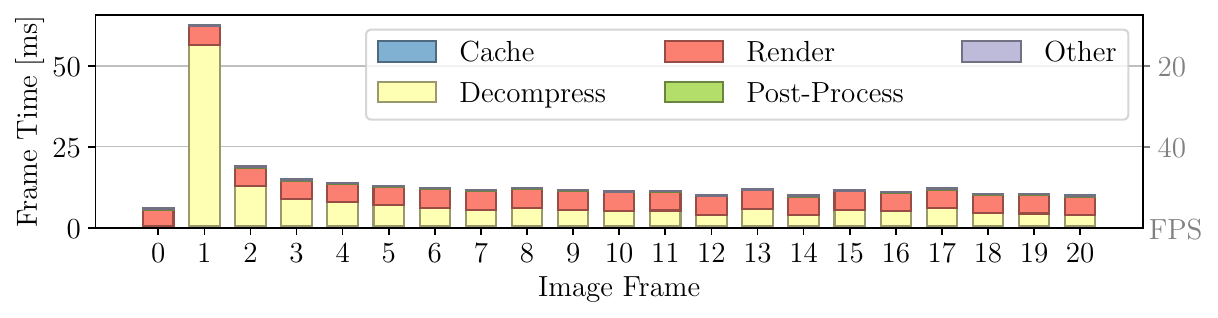}
        
    \end{subfigure}

    \vspace{-0.5cm}

    \begin{subfigure}[t]{\textwidth}        
        {
        \caption{}
        \label{fig:vcnt:gpu-rendering:timing-video}
        
        \pltvideoframes{Motta2019}{shadow}

        \vspace{-0.5em}
        
        \includegraphics[width=\textwidth]{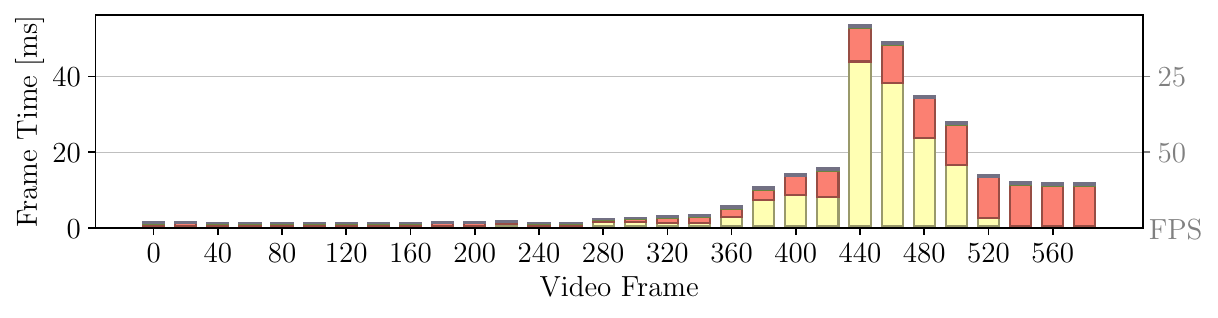}
        }

        \vspace{-0.2cm}
    \end{subfigure}

    \caption{\subref{fig:vcnt:gpu-rendering:pipeline} Volcanite rendering pipeline. Each stage is a separate shader dispatch. Brick cache management is split into several stages (cf. shading-atlas streaming~\cite{Mueller:2018:sas} and~\cite{Piochowiak:2024:csgv}): \textit{Request} frees cache regions of no longer requested bricks onto \gls{lod} free stacks. \textit{Provision} allocates new elements from the back of the cache buffer if the free stack is not sufficient to fulfill all brick requests. \textit{Assign} assigns newly requested bricks to free cache regions after which \textit{Decode} decompresses them into the cache. \textit{Render} executes one ray marching thread per pixel, accessing cached bricks to compose colors, and requests all visible bricks. \textit{Resolve} repeatedly executes our \`{A}-trous filter for unified denoising and upsampling.
    \subref{fig:vcnt:gpu-rendering:timing-image} Frame times for rendering the 927 GB (raw) \cortex{} volume with a stationary camera on a 16 GiB \gls{gpu}. Early decompression demand is high since caches are initially empty. Later frames are dominated by rendering.    
    \subref{fig:vcnt:gpu-rendering:timing-video} Frame times for an animation (camera rotates around and zooms in to \cortex{}) with shadows. Later frames access data in finer resolutions, leading to more costly decompression and rendering.}
    \label{fig:vcnt:result:gpu-rendering}

\end{figure*}

\Cref{fig:vcnt:gpu-rendering:timing-image} shows frame timings for rendering a \honewm{} perspective with shadows and a static camera.
Volcanite's rendering is progressive, where additional accumulated frames improve image quality through anti-aliasing and, in case of global illumination, more Monte Carlo samples.
It alternates between \gls{gpu} rendering and decompression stages where rendering flags all required yet undecoded bricks for decompression in the next frame.
Volcanite decompresses only visible \gls{csgv} bricks into fitting \glspl{lod} so that each decoded voxel maps to roughly one pixel in the image (\Cref{fig:vcnt:visual-abstract:rendering}).
This limits decoded voxel counts for better performance and memory utilization. 
The evaluated close up perspectives therefore put the most stress on the renderer.
For small dataset dimensions, roughly all bricks are decoded to the finest \gls{lod}.
Large datasets (e.g.\ \cortex{} and \hone{}) mostly access coarser \glspl{lod} which can be faster.
The second frame has high timings since it must decode all initially visible bricks.
Timings quickly settle down to highly interactive 2 and 29 ms per frame with rare higher outliers.
Rendering the datasets with VTK achieves significantly worse performance than our renderer due to its ray marching algorithms (\cref{fig:vcnt:tools:vtk-timings}).
\Cref{fig:vcnt:gpu-rendering:timing-video} shows timings for rendering a video where the camera makes a full rotation around the dataset and moves from a far away to a close up perspective.
The initial perspective requests bricks at a coarser resolution leading to a less severe initial decompression workload.
Afterwards, the moving camera creates a constant decompression and caching workload as brick visibilities change.
Timings remain interactive in all cases.
All plots are in \cref{fig:vcnt:all-gpu-timings}, timings in \cref{tab:vcnt:timings}.

\subsection*{Flexible shading modes}
Volcanite generally provides four shading modes (\cref{fig:vcnt:visual-abstract:shading}) from evaluating a local surface reflection model, over computing one shadow ray (\textit{shadow}), or ambient occlusion, to full global illumination computation with path tracing.
\Cref{fig:vcnt:shading:images} shows shading mode renderings for the evaluated camera perspective of \honewm{}.
Shadow rays and ambient occlusion increase perception of spatial structures.
While path tracing offers the highest visual quality, render times increase by factors of up to 5.5 from the first to the latter (\Cref{tab:vcnt:timings}).
At lower directional correlation and deeper path depths (numbered in \cref{fig:vcnt:shading:rays}), more advanced shading accesses more bricks.
Even though the camera remains still, randomized view ray offsets for anti-aliasing and indirect ambient occlusion or path tracing rays nevertheless request new bricks in each frame.
Shading modes and blends in between can be switched anytime in the interactive application with immediate effect, underpinning Volcanite's flexibility to scale between fast exploration and high quality presentation (\supvideocells{}).
Other explorative tools like Neuroglancer~\cite{neuroglancer} or VTK/ParaView~\cite{paraview} only support local shading which limits the perception of spatial structures (\cref{fig:vcnt:toolsrender:vtk-renderings,fig:vcnt:tools:azba-comparison}).
While VTK generally supports single scattering approximations, this feature had no effect for the surface-like rendering of segmentation volumes.

\subsection*{GPU-based rendering pipeline}
\label{sec:vcnt:results:rendering-pipeline}
\Cref{fig:vcnt:gpu-rendering:pipeline} shows the \gls{gpu} frame render pipeline (groundwork previously published~\cite{Piochowiak:2024:csgv,Piochowiak:2025:csgvr}).
It is fully \gls{gpu}-based without any \gls{gpu} to \gls{cpu} synchronization within a frame to reduce timings and latencies.
Similar to the caching from Shading Atlas Streaming~\cite{Mueller:2018:sas}, \textit{cache} stages count newly requested bricks per \gls{lod} (Request), allocate memory (Provision), and assign newly required bricks to cache regions (Assign).
The \textit{Decompress} stage operates in parallel over all newly requested bricks.
The \textit{Render} stage computes light transport along $\leq 1$ ray path per pixel using ray marching through the decoded voxel grid (\cref{fig:vcnt:visual-abstract:rendering,fig:vcnt:shading:rays}).
The multi-pass \textit{Post-Processing} stage performs denoising, upsampling, and color space operations (\sectionref{sec:vcnt:results:postprocess}{upsampling and denoising}).
Since the full volume is stored in \gls{gpu} \gls{vram}, we can eliminate most \gls{cpu} to \gls{gpu} data streaming for better latency reduction and performance.
The only exception being asynchronous construction and uploading of \gls{csgv} detail \gls{lod} encodings if those were separated to reduce \gls{vram} footprints.
Cache management involves many atomic operations in four subsequent stages.
The Request stage additionally checks all brick palettes for visible labels under current shading materials in each frame.
Nevertheless, \textit{Cache} stages have negligible effects on frame timings even for terabyte data (\cref{fig:vcnt:gpu-rendering:timing-image,fig:vcnt:gpu-rendering:timing-video}).
Timings are dominated by \textit{Decompress} and \textit{Render} stages, where the first is strongly affected by camera movement and both depend on accessed brick \glspl{lod}.
As opposed to other tools, our \gls{gpu}-based pipeline ensures minimal latency on user interaction.
VTK struggles with interactivity for large segmentation volume visualization, if rendering is possible at all (\cref{fig:vcnt:tools:vtk-timings}), and Neuroglancer relies on high-latency data streaming through server-client architectures~\cite{neuroglancer}.

\subsection*{Fast combined upsampling and denoising filter}
\label{sec:vcnt:results:postprocess}
Post-processing performance is dominated by kernel-based pixel interpolations:
If pixels receive few paths, their colors may contain noise which is removed through blending with neighboring surface points (\cref{fig:vcnt:resolve:denoising}).
If subsampling is optionally enabled to increase performance, only $\frac{1}{4}$th, $\frac{1}{16}$th, or $\frac{1}{64}$th of pixels are rendered per frame.
Holes in early frames are filled by copying the nearest rendered neighbor (\cref{fig:vcnt:resolve:upsampling}).
Upsampling increases post-processing timing but significantly reduces render timings for faster overall rendering (\cref{fig:vcnt:resolve:results}).
Existing \`{A}-trous filters~\cite{Dammertz:2010:denoise} efficiently perform denoising over large kernels through iterative blending with increased neighbor step sizes of $2^i, i \in \mathbb{N}$ (\cref{fig:vcnt:resolve:atrous}).
As opposed to \cite{Dammertz:2010:denoise} and other \`{A}-trous filter methods~\cite{Schied:2017:svgf}, we do not store all intermediate pass results but directly accumulate final colors in the in/out buffers (\textit{Denoise A/B}) to reduce memory footprints.
We further extend the method to perform upsampling in the same computation:
Rendered pixels blend their colors with kernel neighbors in the \textit{Denoise A/B} buffers, non-rendered pixels store iteratively computed offsets to their closest already rendered neighbor from which the final pass copies its color.
Additionally, we optimize the blending edge-stopping weights for segmentation volumes, and reduce memory consumption with a segmentation volume-specific G-buffer layout (\cref{fig:vcnt:resolve:gbuffer}) which guides kernel weights similarly to bilateral filtering~\cite{Tomasi:1998:bilateral}.
Computation is fast (\cref{fig:vcnt:resolve:results}) at minimal memory overhead (three image-sized textures) %
and produces quality images in early frames when the camera is moved (\cref{fig:vcnt:resolve:upsampling,fig:vcnt:resolve:denoising}).
Once enough pixel samples are accumulated for a static camera perspective, denoising and upsampling fade out converging to an unbiased image.

\subsection*{GPU memory utilization optimization}
\label{sec:vcnt:results:memory-optimization}
\Cref{fig:vcnt:vram:results} lists overall memory usage for rendering the close ups.
For small data, cache memory dominates the overall \gls{vram} consumption due to our strong \gls{csgv} compression of the volume.
For large data, the compressed volume still reaches \gls{vram} limitations.
Through our memory optimizations, rendering still complies with the 16 GiB \gls{vram} limitation of our system:
For \honewm{}, streaming its finest \gls{lod} reduces \gls{csgv} memory by 65.7\%.
Decompressing bricks into a packed palettized format allows using a smaller cache for the same number of decompressed voxels by a packing factor (\cref{fig:vcnt:vram:cache-palette}).
Here, the cache stores packed indices into the brick palettes which are directly accessible from brick encodings, effectively increasing usable cache sizes by a factor when enabled (\textit{Packing} in \cref{fig:vcnt:vram:results}, \gls{vram} utilization in \cref{tab:vcnt:cache-palette}).
If the cache buffer would still not be sufficient to allocate all required bricks, brick requests will automatically be limited to a subset of screen pixels (\cref{fig:vcnt:vram:req-limit}).
At the expense of required higher frame counts for converged images, this ensures correctly rendered results by preventing cache deadlocks in which some screen areas occupy the full cache with their required and thus never released bricks.
Volcanite renders TB data with immediate feedback at a high visual quality.
Other voxel-precise 3D renderers like VTK store the full data in \gls{vram}~\cite{paraview, Pieper:2004:3dslicer, Weissenboeck:2014:FiberScout}, preventing visualization of large-scale datasets on commodity hardware (\cref{fig:vcnt:tools:vtk-timings}).
Our memory-efficient method enables interactive, explorative analysis of large segmentation volumes in 3D which is crucial for generating and validating novel hypotheses.

\section*{Discussion}

Dense segmentation volumes are now a common output of imaging-driven science, but their visual interpretation remains constrained by methods developed for different data types. Volcanite addresses this mismatch by treating dense segmentations as categorical, boundary-sensitive voxel data rather than as scalar fields or precomputed surfaces. This design enables interactive rendering of teravoxel-scale labeled volumes on commodity hardware while preserving voxel-level labels, avoiding mesh generation and retaining spatial cues through shadows and global illumination.

The practical consequence is a shorter path from segmentation output to visual inspection as the same CSGV representation provides both lossless compression and the acceleration structure used for rendering, reducing format conversion and preprocessing overhead. This matters because exploratory segmentation analysis is usually iterative: users must move through the volume, change label selections, adjust opacity and color mappings, and test viewpoints before relevant structures or artifacts become apparent. In densely packed morphologies such as neuronal processes, vasculature and cellular tissues, voxel-precise rendering with depth cues can expose stitching errors, thin structures and segmentation defects that are less apparent in slices or simplified surface representations.

Volcanite also changes how large segmentation data can be discussed and shared in practice. Because selected subvolumes can be compressed and explored locally, researchers can inspect data with collaborators on a workstation or portable conference setup instead of relying on remote visualization infrastructure. Such fully local execution also simplifies handling unpublished, sensitive or proprietary data. This makes Volcanite useful not only for final visualization, but also for quality control, collaborative analysis and iterative interpretation during ongoing projects.

A further strength is that these capabilities are not limited to one domain or label regime. The evaluated datasets span life science, material, and synthetic/computational examples, with object counts ranging from a few labels to millions. Volcanite therefore provides a general approach to dense segmentation visualization rather than a tool tuned to a specific single dataset class and can be integrated into, or, as it is fully open-source, readily adapted to workflows to transfer across data sources, acquisition modalities and segmentation pipelines.

Compared with existing tools, Volcanite changes the practical balance between scale, fidelity and iteration time. Slice-based viewers remain accurate but provide limited three-dimensional context. Mesh-based approaches support 3D views but require costly preprocessing and may hide voxel-scale defects or small structures through geometric simplification. Streaming systems are essential for navigating datasets beyond local hardware limits, including petascale reconstructions, but typically expose smaller regions at a time or rely on approximate 3D representations. Volcanite complements these systems by enabling high-fidelity, voxel-precise inspection of large selected regions with richer shading and substantially faster local iteration.

The current implementation targets local rendering of volumes up to the teravoxel range. This covers many current datasets and large subregions of even the largest biological reconstructions, but complete petascale volumes remain outside the scope of the present system. %
A natural next step is to combine Volcanite’s compressed-brick representation and GPU cache with on-demand streaming from CPU memory, local storage or remote servers. Such an extension would require additional scheduling and latency-hiding mechanisms, but would preserve the central advantage demonstrated here: direct, voxel-precise interaction with dense segmentation volumes. In its current form, Volcanite already lowers a major barrier between producing large segmentation volumes and interpreting their three-dimensional structure.

\section*{Methods}

\subsubsection*{Evaluation setup}
We evaluate rendering performance on commodity hardware, in particular an AMD Ryzen 7 5800X CPU with 48 GiB DDR4 RAM and an RTX 4070 Ti Super \gls{gpu} with 16 GiB \gls{vram} at fixed GPU clock speeds on an Ubuntu 24.04 system.
Input data is read from a Lexar NM790 4 TB NVMe, file output is stored on a Samsung EVO 970 1 TB NVMe.
We provide the python scripts to download all data (except \cells{}, \fiber{}), execute Volcanite in headless mode to generate all results, and create all plots and tables.

\subsubsection*{Compression and rendering parameters}
We use $b=32$ as default \gls{csgv} brick size and $b=64$ for large data (\cortex{}, \honebv{}, \honewm{}, \griesser{}).
For these datasets and \fiber{} and \polyamid{}, the effective brick cache size is increased by decompressing voxels into a compact format (cache paletting) due to high caching demands.
As the compressed \honewm{} still exceeds \gls{vram}, bricks of its finest \gls{lod} (65.7\% of the encoding) are streamed to the \gls{gpu} only on demand (detail streaming).
We enable the new unlimited palette back references in the encoding for \cells{}, \cortexsmall{}, \cortex{}, \liconn{}, and \honewm{}, and disable palette back reference for all other datasets.
All renderings are in 1920 $\times$ 1080 resolution with one sampling ray per pixel (except pixel subsampling for upsampling evaluations).
For the \textit{image} evaluations, 1024 frames are rendered with a static camera and anti-aliasing through temporal supersampling.
In \textit{video} evaluations, the camera rotates around and zooms in to the focal point for 600 frames.
When evaluating different shading modes, only mode-relevant parameters are varied.

\subsubsection*{VTK and Neuroglancer comparisons}
We compare our renderer with VTK~\cite{Schroeder:2006:vtkBook} v9.5.0, the rendering backend of ParaView~\cite{paraview}, 3D Slicer~\cite{Pieper:2004:3dslicer}, and pyVista~\cite{Sullivan:2019:pyvista}.
Since VTK's \gls{gpu}-based volume rendering is designed for grayscale volumes, visualization of segmentation volumes usually relies on surface extractions for each label and mesh-based rendering.
Alternatively, direct voxel-based rendering supports up to 32 categorical labels.
As our use case is precise voxel-based rendering of densely labeled volumes, meshing and categorical labels are not sufficient.
We use VTK's \gls{gpu}-based volume rendering instead.
For each dataset, we translate our Volcanite material configuration into an opacity transfer function mapping over the label domain which is either fully opaque for opaque and semi-transparent labels in Volcanite or fully transparent for invisible ones.
Omitting semi-transparency in VTK only skews performance results in favor of VTK.
This results in mapping the label domain to the transfer function texture size, resulting in undersampling for densely labeled volumes.
Color transfer functions have no effect on performance and are set to a HSV colormap.
We execute VTK's OpenGL volume raycasting mapper through its C++ interface for 1024 frames without camera movements to compute frame time aggregates in Full-HD as in our Volcanite \textit{image} evaluation.
Camera configurations and volume transformations are translated from Volcanite configurations for identical projection mappings, the ray marching sample distance is 0.5 voxels.
As VTK's raycasting does not support shadows or global illumination, we only compute voxel normals for local shading as in Volcanite's \textit{local shading} configurations.

For the Neuroglancer~\cite{neuroglancer} evaluations in Python, volumes are first converted into the Neuroglancer precomputed compressed segmentation format with a chunk size of $64^3$ in single resolution (original $1^3$) and stored on disk with gzip compression.
Meshes for all labels are computed in parallel with Igneous~\cite{Silversmith:2022:igneous} at the original resolution only and simplified with quadratic edge collapse~\cite{Hoppe:1999:meshSimplQM,Garland:2023:meshSimplQEM}.
For \cortex{} and \honewm{}, we used a mesh manifest magnitude of 5 (default 3 otherwise) to split the mesh file creation work over more tasks due to the large number of mesh files.
Mesh manifest stage timings for \honewm{} were computed in a second run after the first combined execution crashed.
Since we do not compute multi-resolution meshes, gzip instead of Draco~\cite{Google:2017:draco} compression is applied to mesh files as proposed by the authors~\cite{Silversmith:2022:igneous}.
Neuroglancer visualization relies on a client-server architecture where content in view space is steadily streamed to the visualization client and meshes pop up one mesh at a time during rendering.
We therefore do not measure any visualization frame times or time-to-first-frame, as it is not possible to obtain these metrics in a precise manner.

\subsection*{Compressed segmentation volumes}\label{sec:vcnt:compression}

\label{sec:vcnt:renderer:compression}
The key component in our renderer to process large scale segmentation volumes ($V: \mathbb{N}^3 \to \mathbb{N}$) on single GPU systems is lossless data compression.
Volume visualization on GPUs is heavily memory limited but we aim to keep as much of the dataset in GPU memory as possible.
To that end, our renderer uses the brick-wise multi-resolution \gls{csgv} encoding from~\cite{Piochowiak:2024:csgv}.
We refer to the encoding overview in the \sectionrefmanual{sec:vcnt:results:csgv}{\hyperref[sec:vcnt:results:csgv]{CSGV results}} and the original method for details~\cite{Piochowiak:2024:csgv, Piochowiak:2025:csgvr}.

We make the following changes to the original \gls{csgv} encoding:
We allow disabling all operations except the indispensable operation to insert and read the next label palette entry during compression.
For example, not including back references can improve compression rates in certain cases (\cref{fig:vcnt:csgv-ablation}) as it introduces a tradeoff by reducing palette duplicates at the expense of more different values in the operation stream (back reference distances $\delta$).
The latter worsens the compression rates of the secondary variable bit-length encoding.
A smaller operation set also reduces branching during decoding and could thus improve decompression performance.
Therefore, we usually only enable palette back references if small palette sizes are preferable and volume label counts are high.
For such back references, \cite{Piochowiak:2024:csgv} limit the following $\delta$ entry to fit into 4 bits to minimize the negative effect on variable bit-length compression rates.
This means that back references cannot be resolved at arbitrary distances and some duplicated 32-bit labels remain in the palette, especially with large brick sizes.
For reasons concerning the rendering (\sectionref{sec:vcnt:rendering:cachepalette}{cache paletting}), we introduce arbitrary length back references instead:
After the operation code, another 4-bit entry is read from the operation stream.
The 3 \glspl{lsb} of this entry denote the first \glspl{msb} of $\delta$.
The last bit denotes if another entry follows for the next lower \gls{msb}.
With each additional entry, $\delta$ is updated as \,
\(\delta \gets (\delta \, \texttt{<<} \, 3) \mathbin{|} (\mathrm{entry} \mathbin{\&} \mathtt{111}_2)\).
This ensures that small $\delta$ will only occupy one additional 4-bit element in the stream while arbitrary lengths can still be encoded, eliminating any palette duplicates.
For a $64^3$ brick, the theoretical worst case number of 4-bit entries would be $\lceil \textrm{log}_2(64^3 + 32^3 + \dots + 1) / 3 \rceil = 7$ which still requires fewer bits (28) than inserting a palette duplicate (32).

Before rendering, the segmentation volume to visualize is compressed with our adapted and configurable \gls{csgv} encoding.
Encoding times are significantly shorter than the time needed to read the volume files from disk (\textit{Compr. only} and \textit{File IO} in \cref{tab:vcnt:tools-preprocess}).
Volcanite can import chunked volumes that are split into multiple smaller files for compressing volumes that would otherwise exceed the RAM.
Compression is executed purely on the CPU and parallelized using OpenMP~\cite{openmp45}.

\subsection*{Volcanite rendering}\label{sec:vcnt:renderer}
For rendering, the compressed segmentation volume is uploaded to GPU memory from which visible regions can be decoded on demand during rendering.
Our renderer leverages multiple properties of the format such as directly accessible label palettes in the encoding to reduce the memory footprint of decompressed volume regions~(\sectionref{sec:vcnt:renderer:memory}{memory management}).
Images are generated performing voxel ray marching through the volume with each voxel of the original volume resolution being one unit large.
For each pixel in the image plane, a view ray is sent through the voxel volume starting from the virtual camera sampling a volume label at each voxel along its path (\cref{fig:vcnt:shading:rays}).
Shading materials are inferred from the label to compose the final pixel color from semi-transparent and solid surfaces (\cref{fig:vcnt:renderer:transferfunction:mapping}).

\paragraph*{Shading Modes}
A surface interaction occurs once a pixel ray hits a solid surface or the accumulated semi-transparent opacity exceeds a certain limit.
The shading mode decides if further reflected ray paths are traced from here on.
\Cref{fig:vcnt:shading:rays} shows resulting possible paths through the volume while \cref{fig:vcnt:shading:images} presents renderings for all modes.
To varying extents, the different modes partially evaluate the rendering equation~\cite{Kajiya:1986:REQ} which determines the outgoing radiance from a point $x$ towards directions $\omega_o$ based on the incoming radiance $L_i$ at $x$ from all directions $\omega_i$ on the positive hemisphere $\Omega^+$:

\ifdoublecol
    \begin{align*}
        L_o(x, \omega_o) = &L_e(x, \omega_o) + \\
         &\int_{\Omega^+} f(x, \omega_i, \omega_o) \, L_i(x, \omega_i) \; \cos\theta_i \; d \omega_i
    \end{align*}
\else
    \begin{equation*}
        L_o(x, \omega_o) = L_e(x, \omega_o) + \int_{\Omega^+} f(x, \omega_i, \omega_o) \, L_i(x, \omega_i) \; \cos\theta_i \; d \omega_i
    \end{equation*}
\fi

where $L_e$ is the emitted radiance at $x$, $\theta_i$ is the surface angle of the incident radiance and $f$ is the \gls{brdf} that determines the portion of the radiance that is reflected from $\omega_i$ towards $\omega_o$ at $x$.
The modes determine the rendering's overall appearance and are listed here from fastest to slowest rendering:
Simple \textit{Local Shading} assumes constant incoming radiance $L_i$ from one direction $\omega_l$ and $L_i(x, \omega_i) = 0$ for $\omega_i \neq \omega_l$. No further rays are traced, which makes it the fastest rendering mode.
One \textit{Shadow Ray} can be cast out towards $\omega_l$ to determine if the surface point is shaded by other structures in the dataset in which case $L_i$ is set to $0$.
\textit{Ambient Occlusion} casts a single ray as well but towards any random direction $\omega_i \in \Omega^+$ to approximate the attenuation of ambient lighting from surrounding occluding geometry~\cite{Zhukov:1998:AO}. No further rays after this first bounce are evaluated.
\textit{Path Tracing} evaluates the full rendering equation capturing complex effects like indirect lighting to create photorealistic images.
It is usually used in offline production renderers~\cite{Keller:2015:PathTracing}, but nowadays applied to real-time visualization as well~\cite{Iglesias-Guitian:2022:DPT, Zellmann:2023:AMRPT}.
The rendering equation is evaluated in one random direction for each recursive $L_i$ respectively.
This yields a single sample for a Monte Carlo process that converges to the evaluation of the full rendering integral.
The recursion terminates when the ray bounces out of the volume where $L_i(x, \omega_i)$ is the incoming environmental radiance from $\omega_i$ or after a path depth threshold.

\paragraph*{Transparency}
In contrast to mesh-based 3D segmentation volume visualizations~\cite{Boergens:2017:webknossos, neuroglancer}, our voxel-based rendering always traverses volume regions and surfaces in the correct order from closest to furthest from the camera.
This allows straightforward integration of semi-transparency in the rendering even though the compositing of semi-transparent surfaces is not commutative (e.g.\ \ara{} in \cref{fig:vcnt:all-gpu-timings}).
Users can assign opacities $\sigma$ to label materials.
If a path enters a semi-transparent material, it does not change its direction but instead adds the contribution of a Lambertian local surface shading model to the accumulated radiance and reduces the throughput of the ray by $\sigma$, occluding following surfaces.

As we always traverse all voxels along a path through the volume, we can process volumetric information inside regions as well.
We experimented with evaluating an emission absorption model~\cite{Engel:2006:volren, Joensson:2014:SVI} for piece-wise homogeneous volumes between consecutive surfaces along ray paths but found that using only surfaces created more effective visualizations.
In the future, we plan to extend the shading with different \glspl{brdf} and improve volumetric light transport, e.g.\ with full multiple-scattering~\cite{Novak:2018:SVPT}.

\subsubsection*{Ray marching}
\label{sec:renderer:raymarching}
We use ray marching to determine the voxels and their respective labels along a given path.
The renderer determines the \gls{lod} for decoding any accessed voxel brick so that one brick voxel maps to approximately one pixel in the image plane.
As labels are constant within voxels, precise first and last hit points of each ray with the respective voxels have to be found.
To that end, we apply multi-resolution \gls{dda} traversal~\cite{Amanatides:1987:DDA,Sung:1991:DDA}.
However, for large volume dimensions this leads to numerical instabilities from floating point operations and aliasing in final renders.
The multi-resolution traversal of Hofmann and Evans~\cite{Hofmann:2021:DDA} solves this by enlarging voxels by half a unit for Monte Carlo volume rendering of density volumes.
As this is not applicable to our case of voxel-constant labels, we propose a different numerically stable solution leveraging decomposition of the voxel position.
The integer vector \textcode{voxel} $\in \mathbb{N}^3$ stores the voxel index of the current traversal position while the floating point vector \textcode{vpos} $\in \mathbb{R}^3$ stores the sub-voxel position in $[0, 1)^3$.
Computing the distances to the next hit points operates in the numerically stable $[0, 1)^3$ interval:

\begin{lstlisting}[style=GLSL]
void stableDDA(inout ivec3 voxel,
               inout vec3 vpos) {
  // -1 or 1 sign per ray direction axis
  vec3 signs = sign(ray_dir);
  
  // axis distance to next voxel surface
  vec3 dist = 1./ray_dir *
               (max(signs,vec3(0)) - vpos);
  // find closest x/y/z axis hit: 0,1, or 2
  int axis = argmin(dist);

  // advance voxel along the axis of the closest next hit point
  voxel[axis] += signs[axis];

  // update the sub-voxel position vpos
  vpos = fract(vpos +
                (signs[axis] * ray_dir));
  vpos[axis] = max(-signs[axis],0.) *
                ONE_MINUS_EPSILON;
}
\end{lstlisting}
where \textcode{signs} $\in \{-1, 1\}^3$ is -1 or 1 per axis depending on the respective sign of \textcode{ray\_dir} and \textcode{fract} returns the decimal part of a number.
Stepping along the ray in coarser \glspl{lod} --- where the voxel size is a power-of-two unit larger than 1 --- can be done by scaling \textcode{voxel} and updating \textcode{vpos} accordingly using the modulo operator \textcode{\%} before computing the next step:

\begin{lstlisting}[style=GLSL]
void lodStableDDA(inout ivec3 voxel,
                  inout vec3 vpos,
                  int voxel_size) {

    // transform from fine to coarse resolution voxel coordinates
    voxel = voxel / voxel_size;
    vpos = (vec3(voxel % voxel_size) + vpos)
             / float(voxel_size);

    // perform one step with the virtual voxel size of 1 unit
    dda(voxel, vpos);

    // revert the transformation
    voxel = voxel * voxel_size +
             ivec3(floor(vpos * voxel_size));
    vpos = fract(vpos * voxel_size);
}
\end{lstlisting}

The same GPU ray marching loop evaluates primary rays originating from the camera and the secondary shadow, ambient occlusion or indirect path tracing rays.
This optimizes thread occupancy for the \gls{simt} architecture of \glspl{gpu}.
We use stack-less tracking of a pixel path's throughput and accumulated RGB radiance in the marching loop as is common in interactive path tracing~\cite{Boksansky:2021:referencePT}.
Path tracing uses Russian roulette~\cite{Boksansky:2021:referencePT} to terminate rays early with up to 32 bounces.
In a typical workflow, users can iteratively explore the data with fast rendering modes like local shading and, once a good rendering configuration is found, render high-quality images with ambient occlusion or path tracing.
We perform temporal anti-aliasing~\cite{Yang:2020:TAA}: The sub-pixel offsets of dispatched rays are jittered based on an approximated Blackman-Harris filter~\cite{Harris:1978:BlackmanHarris}.
As long as the camera does not move, this averages incoming radiance over pixel footprints over subsequent frames to remove aliasing from the undersampling of voxels.

Note that it is not possible to pre-filter voxel values in segmentation volumes (cf.\ Mip-Mapping in classic volume rendering of density fields with linear pre-filtering~\cite{Engel:2006:volren}) since labels are not filterable.
In the case of minification, i.e.\ a pixel footprint from the image plane covering many voxels at sampling points with far-away depths, a ground truth solution would need to access all covered voxels, apply color and opacity values on all extracted labels, and filter the resulting values.
This is not feasible for performance reasons.
The mixture graph~\cite{Al-Thelaya:2021:TMG} solves this filtering problem by storing sets of existing labels per spatial region in a compressed graph which works for smaller volumes.
An alternative would be to stochastically filter a single voxel per pixel footprint per frame in combination with a spatio-temporal accumulation filter~\cite{Pharr:2024:filtering}.
Volcanite relies on the multi-resolution hierarchy in the \gls{csgv} encoding instead which is comparable to a Mipmap hierarchy that accumulates the mode instead of the average of $2^3$ voxels per coarser voxel.
This is necessary for limiting cache memory as well, as it results in distant bricks being decoded in coarser resolutions, but can create bias in renderings.
Apart from temporal anti-aliasing, renderings with local shading or shadow rays create converged renderings in the first frame.
The accumulation of ray samples for the Monte Carlo evaluation of ambient occlusion and path tracing is distributed over multiple frames.
Initial renderings are noisy and progressively converge towards a noise-free image.
In any frame, either none or one sample of $L_i$ is evaluated per pixel.

\subsubsection*{Post-processing}
The noise in the initial frames with Ambient Occlusion or Path Tracing can be reduced through denoising post-processing methods~\cite{Huo:2021:SurveyDenoising, Dammertz:2010:denoise, Iglesias-Guitian:2022:DPT}.
Volcanite uses an edge-avoiding wavelet denoiser~\cite{Dammertz:2010:denoise} which operates hierarchically.
If images are rendered in a high resolution or GPU performance is limited, rendering can be limited to a subset of pixels per frame.
Values for non-rendered pixels in early frames are filled via upsampling (\cref{fig:vcnt:resolve:upsampling}).
Our denoising and upsampling is carried out in a shared post-processing pipeline for efficiency.
It consists of multiple compute shader dispatches with one thread per pixel in the current output buffer (\Cref{fig:vcnt:resolve-pipeline}).
The first iteration takes the rendering output --- the pixel-wise accumulation buffers for color and opacity (RGBA) and a G-buffer --- and the number of rendered samples per pixel (sample count) as input.
Later iterations operate on the output of their prior iteration.
An additional G-buffer in which the renderer stored additional information for each pixel guides the denoising.
The G-buffer is constant over all iterations and contains additional pixel information to guide the denoising.
The last iteration performs optional tone mapping~\cite{Narkowicz:2016:aces}, brightness, contrast, gamma corrections and blending with the background color.

\paragraph*{G-Buffer}
The G-Buffer is written by the renderer stage and only contains information about the most recent frame in 48 bits per pixel.
For the point at which the first surface interaction was detected, the G-buffer stores (\cref{fig:vcnt:resolve:gbuffer}):
The surface normal (3 bits).
Normals in our voxel-precise rendering are oriented in one of six discrete directions only.
The quantized surface depth (13 bits).
To maximize quantization precision we compute the depth as the normalized distance from the surface to a sphere that encapsulates the full dataset (\cref{fig:vcnt:resolve-depth}).
The remaining 32 bits store the label at the hit point, from which surface materials and albedo colors can be inferred.
Normal bits \texttt{011} denote that no surface but only volumetric interaction occurred along the ray (\textit{no hit}), \texttt{111} denotes that the pixel was not yet rendered (\textit{invalid}).

\paragraph*{Denoising}
Rendering accumulates at most one sample per frame per pixel.
Two ping-pong buffers track the accumulated colored radiance and opacity (\textit{Accumulation} in \Cref{fig:vcnt:resolve-pipeline}) and rendered sample count per pixel (\textit{Sample Count}):
Each execution reads the previously accumulated pixel from buffers A, writes the result after adding the new sample to the buffers B, and flips A and B.
We currently implement no temporal projection and reset the accumulation when parameters change.
With Ambient Occlusion or Path Tracing, while later frames converge to noise free averages, early frames are noisy.
The post-processing pipeline reduces this early noise by blurring pixel colors via an edge-avoiding \`{A}-trous wavelet transform similar to Dammertz et al.~\cite{Dammertz:2010:denoise} (\cref{fig:vcnt:resolve:denoising}).

In each post-processing pipeline iteration $i$, a respective pixel operates on $3 \times 3$ pixels (or $5 \times 5$ for higher quality) of the previous iteration's output (\Cref{fig:vcnt:resolve:resolve-ours}).
The distance between accessed pixels is $2^i$ in each dimension.
This \`{A}-trous filtering creates large receptive fields ($15 \times 15$ footprint after 3 iterations) at identical work and memory accesses per dispatch~\cite{Dammertz:2010:denoise} (\Cref{fig:vcnt:resolve:atrous}).
In contrast to Dammertz et al.~\cite{Dammertz:2010:denoise}, we do not use a final up-pass (increasing $i$) to gather the final output from  down-passes (decreasing $i$) but directly output the final down-pass result instead.
This reduces memory consumption as intermediate results are not tracked.

In each iteration, the $3 \times 3$ input pixels $\mathbf{q}$ from the previous output $c_i$ are weighted together for the current output pixel $\mathbf{p}$ in $c_{i+1}$ as in bilateral filtering~\cite{Tomasi:1998:bilateral}:
\[c_{i+1}(\mathbf{p}) = \frac{1}{k} \sum_{\mathbf{q}} h_i(\mathbf{q}) \cdot w(\mathbf{q}, \mathbf{p}) \cdot c_i(\mathbf{q}) \; .\]
Here $h_i$ spatially weights $\mathbf{q}$ based on its distance to $\mathbf{p}$, and $w$ is an edge-stopping weight in $[0, 1]$ that is high for pixels with similar G-buffer entries and low when G-buffers differ~\cite{Dammertz:2010:denoise}.
Values are normalized by $k$ which is the sum of all $h \cdot w$ products for the output pixel.
For $h$, we use a Gaussian-like filter $(1,2,1) \otimes (1,2,1)$ where $\otimes$ is the outer product.
$w$ is computed based on the G-buffer differences between $p$ and $q$.
Given our discretized normals, we set $w = 0$ if the normal vectors $\mathbf{n}$ of $\mathbf{p}$ and $\mathbf{q}$ differ.
Otherwise, $w$ is computed based on the distance of $\mathbf{p}$'s and $\mathbf{q}$'s albedo colors $\mathbf{\rho}^{hsv}$ in HSV space and depth values $d$ of the first visible surface:
\begin{equation*}
w = 
     \begin{cases}
        0 & \textrm{if} \; \mathbf{n}_{\mathbf{p}} \neq \mathbf{n}_{\mathbf{q}},\\
        \frac{1}{1 + \langle \rho_{\Delta}, \rho_{\Delta} \rangle} \cdot e^{-\lvert d_{\mathbf{p}} - d_{\mathbf{q}} \rvert \phi_d} & \textrm{otherwise}
     \end{cases}
\end{equation*}
where $\rho_{\Delta}$ is $(\mathbf{\rho}^{hsv}_{\mathbf{p}} - \mathbf{\rho}^{hsv}_{\mathbf{q}}) \phi_{\rho}$ and $\phi_d = \textrm{voxel size}$ and $\phi_{\rho} = 8$ control the influence of depth and albedo differences.
Intuitively, the denoiser blurs neighbor pixels with a Gauss kernel only if they belong to a similar surface.

\paragraph*{Upsampling}
To retain interactivity with expensive rendering on less capable systems, the internal rendering resolution can be reduced to only compute 1/4th, 1/16th, or 1/64th of all pixel colors.
The gaps between rendered pixels are filled using a simple upsampling method (\cref{fig:vcnt:resolve:upsampling}).
We leverage the hierarchical accumulation of the denoiser to simultaneously perform this hole filling.

For pixels that were not yet rendered (normal bits \texttt{111}), no denoising is performed in the hierarchical accumulation.
Instead, the three color channels of the \textit{Denoise} A/B ping-pong buffers store a 2D offset to the closest neighbor pixel that was already sampled, as well as the distance to this pixel (\cref{fig:vcnt:resolve:resolve-ours}).
In each pass $i$, pixel $\mathbf{p}$'s neighbors with distance $2^i$ (\cref{fig:vcnt:resolve:atrous}) are searched for a pixel $\mathbf{q}$ that either stores an offset to another rendered pixel $\mathbf{o}_\mathbf{q}$ or rendered a sample itself ($\mathbf{o}_\mathbf{q}=\mathbf{0}$).
The closest already rendered pixel to $\mathbf{p}$ is found by updating $\mathbf{o}_{\mathbf{p}} \gets \min_{\mathrm{length}}[\mathbf{o}_{\mathbf{p}},\; \mathbf{o}_{\mathbf{q}} + (\mathbf{q} - \mathbf{p})]$.
After finishing the full kernel evaluation, the (possibly denoised) color value from the rendered neighbor referenced by the offset is directly copied to $\mathbf{p}$.
The upsampled result creates a Voronoi pattern that progressively dissolves the more pixels are rendered (\cref{fig:vcnt:resolve:upsampling}).
In fact, the hierarchical offset computation closely resembles the Jump-Flood-Algorithm~\cite{Rong:2006:JFA} computation for rasterizing approximate Voronoi diagrams.
We use stratified sampling to ensure that one pixel is rendered per respective $2^2$, $4^2$, or $8^2$ pixel block after a minimal number of frames was rendered. This limits offset distances in upsampling.
Sequences between blocks are blue-noise displaced~\cite{Peters:2016:BlueNoise} (\cref{fig:vcnt:resolve:upsampling}).
Since post-processing accesses identical neighbor pixels for both denoising (pixel already rendered) and upsampling (otherwise), we can implement both filters within a single loop in the shader for high efficiency.

\subsection*{Volcanite application}
\label{sec:vcnt:application}
Apart from the graphical application, Volcanite can be executed through a \gls{cli} for task automation or on headless systems with full access to all parameters (\cref{fig:vcnt:visual-abstract}).
In the graphical application, parameters are interactively controlled through a \gls{gui}. 

\paragraph*{Material editor and transfer functions}
\label{sec:vcnt:renderer:transferfunction}

\noindent
The right \gls{gui} window in the application (\cref{fig:vcnt:visual-abstract:gui}) contains the material editor.
It controls the transfer function for mapping labels to optical properties.
Similar to multi-material shading user interfaces~\cite{Igouchkine:2018:MultiMat, Ljung:2016:TF} or the more complex attribute selection in~\cite{Lesar:2022:VC}, this is a multi-indirection mapping (\cref{fig:vcnt:renderer:transferfunction:mapping}):
Materials are assigned through a minimum and maximum \textit{filter} attribute value, e.g.\ all labels whose \textit{type} is in $[5.2, 27]$ for the left material in \cref{fig:vcnt:renderer:transferfunction:editor}.
For a given label, the first in the list of materials with a matching \textit{filter} interval is selected.
This material then determines the label's color, opacity, and emission.
Colors can be either constant or based on a color map from~\cite{Smith:2015:viridis,Moreland:2009:diverging}.
For color maps, a secondary visualization attribute and interval map labels to colors; these are independent of the \textit{filter} attribute (Attribute in \cref{fig:vcnt:renderer:transferfunction:editor}).
Values are linearly mapped to the color map, either clamping its end points to values outside the mapping interval (Clamp) or repeating the color map for such values (Repeat) (\cref{fig:vcnt:renderer:transferfunction:editor}).
Alternatively, the labels of the material can be randomly assigned to colors in the color map (Random) which is particularly useful for categorical attributes.
Emission and opacity factors are constant for all labels of a respective material.

Label attributes $A_i: \mathbb{N} \to \mathbb{R}$ are managed in an SQLite3 database on the \gls{cpu} side which allows adding derived attributes through SQL queries.
The dummy attribute $A_0(l) = l$ for labels $l$ is always available.
Additionally, Volcanite can compute a number of voxel attributes like volume, or surface area for each label for a given \gls{csgv} volume, even if no user provided attributes exist.
Attributes are uploaded to the \gls{gpu} as floating point buffers on demand where the label directly indexes into the buffer.
As the label count in a volume is typically orders of magnitude smaller than the number of voxels, this is acceptable.
Unfortunately, label IDs in a segmentation volume do not necessarily span a continuous region, possibly creating high maximum labels and thus attribute buffer sizes.
When attributes other than $A_0$ are used, we therefore pre-process each volume by traversing all voxels along a 3D Morton curve~\cite{Morton:1966:curve} and replacing voxel labels with ascending indices in the order of occurrence.
The mapping is stored in the database to retain the original labels as well.
After this step, the maximum label in the volume is one before the total label count and the labels span a continuous index region.

\glsreset{cli}
\paragraph*{Command-line interface}\label{sec:vcnt:headless}
Volcanite features a headless mode through its \gls{cli}.
Volcanite provides \gls{cli} arguments for fine-grained control of all features, including compressing or decompressing segmentation volumes and rendering images or video sequences.
Additionally, the arguments expose advanced settings not present in the GUI.
On systems that do not provide a windowing system like headless servers, Volcanite can be built without windowing dependencies.
While we currently do not support direct python integration, the comprehensive \gls{cli} allows using Volcanite from python and bash scripts by passing input volumes and output artifacts through the file system. %

\subsection*{GPU rendering pipeline}
\label{sec:vcnt:pipeline}
Volcanite is implemented in C++ using the Vulkan API for \gls{gpu} interaction.
Volume compression is implemented on the \gls{cpu} host and parallelized with OpenMP~\cite{openmp45} while rendering and all accompanying decoding and caching systems are implemented in GLSL \gls{gpu} shaders.
Shaders are compiled just-in-time for a given compressed volume which allows passing many parameters and code paths as compile time constants, e.g.\ brick and cache sizes, enabled \gls{csgv} operations and features in use, or the used \gls{csgv} encoder.
To that end, a \gls{csgv} volume and its accompanying encoder object (as we provide different encoder implementations for plain 4-bit, range \gls{ans}, or random access wavelet tree~\cite{Piochowiak:2025:csgvr} encoded \gls{csgv} volumes) implement a virtual function that provides constant definitions for the shader compiler.
This eliminates unused code paths and lets the compiler use instruction-level optimizations, especially since many values are power-of-two constants, and maximize \gls{gpu} register usage.
Regression tests ensure that rendering under different configurations always produces correct results.

For the renderer to access the visible voxel data along traced rays, all visible \gls{csgv} bricks must be decompressed into a brick cache.
Our cache management adapts the request, assign, and provision stages from shading-atlas streaming~\cite{Mueller:2018:sas} as in \cite{Piochowiak:2024:csgv} to assign visible bricks to free regions in the cache buffer (\cref{fig:vcnt:gpu-rendering:pipeline}).
We make two adaptions to the pipeline:
First, the brick decompression is separated from the assign stage for switching efficiently between different brick encoding types.
Second, our new multi-pass resolve stage which computes denoising and upsampling in a single \`{A}-trous filter is added at the end of the pipeline.
In particular, the pipeline operates as follows:
During rendering (\textit{render}), the rays that are traced through the volumes either access voxels in already decoded bricks or mark undecoded bricks as newly visible.
Bricks that do not contain any visible labels are skipped completely.
The \textit{request} stage executes one thread per brick to mark invisible bricks (based on the labels in their palettes and the current transfer function), to push newly visible bricks on a request stack for their respective LOD, and to push bricks that are in the cache but no longer visible onto a free stack.
As opposed to the original method, we no longer compute the requested \glspl{lod} for newly visible bricks in the rendering stage but in the next frame's request stage instead.
This is possible as \glspl{lod} only depend on brick center positions~\cite{Piochowiak:2024:csgv}.
It removes the latency in the \gls{lod} computation which previously could lead to bricks being decoded in a deprecated \gls{lod} before rendering in the next frame which caused visible artifacts when the camera moves.
The \textit{provision} stage checks if each \gls{lod}'s free stack is sufficient to fulfill all brick location requests from the respective request stack.
If this is not the case, it allocates new memory space at the end of the brick cache by atomically increasing the index pointing to the current top element in the cache buffer.
The \textit{assign} stage can then assign newly visible bricks to free slots in the cache (either retrieved from the free stacks or from the cache top) after which the \textit{decode} stage will decompress these bricks.
All of this happens for each rendered frame.
In the first frames, the brick cache is completely empty and all initially visible bricks have to be decoded once, but in later frames, runtimes of stages apart from \textit{render} are generally short ($<2$ ms).

\subsection*{Memory management}\label{sec:vcnt:renderer:memory}
Raw segmentation volume sizes easily exceed \gls{vram} limits for rendering on consumer \glspl{gpu}.
For example, a $2048^3$ volume of 32-bit integer labels already occupies 32 GiB of memory.
We use our \gls{csgv} compression and optional \gls{cpu} to \gls{gpu} data streaming to minimize memory consumption.
Bricks are decoded into the cache only on demand.
Further, the renderer requests bricks in coarser \glspl{lod} if they are further away from the camera which reduces the memory footprint of the brick in the cache by a factor of 8 per \gls{lod}.
Nevertheless, the overall \gls{vram} consumption can be high as buffers other than the large \textit{brick cache} have to be present as well:
the compressed \textit{brick encodings}, %
\textit{cache information} denoting where and if bricks are cached, \textit{attributes} for the labels, as well as \textit{material} and \textit{rendering parameters}.
\Cref{fig:vcnt:renderer:memory} gives an overview of \gls{gpu} buffers and data flows.
While total memory consumption is usually significantly smaller than the uncompressed volume, it may still exceed the \gls{vram}.
Most large buffer sizes are fixed or directly determined by the dataset.
Therefore, the brick cache size must be reduced in such cases.
In the worst case, some requested bricks might not fit into the smaller cache and can only be accessed in the coarsest \gls{lod} directly from the encoding. 
We therefore lay out multiple optional strategies to reduce the strain on the caching system and enable the rendering of larger volumes in the following.

\paragraph*{Encoding buffer optimizations}\label{sec:vcnt:renderer:detail-streaming}
With compression rates of 1\% to 3\%, most \gls{csgv} encoded volumes fit within the \gls{vram} of commodity \glspl{gpu}.
First, as our implementation uses 32-bit indexing for brick offsets within the contiguous encoding buffers (\textit{Brick Starts} in \cref{fig:vcnt:renderer:memory}), encodings would be limited to 16 GiB size. %
Since \gls{gpu} APIs cannot allocate a contiguous buffer larger than 4 GiB, practical encoding size limits would be even smaller.
Second, encodings may be larger than \gls{vram} overall, making fully GPU-based rendering impossible.

We address the first problem by distributing bricks over multiple split encoding buffers.
At compression time, the number of bricks whose encodings surpass the target buffer size is determined as $\eta$.
Brick indices $b_i$ are then mapped to buffer indices as $\lfloor \frac{b_i}{\eta} \rfloor$ in constant time.
Usually, pointers are not supported by \gls{gpu} APIs but Vulkan introduced references with the \texttt{buffer device address} extension. %
This enables our \gls{gpu} brick decompression to be agnostic towards the actual buffer in which an input encoding is stored, without introducing any performance hazards like branches.

For the second challenge, we use \gls{cpu} to \gls{gpu} streaming as in~\cite{Piochowiak:2024:csgv}.
The finest \gls{lod} of each brick's operation stream is stored in separate buffers (\textit{Brick Detail Encodings} in \cref{fig:vcnt:renderer:memory}).
Coarser \glspl{lod} constitute only 32-55\% of overall memory (\cref{fig:vcnt:csgv:results}) and are uploaded fully to the GPU. 
The finest \gls{lod} is only accessed for bricks closest to the camera.
The renderer requests those brick indices on demand (\textit{Detail Requests} buffer in \cref{fig:vcnt:renderer:memory}) in a buffer that is regularly downloaded to the \gls{cpu}.
On the \gls{cpu}, an asynchronous pipeline copies requested detail encodings to a consecutive buffer for the \gls{gpu} (\cref{fig:vcnt:renderer:detail}).
As opposed to the original pipeline~\cite{Piochowiak:2024:csgv}, this process can now take several frames with a higher CPU parallelization to enable the construction of larger buffers.
Once finished, an asynchronous GPU upload for the constructed detail encoding buffer is scheduled during which this buffer is flagged as inconsistent on the GPU side to prevent invalid accesses by the renderer during the transaction.

\paragraph*{Cache paletting.}\label{sec:vcnt:rendering:cachepalette}
By default, the brick cache stores uncompressed 32-bit labels, yet segmentation volumes contain only a few unique labels per brick.
We optionally decompress bricks into a palettized format, similar to Neuroglancer precomputed volumes~\cite{neuroglancer}:
As brick encodings already store a label palette, we store packed indices into these palettes instead of raw labels (\cref{fig:vcnt:vram:cache-palette}).
Given a volume's maximum palette size $\lvert P \rvert{\mathrm{max}}$ it would be sufficient to store $I = \lceil \log_2(\lvert P \rvert{\mathrm{max}}) \rceil$ bits per voxel.
To prevent expensive memory fetches over \gls{vram} word borders, we only store $\lfloor \frac{32}{I} \rfloor$ entries per 32-bit cache entry (\cref{fig:vcnt:vram:cache-palette}).
Cache paletting increases the effective cache size by a packing factor (\cref{tab:vcnt:cache-palette,fig:vcnt:vram:cache-palette}) at the expense of an additional global memory fetch from the added indirection.
The original CSGV~\cite{Piochowiak:2024:csgv} could cause label duplicates in brick palettes due to the limited distances for palette back reference operations.
This leads to higher $\lvert P \rvert{\mathrm{max}}$, possibly worsening the cache utilization with cache paletting as opposed to our adapted encoding operation that allows arbitrary-length back references which never creates palette duplicates (\cref{fig:vcnt:vram:cache-palette-result}).

\paragraph*{Cache defragmentation}
Cache regions strictly belong to a specific \gls{lod} once allocated even after deallocation (pushed to their \gls{lod}'s free stack \cref{fig:vcnt:gpu-rendering:pipeline}) which can cause fragmentation~\cite{Piochowiak:2024:csgv}.
As in~\cite{Piochowiak:2024:csgv, Mueller:2018:sas}, we reset the brick cache once if its full capacity is exceeded for defragmentation.
This is usually not visible as temporal framebuffer accumulation halts while repopulating the cache. 
We deliberately introduce another source of fragmentation:
When camera distances change, bricks must switch \glspl{lod} with a simultaneous deallocation in \gls{lod}$_{\mathrm{prev}}$ and allocation in \gls{lod}$_{\mathrm{cur}}$. 
If the free stack (capacity of 1,048,576 bricks each) of \gls{lod}$_{\mathrm{prev}}$ is full, deallocation would not be possible, causing a deadlock in the original implementation~\cite{Piochowiak:2024:csgv}.
We therefore switch the \gls{lod} in this situation without pushing the freed region on the stack for later use.
This deliberately creates dead cache space until the next defragmentation.

\paragraph*{Brick request limitation}
Cache defragmentation is sufficient as long as the cache is able to hold all visible decoded bricks in their required \glspl{lod} for a rendering configuration.
Otherwise, some rendered pixels may never successfully request space in the cache as all space is held occupied by rays from other pixels.
We solve this with \textit{brick request limitation} that allows only an area of image pixels (green in \cref{fig:vcnt:vram:req-limit}) to request bricks during rendering, concentrating cache ownership on a smaller subset of rays.
Consequently, bricks that are only required for rendering pixels outside of this area will be deallocated, freeing up space in the cache.
Brick request limitation is automatically enabled by the renderer if the difference between the minimum and maximum sample counts of any pixel reaches a threshold.
For a certain duration, the area covers the pixel with the lowest sample count while a feedback control loop increases or decreases its size depending on if the pixel's sample count increases.

\Cref{fig:vcnt:shading:rays} shows how more volume bricks are accessed the higher the path depth (i.e.\ the number of indirect ray segments after subsequent surface hit points) is.
The effect is limited for view, shadow, and ambient occlusion rays (depth $\leq 1$) but long path tracing rays may request decompression of full bricks that are only relevant for single samples.
We therefore provide a user-controlled ray depth limit for requests and set it to 1 by default.
Note that most path tracing rays at higher depths still access already decoded bricks.
If an undecoded brick --- and therefore the most frequent label from the coarsest brick \gls{lod} --- is accessed instead, we do not invalidate the gathered pixel color.
The bias this introduces is usually small since such indirect illumination is typically low frequency.

\paragraph*{Random number generation}
At multiple occasions during rendering, random numbers determine which bricks are accessed.
These are mainly the ray directions that are sampled after a surface interaction was detected but also include the random sub-pixel offsets of view rays for anti-aliasing.
Recall that the brick cache has a delay of at least one frame as newly accessed bricks are only decoded in the next frame (\cref{fig:vcnt:gpu-rendering:pipeline}).
If random numbers change in this next frame, other (again not yet decoded bricks) might be accessed instead.
For this reason, we run one deterministic random number generator per pixel thread which is initialized from the number of valid samples accumulated for this pixel so far.
This ensures that the pixel will evaluate the exact same ray path until all bricks along this path are available and a valid sample is computed.
To prevent deadlocks, we always change the number every 256 frames.
The pixel coordinate dependent initialization is done according to a tilable $32^2$ blue noise pattern~\cite{Peters:2016:BlueNoise,Ulichney:1993:voidcluster} to optimize sampling noise.
We use 2D PCG hashing~\cite{Jarzynski:2020:gpuhash} as the generator.
The complete initialization seed per pixel $\mathbf{p}$ with currently accumulated valid sample count $s$ after $f$ total rendered frames is
\[\textrm{PCG}_{\textrm{2D}}(\textrm{blueNoise}_{32 \times 32}(\mathbf{p}), (2^8 \lfloor \frac{f}{256} \rfloor) \oplus s)\]
where $\oplus$ is the bit-wise XOR operator.

\vfill

\section*{Declarations}

\paragraph*{Funding}
This work has been supported by the Helmholtz Association (HGF) under the joint research school ``HIDSS4Health -- Helmholtz Information
and Data Science School for Health'' and through the Pilot Program Core Informatics.
A.S., and J.H. recognize support by the Helmholtz Association\textquoteright s Initiative and Networking Fund (INF) under the Helmholtz AI platform grant and the grant ZebraTwin. A.S. recognizes support by the Helmholtz Foundation Model Initiative (HFMI) of the Helmholtz Association under the grants PROFOUND and Virtual Cell.
The funders had no role in study design, data collection, analysis, decision to publish, or preparation of the manuscript.

\paragraph*{Data availability}
All datasets evaluated in this work are publicly available except \cells{} {\small \texttt{Big01/output\_cells-00055.vti}} and \fiber{} {\small \texttt{glassfibrereinforcedpolymer\-\_unloaded\-\_1579x1092x1651\-\_2umVS\-\_labeled\-\_16bit.raw}} which are available on request from the authors in~\cite{Rosenbauer:2020:ETD,Maurer:2022:Fiber}.
We provide a python script to download all public datasets in the evaluation repository at \url{https://github.com/max-pio/volcanite-evaluation}.

\newpage
\paragraph*{Code availability}
Volcanite's code is publicly available under a GPLv3 license at \url{https://github.com/max-pio/volcanite}.
All Volcanite evaluation scripts are publicly available in the main evaluation repository \url{https://github.com/max-pio/volcanite-evaluation} which includes references to the VTK evaluations in \url{https://github.com/max-pio/volcanite-evaluation-vtk} and Neuroglancer evaluations in \url{https://github.com/max-pio/volcanite-evaluation-ng}.

\paragraph*{Author contribution}
Volcanite algorithms and implementation M.P.;
main evaluation scripting and dataset curation M.P.;
Vulkan/C++ backend implementation M.P., R.D.;
Neuroglancer scripting J.H., M.P.;
domain expert feedback J.H., E.B., A.S.;
manuscript writing, M.P., J.H., A.S., C.D.;
funding C.D., A.S.

\paragraph*{Conflict of interest/Competing interests}
The authors declare no competing interests.

\paragraph*{Materials availability}
See data and code availability above.
The Volcanite repository contains precompiled binary releases for Windows and Ubuntu Linux~\url{https://github.com/max-pio/volcanite}.
A data storage containing a snapshot of all code repositories, the precompiled binaries, and the supplemental videos is available at: \url{https://doi.org/10.17605/OSF.IO/K25AY}

\vfill
\FloatBarrier
\hrule
\section*{Extended Data}
\label{sec:extended-data}

\begin{appendices}

\setcounter{figure}{0}
\setcounter{table}{0}
\renewcommand{\thefigure}{A\arabic{figure}}
\renewcommand{\thetable}{A\arabic{table}}

\noindent
\textbf{Video A1}: \href{https://doi.org/10.17605/OSF.IO/K25AY}{\cortex{} offline rendering.}\\[0.5em]
\textbf{Video A2}: \href{https://doi.org/10.17605/OSF.IO/K25AY}{\cells{} interactive exploration.}

\begin{figure*}[ht]   
    \caption{%
    \subref{fig:vcnt:resolve:atrous} \`{A}-trous filters evaluate large kernels through successive iterations with parting support pixels.
    \subref{fig:vcnt:resolve:gbuffer} G-buffer containing per pixel label, depth, and surface normal as filter guide.
    \subref{fig:vcnt:resolve:resolve-ours} Our adapted \`{A}-trous wavelet denoiser and upsampler simplified in 1D . Displayed are all paths contributing to the center pixel in the final image. Per iteration $i$, each pixel of the current denoising buffer read 3 pixels (9 in 2D) from the previous buffer. Step sizes increase as $2^i$, creating large receptive fields (blue). G-Buffer information (colored) guides the denoising. For not yet sampled pixels, the same pipeline computes an offset vector to the closest sampled pixel (dashed). The denoised color of that pixel is copied in the final pass.
    \subref{fig:vcnt:resolve:results} Frame times for \cells{} video animation renderings (denoising and upsampling in \textit{Post-Process}).
    \subref{fig:vcnt:resolve:upsampling} Subsampling of \cells{} renderings (4K output). To retain interactivity on weaker systems, rendering can be limited to one pixel per $2^2$, $4^2$ or $8^2$ block (left). For non-rendered pixels, our upsampling copies their closest rendered pixel (middle).
    \subref{fig:vcnt:resolve:denoising} Path tracing rendering with one radiance sample per pixel. Our denoising removes Monte Carlo noise from early frames.
    }
    \label{fig:vcnt:resolve}

    \let\oldspylenstextoptions\spylenstextoptions
    \renewcommand{\spylenstextoptions}[1]{{\small\sffamily\textbf{\contour{black}{#1}}}}

    \begin{subfigure}[t]{0.49\textwidth}
        \caption{}
        \label{fig:vcnt:resolve:atrous}

        \vspace{-.8em}
        
        \includegraphics[width=\linewidth,trim={0cm 0.2cm 0cm 0.2cm},clip]{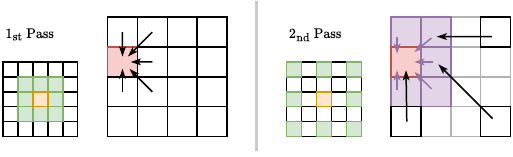}

        \vspace{0.5cm}
        \includegraphics[width=\linewidth]{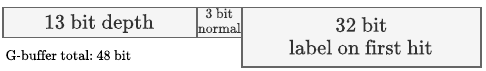}
    \end{subfigure}
    \hfill
    \begin{subfigure}[t]{0.48\textwidth}
        \caption{}
        \label{fig:vcnt:resolve:gbuffer}

        \vspace{-0.8em}

        \contentCmpSplitFour{\includegraphics[trim={20cm 10cm 20cm 15cm},clip]{paper/volcanite/results/gbuffer-eval/cells_depth.jpg}}{Depth}
                            {\includegraphics[trim={20cm 10cm 20cm 15cm},clip]{paper/volcanite/results/gbuffer-eval/cells_normal.jpg}}{Normal}
                            {\includegraphics[trim={20cm 10cm 20cm 15cm},clip]{paper/volcanite/results/gbuffer-eval/cells_label.jpg}}{Label}
                            {\includegraphics[trim={20cm 10cm 20cm 15cm},clip]{paper/volcanite/results/gbuffer-eval/cells_albedo.jpg}}{Albedo}
    \end{subfigure}

    \begin{subfigure}[t]{0.59\textwidth}
        \caption{}
        \label{fig:vcnt:resolve:resolve-ours}

        \vspace{-0.3cm}
        {
        \centering
        \includegraphics[width=\textwidth]{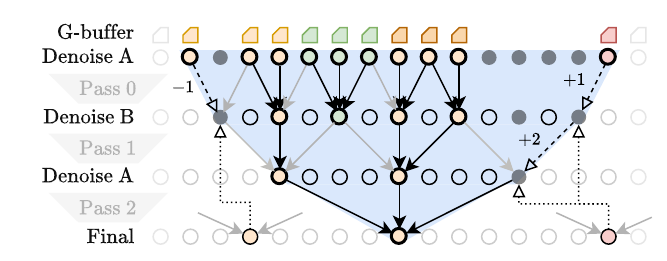}
        }

    \end{subfigure}
    \hfill
    \begin{subfigure}[t]{0.4\textwidth}
        \caption{}
        \label{fig:vcnt:resolve:results}
        
        \vspace{-0.3cm}

        \includegraphics[width=\textwidth]{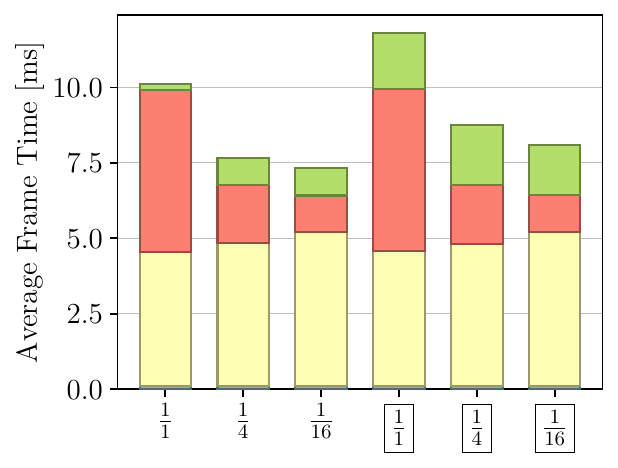}

        \vspace{-0.9cm}
        \makebox[0pt][l]{%
        \hspace*{-8.5cm}
            \hfill
            \includegraphics[width=1.4\textwidth]{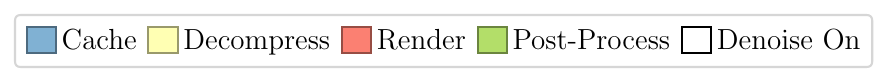}
        }

    \end{subfigure}

    \vspace{0.15cm}

    \begin{subfigure}[t]{\textwidth}
        \caption{}
        \label{fig:vcnt:resolve:upsampling}

        \vspace{-1.5em}
        \begin{minipage}{0.325\textwidth}
            \centering
            \small
            $1/16$ pixels
            \magnifyBottom{
                \includegraphics[width=\columnwidth,trim={30cm 20cm 30cm 30cm},clip]{paper/volcanite/fig/resolve/shadows/4k/cells-sub2-raw.jpg}
            }{8}{0.3}{0.6}{0.6}{0.3}{}{}
        \end{minipage}
        \hfill
        \begin{minipage}{0.325\textwidth}
            \centering
            \small
            upsampled
            \magnifyBottom{
                \includegraphics[width=\columnwidth,trim={30cm 20cm 30cm 30cm},clip]{paper/volcanite/fig/resolve/shadows/4k/cells-sub2.jpg}
            }{8}{0.3}{0.6}{0.6}{0.3}{}{}
        \end{minipage}
        \begin{minipage}{0.325\textwidth}
            \centering
            \small
            ground truth
            \magnifyBottom{
                \includegraphics[width=\columnwidth,trim={30cm 20cm 30cm 30cm},clip]{paper/volcanite/fig/resolve/shadows/4k/cells-sub0.jpg}
            }{8}{0.3}{0.6}{0.6}{0.3}{}{}
        \end{minipage}
    \end{subfigure}

    \vspace{-0.1cm}

    \begin{subfigure}[t]{\textwidth}
        \caption{}
        \label{fig:vcnt:resolve:denoising}

        \vspace{-1.5em}
        \begin{minipage}{0.325\textwidth}
            \centering
            \small
            $1$ sample / pixel
            \magnifyBottom{
                \includegraphics[width=0.9\columnwidth,trim={30cm 20cm 30cm 30cm},clip]{paper/volcanite/fig/resolve/pt/4k/cells-sub0.jpg}
            }{8}{0.3}{0.6}{0.75}{0.3}{}{}
        \end{minipage}
        \hfill
        \begin{minipage}{0.325\textwidth}
            \centering
            \small
            denoised
            \magnifyBottom{
                \includegraphics[width=0.9\columnwidth,trim={30cm 20cm 30cm 30cm},clip]{paper/volcanite/fig/resolve/pt/4k/cells-denoise-sub0.jpg}
            }{8}{0.3}{0.6}{0.75}{0.3}{}{}
        \end{minipage}
        \hfill
        \begin{minipage}{0.325\textwidth}
            \centering
            \small
            ground truth
            \magnifyBottom{
                \includegraphics[width=0.9\columnwidth,trim={30cm 20cm 30cm 30cm},clip]{paper/volcanite/fig/resolve/pt/4k/cells-4000spp.jpg}
            }{8}{0.3}{0.6}{0.75}{0.3}{}{}
        \end{minipage}

    \end{subfigure}

    \let\spylenstextoptions\oldspylenstextoptions

\end{figure*}

\begin{figure*}[ht]

\caption{%
\gls{gpu} memory management.
\subref{fig:vcnt:vram:cache-palette-result} The \opdelta operation is crucial to limit max. brick palette sizes $|P|$, affecting cache packing factors. Our new (unlimited) \opdelta eliminates palette duplicates.
\subref{fig:vcnt:vram:req-limit} Brick request limitation only lets certain pixels (green) request brick cache. This prevents deadlocks on tight cache budgets by concentrating resources around the pixel with the lowest number of rendered samples (red). \subref{fig:vcnt:vram:results} \gls{gpu} memory usage when rendering an image with shadow rays.
For small data, the cache dominates the memory consumption on our 16 GiB \gls{vram} \gls{gpu}.
For larger data, where memory efficiency is imperative, our method significantly limits memory consumption compared to raw data sizes.
For \polyamid{}, \fiber{}, \honebv{}, \griesser{}, \cortex{}, and \honewm{} we use our cache paletting to increase the effective cache size by a \textit{packing} factor.
Total \gls{vram} consumption includes other data like framebuffers, materials, and parameters.}
\label{fig:vcnt:vram}

\begin{subfigure}[t]{0.6\textwidth}
    \raggedright
    \caption{}

    \vspace{-0.5cm}

    {
    \label{fig:vcnt:vram:cache-palette}
    \centering
    \includegraphics[width=\linewidth]{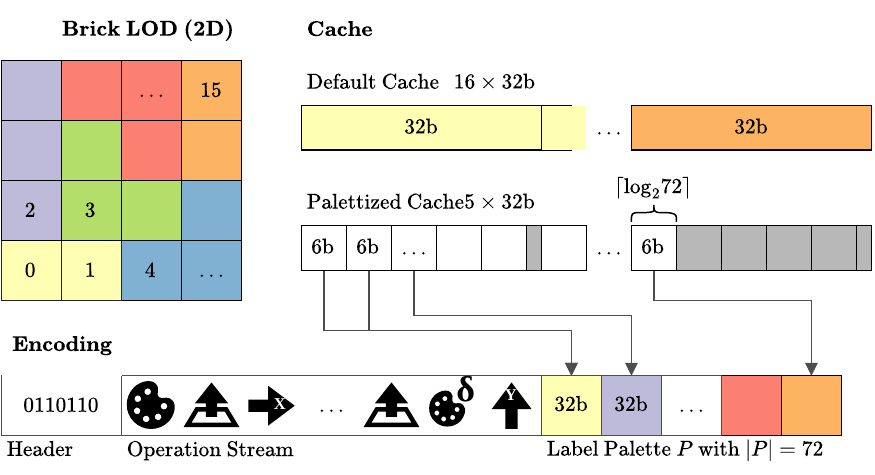}
    }

\end{subfigure}
\hfill
\begin{subfigure}[t]{0.39\textwidth}
    \caption{}
    \label{fig:vcnt:vram:cache-palette-result}

    \vspace{-0.2cm}

    {
    \centering
    \includegraphics[width=\textwidth]{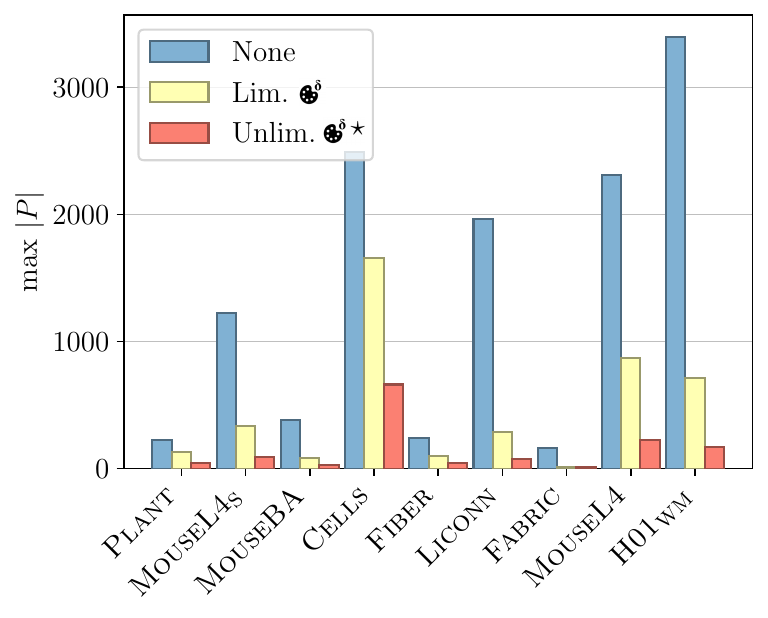}
    }

\end{subfigure}

\vspace{-0.6cm}

\begin{subfigure}[t]{\textwidth}
    \caption{}
    \label{fig:vcnt:vram:req-limit}

    \vspace{-0.3cm}
    
    {
    \centering
    \begin{minipage}{0.495\textwidth}
        \magnifyBottom{\includegraphics[width=\linewidth,trim={0 0 0 1cm},clip]{paper/volcanite/fig/req_limit/request_limit.jpg}}{2}{0.35}{0.8}{0.55}{0.4}{}{}
    \end{minipage}
    \hfill
    \begin{minipage}{0.495\textwidth}
        \magnifyBottom{\includegraphics[width=\linewidth,trim={0 0 0 1cm},clip]{paper/volcanite/fig/req_limit/request_limit_rgb.jpg}}{2}{0.35}{0.8}{0.55}{0.4}{}{}
    \end{minipage}
    }

\end{subfigure}

\vspace{-0.2cm}

\begin{subfigure}[t]{\textwidth}
    \caption{}
    \label{fig:vcnt:vram:results}

    \vspace{-0.5cm}
    
    {
    \footnotesize
    \centering
    \hspace{1cm}
    \begin{filecontents*}{paper/volcanite/results/vram/vram-eval.csv}
#fmt:{orig_gb},{mem_total_mb},{mem_encoding_mb},{mem_cache_mb},{mem_cache_fillrate_pcnt},{mem_materials_mb},{mem_cache_packing_factor}
Data Set,Uncompressed [GB],Total VRAM [MB],Compressed Encoding [MB],Cache [MB],Cache Usage [Pcnt],Attributes / Materials [MB],Cache Packing Factor,Packing Enabled
# 2026.02.05-20:38:05
# pa66 -----------------
# ./data/eval/pa66.csgv --config ./volcanite/eval/volcanite-evaluation/config/pa66.vcfg --resolution 1920x1080 --config global-shadows --video-cfg d1s1024r0:0z0:0 --config [Display] Accumulation Frames: 1024 --verbose --cache-palette --cache-size 4095
pa66,2.5688,1884.028928,92.75383599999999,1542.705692,33.99273753166199,67.125536,8,
# Griesser2022-sample -----------------
# ./data/eval/Griesser2022-sample.csgv --config ./volcanite/eval/volcanite-evaluation/config/Griesser2022-sample.vcfg --resolution 1920x1080 --config global-shadows --video-cfg d1s1024r0:0z0:0 --config [Display] Accumulation Frames: 1024 --verbose --cache-palette --cache-size 2048
Griesser2022-sample,256.97751552,2567.962624,138.792,2180.538956,10.071424394845963,67.125536,4,
# Ara2016 -----------------
# ./data/eval/Ara2016.csgv --config ./volcanite/eval/volcanite-evaluation/config/Ara2016.vcfg --resolution 1920x1080 --config global-shadows --video-cfg d1s1024r0:0z0:0 --config [Display] Accumulation Frames: 1024 --verbose --cache-size 4095
Ara2016,2.40768,4580.966399999999,16.813688,4315.495132,15.839453041553497,67.125536,1,(disabled)
# cells -----------------
# ./data/eval/cells.csgv --config ./volcanite/eval/volcanite-evaluation/config/cells.vcfg --resolution 1920x1080 --config global-shadows --video-cfg d1s1024r0:0z0:0 --config [Display] Accumulation Frames: 1024 --verbose --cache-size 4095
cells,4,4682.416128,118.405068,4315.41462,10.500620305538177,67.125536,1,(disabled)
# xtm-battery -----------------
# ./data/eval/xtm-battery.csgv --config ./volcanite/eval/volcanite-evaluation/config/xtm-battery.vcfg --resolution 1920x1080 --config global-shadows --video-cfg d1s1024r0:0z0:0 --config [Display] Accumulation Frames: 1024 --verbose --cache-size 4095
xtm-battery,0.21952000000000002,1348.5342719999999,10.358372,1089.584604,18.449531495571136,67.125536,1,(disabled)
# azba -----------------
# ./data/eval/azba.csgv --config ./volcanite/eval/volcanite-evaluation/config/azba.vcfg --resolution 1920x1080 --config global-shadows --video-cfg d1s1024r0:0z0:0 --config [Display] Accumulation Frames: 1024 --verbose --cache-size 4095
azba,0.38543760000000005,2111.8976,1.917176,1861.419052,8.54702740907669,67.125536,1,(disabled)
# H01-bloodvessel -----------------
# ./data/eval/H01-bloodvessel.csgv --config ./volcanite/eval/volcanite-evaluation/config/H01-bloodvessel.vcfg --resolution 1920x1080 --config global-shadows --video-cfg d1s1024r0:0z0:0 --config [Display] Accumulation Frames: 1024 --verbose --cache-palette --cache-size 4095
H01-bloodvessel,237.01023225,4685.824,109.66756799999999,4327.52766,7.902970165014267,67.125536,4,
# H01-wm -----------------
# ./data/eval/H01-wm.csgv --config ./volcanite/eval/volcanite-evaluation/config/H01-wm.vcfg --resolution 1920x1080 --config global-shadows --video-cfg d1s1024r0:0z0:0 --config [Display] Accumulation Frames: 1024 --verbose --cache-palette --stream-lod --cache-size 4095
H01-wm,2220.0451072,15621.48864,11564.740192,4353.085548,3.8945529609918594,67.125536,4,
# Wolny2020 -----------------
# ./data/eval/Wolny2020.csgv --config ./volcanite/eval/volcanite-evaluation/config/Wolny2020.vcfg --resolution 1920x1080 --config global-shadows --video-cfg d1s1024r0:0z0:0 --config [Display] Accumulation Frames: 1024 --verbose --cache-size 4095
Wolny2020,0.844562432,2270.0359679999997,3.73196,2017.673564,9.926491230726242,67.125536,1,(disabled)
# liconn -----------------
# ./data/eval/liconn.csgv --config ./volcanite/eval/volcanite-evaluation/config/liconn.vcfg --resolution 1920x1080 --config global-shadows --video-cfg d1s1024r0:0z0:0 --config [Display] Accumulation Frames: 1024 --verbose --cache-size 4095
liconn,59.236245000000004,5120.589824,549.710416,4322.234332,15.788371860980988,67.125536,1,(disabled)
# fiber -----------------
# ./data/eval/fiber.csgv --config ./volcanite/eval/volcanite-evaluation/config/fiber.vcfg --resolution 1920x1080 --config global-shadows --video-cfg d1s1024r0:0z0:0 --config [Display] Accumulation Frames: 1024 --verbose --cache-palette --cache-size 4095
fiber,5.693532936,3734.7000319999997,55.746576,3430.299612,35.780540108680725,67.125536,4,
# Motta2019-small -----------------
# ./data/eval/Motta2019-small.csgv --config ./volcanite/eval/volcanite-evaluation/config/Motta2019-small.vcfg --resolution 1920x1080 --config global-shadows --video-cfg d1s1024r0:0z0:0 --config [Display] Accumulation Frames: 1024 --verbose --cache-size 4095
Motta2019-small,2.147483648,4665.114624,101.11381999999999,4315.41462,22.35197424888611,67.125536,1,(disabled)
# Griesser2022-validation -----------------
# ./data/eval/Griesser2022-validation.csgv --config ./volcanite/eval/volcanite-evaluation/config/Griesser2022-validation.vcfg --resolution 1920x1080 --config global-shadows --video-cfg d1s1024r0:0z0:0 --config [Display] Accumulation Frames: 1024 --verbose --cache-size 4095
Griesser2022-validation,0.07200000000000001,650.248192,2.0516199999999998,399.54798,59.00340676307678,67.125536,1,(disabled)
# Motta2019 -----------------
# ./data/eval/Motta2019.csgv --config ./volcanite/eval/volcanite-evaluation/config/Motta2019.vcfg --resolution 1920x1080 --config global-shadows --video-cfg d1s1024r0:0z0:0 --config [Display] Accumulation Frames: 1024 --verbose --cache-palette --cache-size 1024
Motta2019,927.712935936,15863.775232,14501.890828,1113.063532,18.78455877304077,67.125536,4,
\end{filecontents*}
\pgfplotstabletypeset[
        percent is letter=true,
        col sep=comma,
        header=true,
        column type={r},
        sort,
        sort key={Uncompressed [GB]},
        every head row/.style={after row=\midrule},
        columns={{Data Set},{Uncompressed [GB]},{Total VRAM [MB]},{Compressed Encoding [MB]},{Cache [MB]},{Cache Usage [Pcnt]},{Cache Packing Factor},{Packing Enabled}},
        columns/{Data Set}/.style={string type, column type={l}, postproc cell content/.code={
                \pgfkeyssetvalue{/pgfplots/table/@cell content}{\dataNameFromCSV{##1}}
            }},
        columns/{Uncompressed [GB]}/.style={column name={Orig [GB]}, column type={r|},
                preproc cell content/.code={
                \pgfmathparse{##1}%
                \pgfkeyslet{/pgfplots/table/@cell content}{\pgfmathresult}%
            }, fixed, precision=2, fixed zerofill},
        columns/{Total VRAM [MB]}/.style={column name={Total [GB]}, column type={l}, fixed, precision=2, fixed zerofill,
            preproc cell content/.code={
                \pgfmathparse{##1/1000}%
                \pgfkeyslet{/pgfplots/table/@cell content}{\pgfmathresult}%
            }},
        columns/{Compressed Encoding [MB]}/.style={column name={CSGV [MB]}, fixed, precision=2, fixed zerofill},
        columns/{Cache [MB]}/.style={column name={Cache [MB]}, fixed, precision=2, fixed zerofill},
        columns/{Cache Usage [Pcnt]}/.style={column name={Usage [\%]}, fixed, precision=2, fixed zerofill},
        columns/{Cache Packing Factor}/.style={column name={Packing}, fixed, precision=1},
        columns/{Attributes / Materials [MB]}/.style={column name={Attributes [MB]}, fixed, precision=2, fixed zerofill},
        columns/{Packing Enabled}/.style={column name={}, string type},
        empty cells with={},
    ]{paper/volcanite/results/vram/vram-eval.csv}
    }

\end{subfigure}

\end{figure*}

\begin{figure*}

    \caption{\subref{fig:vcnt:resolve-pipeline} Post-processing pipeline for denoising and upsampling of accumulated rendering frames. Two alternating ping-pong buffers contain the previous (input) and current (output) RGBA radiance (\textit{Accum.}) and sample counts (\textit{SampleCnt}) per pixel. The following resolve pipeline operates on two alternating 16-bit RGBA textures and the G-buffer renderer output which remains unchanged through the pipeline. \subref{fig:vcnt:resolve-depth} In world space, the dataset is scaled to a size of 1 along its largest dimension and centered around the origin. Depth values for the G-buffer are computed as $\lVert \mathbf{h} - \mathbf{p} \rVert$ where $\mathbf{p}$ is the first surface hit point $\mathbf{h}$ projected onto the sphere that encloses the unit cube in direction of the camera $\mathbf{c}$.
    \subref{fig:vcnt:renderer:memory} Relevant memory buffers and data flow within the rendering architecture. Brick encodings are stored in multiple large buffers. For very large datasets, the finest resolution of bricks (detail) is streamed to the GPU on demand. During rendering, visible bricks are decompressed into a \gls{gpu} cache, possibly in a packed format. \subref{fig:vcnt:renderer:detail} CPU Pipeline for constructing encoding buffers for the finest \gls{lod}. Pipeline states in white, procedures in blue boxes.}
    \label{fig:vcnt:pipelines}

    \centering

    \begin{subfigure}[t]{0.69\textwidth}
        \caption{}
        \label{fig:vcnt:resolve-pipeline}
        \centering
        \includegraphics[width=\textwidth]{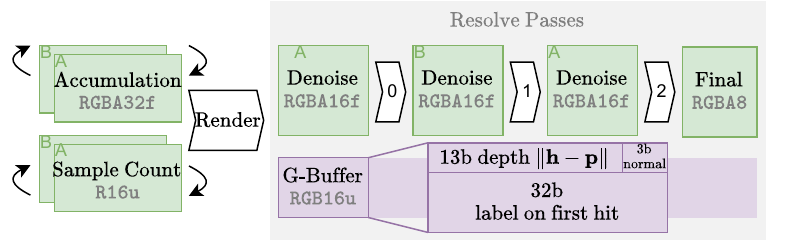}
    \end{subfigure}    
    \hfill
    \begin{subfigure}[t]{0.3\textwidth}
        \caption{}
        \label{fig:vcnt:resolve-depth}
        \centering
        \includegraphics[width=\textwidth]{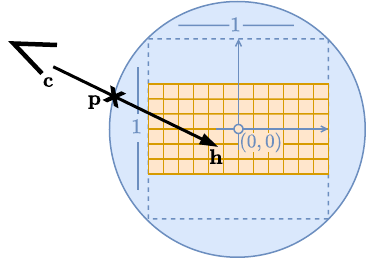}
    \end{subfigure}

    \begin{subfigure}[t]{\textwidth}
        \caption{}
        \label{fig:vcnt:renderer:memory}
        
        \centering
        \includegraphics[width=0.8\textwidth]{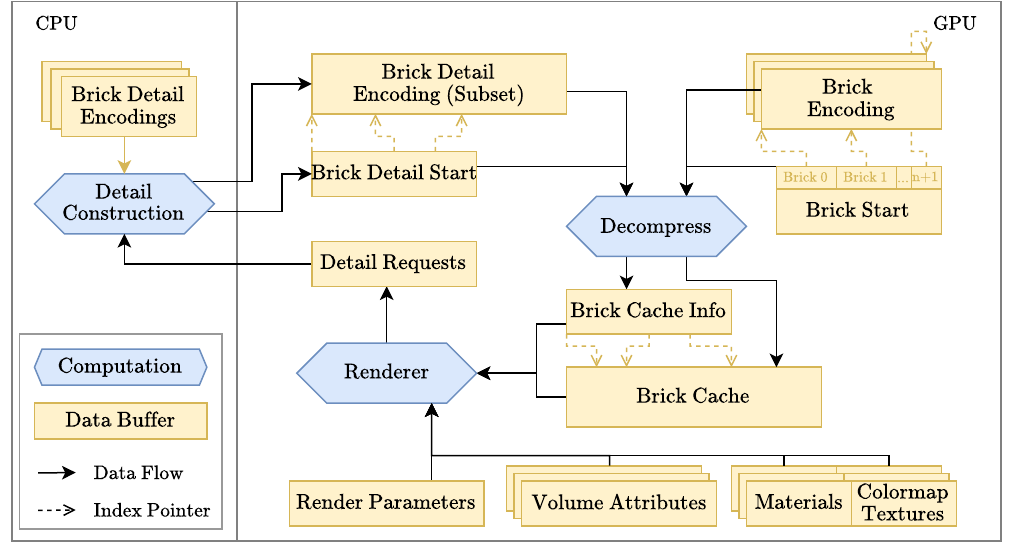}
    \end{subfigure}

    \begin{subfigure}[t]{\textwidth}
        \caption{}
        \label{fig:vcnt:renderer:detail}
        \centering
        \includegraphics[width=0.8\textwidth]{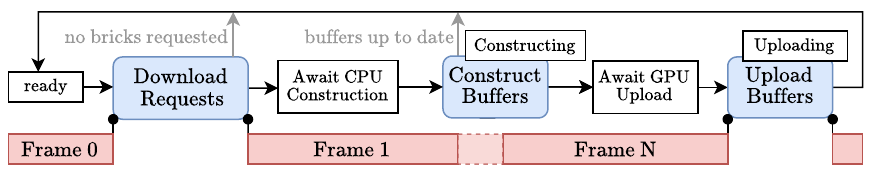}
    \end{subfigure}

\end{figure*}

\begin{figure*}[ht]
    \caption{\subref{fig:vcnt:csgvplots:compression-rate} \glsfirst{csgv} lossless compression rates in \% of the original uncompressed volume (8, 16, or 32 bits per voxel) for all evaluated datasets in our default configurations. \subref{fig:vcnt:csgvplots:csgv-size} Datasets are in ascending order of their original size, reaching from 72 MB to over 2 TB. Our respective \gls{csgv} sizes only reach from 1 MB to 30 GB. \subref{fig:vcnt:csgvplots:label-count} \gls{csgv} is largely independent of dataset label counts (\honewm{} has over 13 million labels). This effectively results in worse compression rates for data stored with 8 (blue) or 16 bits per voxel (orange) as it is originally packed more tightly compared to 32-bit data (red). \subref{fig:vcnt:csgvplots:label-density} Compression rates marginally dependend on local label \textit{densities}, i.e.\ the number of unique labels per brick.}
    \label{fig:vcnt:csgvplots}
    \centering

    \begin{subfigure}{\textwidth}        
        \caption{}
        \label{fig:vcnt:csgvplots:compression-rate}

        \vspace{-0.6cm}

        \hfill\includegraphics[width=0.94\textwidth]{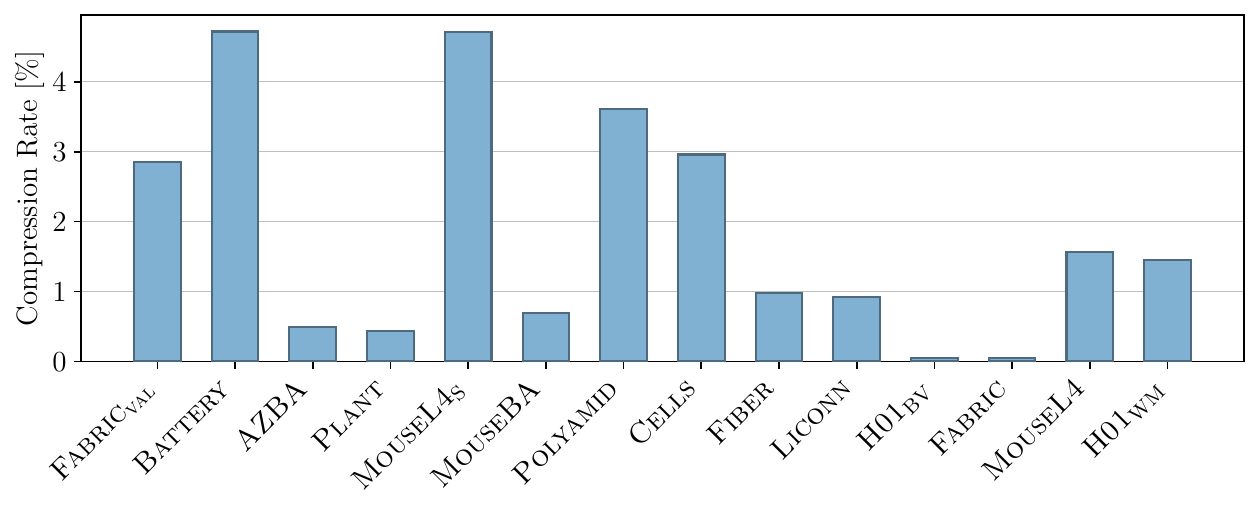}

    \end{subfigure}

    \vspace{-0.5cm}

    \begin{subfigure}{\textwidth}
        \caption{}
        \label{fig:vcnt:csgvplots:csgv-size}

        \vspace{-0.3cm}

        \hfill\includegraphics[width=0.97\textwidth]{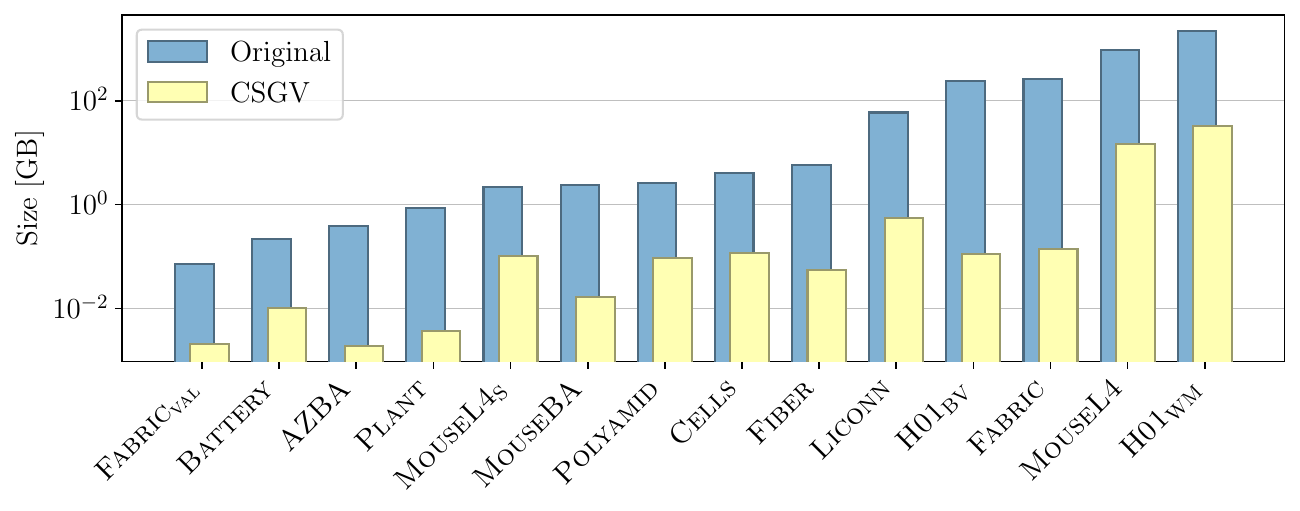}

    \end{subfigure}

    \vspace{-0.6cm}
    
    \begin{subfigure}[t]{0.49\linewidth}
        \caption{}
        \label{fig:vcnt:csgvplots:label-count}

        \vspace{-0.3cm}
        
        \includegraphics[width=\textwidth]{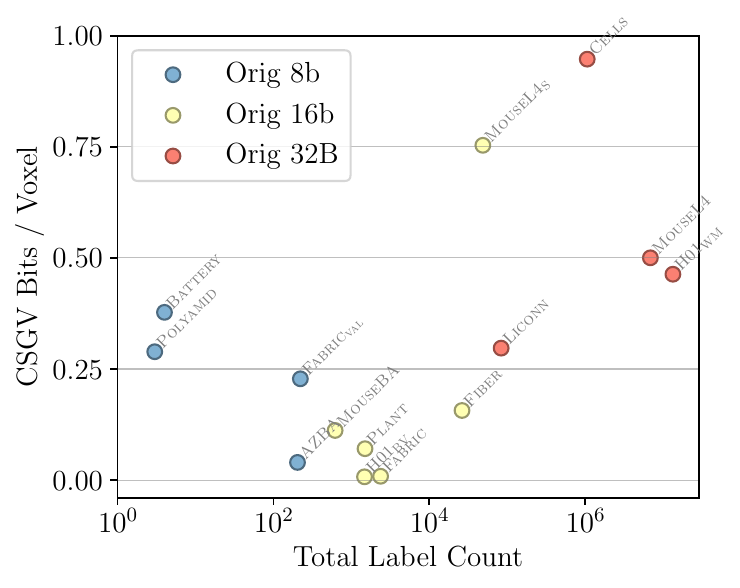}

    \end{subfigure}
    \hfill
    \begin{subfigure}[t]{0.49\linewidth}
        \caption{}
        \label{fig:vcnt:csgvplots:label-density}

        \vspace{-0.3cm}

        \includegraphics[width=\textwidth]{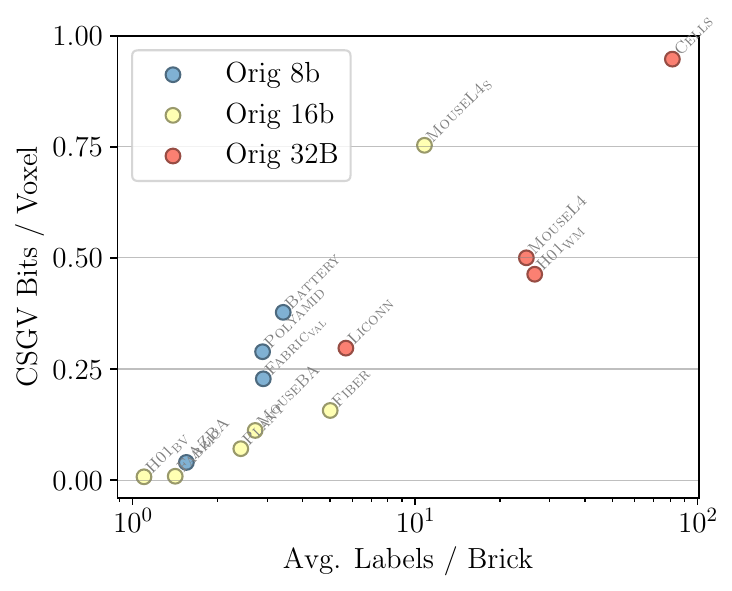}

    \end{subfigure}
  
\end{figure*}

\newcommand{\opimagefromprefix}[1]{%
    \IfStrEqCase{#1}{%
    {p}{\opparent}
    {px}{\opxneighbor}
    {pxy}{\opxneighbor}
    {pxyz}{\opzneighbor}
    {pxyzl}{\oplast}
    {pxyzld-}{\opdeltaold}
    {pxyzld}{\opdelta}
    {ps}{\overbar{\opparent}}
    {pxs}{\overbar{\opxneighbor}}
    {pxys}{\overbar{\opxneighbor}}
    {pxyzs}{\overbar{\opzneighbor}}
    {pxyzls}{\overbar{\oplast}}
    {pxyzld-s}{\overbar{\opdeltaold}}
    {pxyzlds}{\overbar{\opdelta}}
    }[UNKNOWN OPERATION PREFIX]
}

\NewDocumentCommand\supfigurecsgvablation{s O{} m m}{%
    \begin{figure*}[ht]
        \IfBooleanTF{#1}{\vspace{-0.5cm} \ContinuedFloat}{}

        \if\relax\detokenize{#2}\relax
        \else
            \caption{#2}
            \IfBooleanTF{#1}{}{\label{fig:vcnt:csgv-ablation}}
            \vspace{0.5em}
        \fi

        \centering
        \begin{subfigure}{0.49\textwidth}
            \includegraphics[width=\columnwidth]{results/plots/csgv-ablation_#3}
        \end{subfigure}
        \hfill
        \begin{subfigure}{0.49\textwidth}
            \includegraphics[width=\columnwidth]{results/plots/csgv-ablation_#4}
        \end{subfigure}
        
    \end{figure*}
}

\NewDocumentCommand\supfigurecsgvablationlast{s O{} m m}{%
    \begin{figure*}[ht]
        \IfBooleanTF{#1}{\vspace{-0.5cm} \ContinuedFloat}{}

        \if\relax\detokenize{#2}\relax
        \else
            \caption{#2}
            \IfBooleanTF{#1}{}{\label{fig:vcnt:csgv-ablation}}
            \vspace{0.5em}
        \fi

        \centering
        \begin{subfigure}{0.49\textwidth}
            \includegraphics[width=\columnwidth]{results/plots/csgv-ablation_#3}
        \end{subfigure}
        \hfill
        \begin{subfigure}{0.49\textwidth}
            \includegraphics[width=\columnwidth]{results/plots/csgv-ablation_#4}
        \end{subfigure}

        \vspace{4cm}
    \end{figure*}
}

\supfigurecsgvablation[Ablation study showing the effectiveness of each of the operations used in our \gls{csgv} compression~\cite{Piochowiak:2024:csgv} for several datasets.
Each bar shows compression rate (lower is better) for the dataset compressed with the operation set consisting of \oppalette and a respective prefix of the ordered set $\{\opparent,\opxneighbor,\opyneighbor\,\opzneighbor,\oplast,\opdelta[^\star]\}$.
\gls{csgv} encodes each multi-resolution brick of the volume as an ordered stream of one compact operation per voxel where an operation:
\oppalette reads the next label from the palette,
\opparent copies the label from the voxel's parent,
\opxneighbor, \opyneighbor, \opzneighbor copies the label from the voxel's respective X/Y/Z neighbor that has a different parent,
\oplast re-reads the previous label from the palette ($\delta = 1$),
\opdelta re-reads the label from $\delta$ entries ago.
\opdelta in its original variant (\opdeltaold) limits $1 < \delta \leq 16$ while our improved variant (\opdelta) allows unlimited $\delta$ which effectively removes any duplicate entries in brick's label palettes.
Shorter palette sizes allow more efficient cache packing and faster processing, e.g.\ identifying bricks that contain any visible labels.]{Griesser2022-validation}{xtm-battery}

\supfigurecsgvablation*{azba}{Wolny2020}    
\supfigurecsgvablation*{Motta2019-small}{Ara2016}
\supfigurecsgvablation*{pa66}{cells}
\supfigurecsgvablation*{fiber}{liconn}
\supfigurecsgvablationlast*{H01-bloodvessel}{Griesser2022-sample}

\begin{table*}[pt]
\centering
\caption{Preprocessing times of Volcanite, VTK (C++ API), and Neuroglancer with file size on disk (incl. secondary \textit{gzip} compression). Volcanite compresses volumes into CSGV files: Raw compression (\textit{Compr. only}) is faster than the additional time to read the input from disk (\textit{File IO}). The overhead (\gls{gpu} context creation, rendering) until the first frame finished (\textit{\gls{ttff}}) is negligible. VTK renders larger uncompressed (\textit{Direct}) data: preprocessing depends on loading the gzip compressed hdf5 files (\textit{gzip}) from disk. The \gls{ttff} is longer due to larger buffer copying. Neuroglancer compresses voxel segmentations before computing 3D mesh approximations (in parallel with Igneous, all timings include IO). \gls{ttff} is not given due to its client/server streaming architecture.}
\label{tab:vcnt:tools-preprocess}

\small

\begin{tabular}{cl|rrrr|rr}
& & \multicolumn{4}{l|}{Preprocessing Times [s]} & \multicolumn{2}{l}{File Sizes [GB]} \\
& Data Set & Compr. only& File IO& Total with IO& TTFF& Direct & gzip \\
\midrule
\multirow{14}{*}{\rotatebox[origin=c]{90}{Volcanite}} & \dataNameFromCSV{Griesser2022-validation} & 00.1s & 01.3s & 01.4s & 06.3s & 0.002 & 0.002 \\
 & \dataNameFromCSV{xtm-battery} & 00.3s & 00.8s & 01.1s & 02.0s & 0.010 & 0.010 \\
 & \dataNameFromCSV{azba} & 00.3s & 02.2s & 02.5s & 03.5s & 0.002 & 0.001 \\
 & \dataNameFromCSV{Wolny2020} & 00.3s & 02.3s & 02.6s & 03.5s & 0.004 & 0.003 \\
 & \dataNameFromCSV{Motta2019-small} & 01.7s & 07.5s & 09.2s & 10.3s & 0.101 & 0.100 \\
 & \dataNameFromCSV{Ara2016} & 01.1s & 11.3s & 12.4s & 17.4s & 0.017 & 0.015 \\
 & \dataNameFromCSV{pa66} & 02.9s & 09.1s & 12.0s & 13.5s & 0.093 & 0.090 \\
 & \dataNameFromCSV{cells} & 01.8s & 06.3s & 08.1s & 09.1s & 0.118 & 0.116 \\
 & \dataNameFromCSV{fiber} & 02.4s & 20.8s & 23.2s & 24.7s & 0.056 & 0.053 \\
 & \dataNameFromCSV{H01-bloodvessel} & 33.0s & 5m 03.6s & 5m 36.6s & 5m 41.5s & 0.110 & 0.069 \\
 & \dataNameFromCSV{liconn} & 15.0s & 3m 34.3s & 3m 49.3s & 3m 56.0s & 0.550 & 0.535 \\
 & \dataNameFromCSV{Griesser2022-sample} & 2m 11.3s & 13m 46.7s & 15m 58.0s & 16m 03.1s & 0.139 & 0.117 \\
 & \dataNameFromCSV{Motta2019} & 7m 17.0s & 35m 23.6s & 42m 40.6s & 43m 41.4s & 14.502 & 14.482 \\
 & \dataNameFromCSV{H01-wm} & 19m 02.8s & 1h 46.1s & 1h 19m 49.0s & 1h 24m 06.4s & 32.138 & 31.831 \\
\multicolumn{8}{c}{}\\
& & \multicolumn{4}{l|}{Preprocessing Times [s]} & \multicolumn{2}{l}{File Sizes [GB]} \\
& Data Set & & & Total with IO & TTFF & Direct & gzip \\
\midrule
\multirow{9}{*}{\rotatebox[origin=c]{90}{VTK}} & \dataNameFromCSV{Griesser2022-validation} &  &  & 01.5s & 03.8s & 0.072 & 0.002 \\
 & \dataNameFromCSV{xtm-battery} &  &  & 02.8s & 09.5s & 0.220 & 0.016 \\
 & \dataNameFromCSV{azba} &  &  & 05.7s & 17.6s & 0.385 & 0.005 \\
 & \dataNameFromCSV{Wolny2020} &  &  & 06.2s & 19.1s & 0.845 & 0.209 \\
 & \dataNameFromCSV{Motta2019-small} &  &  & 25.7s & 58.1s & 2.147 & 0.202 \\
 & \dataNameFromCSV{Ara2016} &  &  & 25.0s & 1m 02.1s & 2.408 & 0.049 \\
 & \dataNameFromCSV{pa66} &  &  & 32.5s & 1m 49.5s & 2.569 & 0.133 \\
 & \dataNameFromCSV{cells} &  &  & 24.9s & 55.7s & 4.000 & 0.248 \\
 & \dataNameFromCSV{fiber} &  &  & 1m 01.5s & 2m 27.0s & 5.694 & 0.143 \\
& \dots & \multicolumn{4}{r|}{{\footnotesize \dataNameFromCSV{Griesser2022-sample}, \dataNameFromCSV{H01-bloodvessel}, \dataNameFromCSV{H01-wm}, \dataNameFromCSV{liconn}, \dataNameFromCSV{Motta2019}: out of memory}} & & \\\multicolumn{8}{c}{}\\
& & \multicolumn{4}{l|}{Preprocessing Times [s]} & \multicolumn{2}{l}{File Sizes [GB]} \\
& Data Set & Compr. Segm. & Meshing & Total with IO & & Direct & gzip \\
\midrule
\multirow{14}{*}{\rotatebox[origin=c]{90}{Neuroglancer}} & \dataNameFromCSV{Griesser2022-validation} & 01.4s & 04.8s & 06.2s &  & 0.014 & 0.006 \\
 & \dataNameFromCSV{xtm-battery} & 02.5s & 1m 06.2s & 1m 08.7s &  & 0.169 & 0.093 \\
 & \dataNameFromCSV{azba} & 03.2s & 16.0s & 19.2s &  & 0.037 & 0.016 \\
 & \dataNameFromCSV{Wolny2020} & 03.5s & 22.0s & 25.5s &  & 0.066 & 0.028 \\
 & \dataNameFromCSV{Motta2019-small} & 18.8s & 5m 34.8s & 5m 53.5s &  & 1.213 & 0.599 \\
 & \dataNameFromCSV{Ara2016} & 21.0s & 2m 10.2s & 2m 31.3s &  & 0.337 & 0.158 \\
 & \dataNameFromCSV{pa66} & 27.3s & 9m 17.3s & 9m 44.6s &  & 1.395 & 0.746 \\
 & \dataNameFromCSV{cells} & 24.3s & 7m 18.3s & 7m 42.6s &  & 3.505 & 1.740 \\
 & \dataNameFromCSV{fiber} & 34.0s & 1m 53.6s & 2m 27.6s &  & 0.802 & 0.350 \\
 & \dataNameFromCSV{H01-bloodvessel} & 36m 01.6s & 5m 16.5s & 41m 18.1s &  & 2.253 & 0.151 \\
 & \dataNameFromCSV{liconn} & 5m 04.1s & 26m 50.4s & 31m 54.5s &  & 8.218 & 3.932 \\
 & \dataNameFromCSV{Griesser2022-sample} & 38m 20.0s & 10m 41.4s & 49m 01.4s &  & 3.663 & 0.796 \\
 & \dataNameFromCSV{Motta2019} & 1h 26m 23.1s & 12h 12m 21.8s & 13h 38m 44.9s &  & 177.108 & 84.029 \\
 & \dataNameFromCSV{H01-wm} & 2h 47m 35.5s & 7d 12m 25.8s & 7d 3h 01.3s &  & 505.080 & 230.897 \\
\end{tabular}

\end{table*}

\FloatBarrier

\newcommand{\supgputimingplot}[2]{%
    \begin{subfigure}[t]{\textwidth}        
        \pltvideoframes{#1}{#2}

        \vspace{-0.2cm}

        \includegraphics[width=0.9\textwidth]{results/plots/gpu-video-timings_#1_#2.pdf}
    \end{subfigure}

    \begin{subfigure}[t]{\textwidth}        
        \begin{minipage}[t]{0.73\textwidth}\vspace{0pt}%
            \vspace{-0.1cm}

            \hfill\includegraphics[width=0.93\columnwidth]{results/plots/gpu-image-timings_#1_#2.pdf}
        \end{minipage}%
        \pltimageframe{#1}{#2}
    \end{subfigure}
}

\NewDocumentCommand{\subtimingplotfigure}{s O{} m m m m}{
    \begin{figure*}[ht]
        \IfBooleanTF{#1}{\ContinuedFloat}{}
    
        \if\relax\detokenize{#2}\relax
        \else
            \caption{#2}
            \IfBooleanTF{#1}{}{\label{fig:vcnt:all-gpu-timings}}
        \fi

        \centering
        
        \supgputimingplot{#3}{#4}

        \vspace{-0.1cm}
        \noindent\textcolor{gray}{\rule{\textwidth}{0.4pt}}
        
        \vspace{-0.5cm}
        \includegraphics[width=0.9\textwidth]{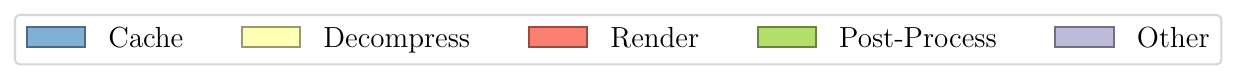}
        \vspace{0.2cm}
        
        \supgputimingplot{#5}{#6}
    \end{figure*}
}

\subtimingplotfigure[Frame timings for image (first 21 frames, static camera) and video renderings with shadow rays. For closeups, rendering dominates timings after the results of the initial decompression overhead are cached.]{Griesser2022-validation}{shadow}{xtm-battery}{shadow}

\subtimingplotfigure*{azba}{shadow}{Wolny2020}{shadow}
\subtimingplotfigure*{Motta2019-small}{shadow}{Ara2016}{shadow}

\subtimingplotfigure*{pa66}{shadow}{cells}{shadow}

\subtimingplotfigure*{fiber}{shadow}{liconn}{shadow}
\subtimingplotfigure*{H01-bloodvessel}{shadow}{Griesser2022-sample}{shadow}

\subtimingplotfigure*{Motta2019}{shadow}{H01-wm}{shadow}

\FloatBarrier

\begin{figure*}[pt]
    \caption{
    \subref{fig:vcnt:shading:rays} Shading modes evaluate light transport along different ray path depths (numbered) and directions. \textit{Local Shading} traces depth $0$ view rays only (black). Optionally, secondary \textit{Shadow Rays} (yellow) are traced to the light source. \textit{Ambient Occlusion} traces $1$ indirection in arbitrary directions (blue). \textit{Path Tracing} evaluates full transport along arbitrary depths (red). Shading modes access an increasing number of bricks in this order (colored).
    \subref{fig:vcnt:shading:performance} Frame times for the 2.2 TB \honewm{} stationary image. Since more advanced shading accesses more volume bricks, render performance decreases.
    \subref{fig:vcnt:shading:images} \honewm{} image rendering after 1024 convergence frames. Shading modes can be changed with immediate feedback. While local shading is fast, shadows or ambient occlusion improve visual perception of structures. Path tracing offers highest visual fidelity, e.g.\ evaluating indirect light bounces from orange neurons.
    The high-frequency label colors at the bottom of the volume indicate the immense label density in the data.
    }
    \label{fig:vcnt:shading}

    \let\oldspylenstextoptions\spylenstextoptions
    \renewcommand{\spylenstextoptions}[1]{{\small\text{\textcolor{black}{#1}}}}

    \begin{subfigure}[t]{0.49\textwidth}
        \phantomcaption{}
        \label{fig:vcnt:shading:rays}
        {
        \centering
        \hspace{0.8cm}\includegraphics[width=0.8\linewidth]{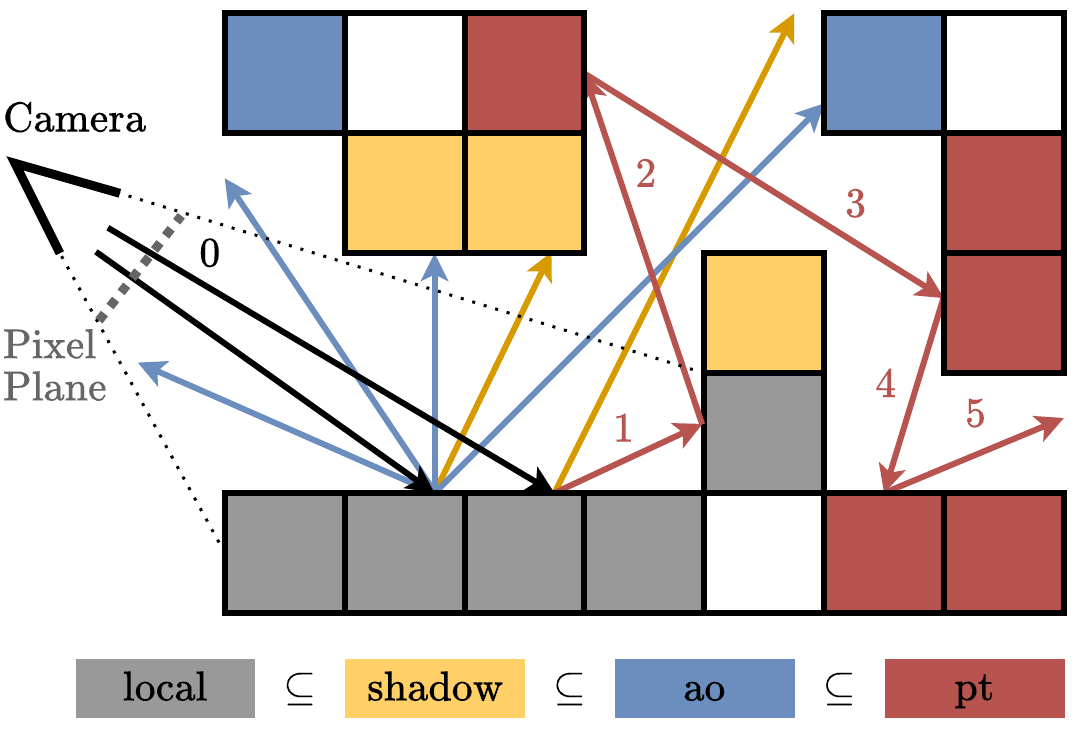}
        }

        \subfiglabel[-4.3cm]{a}
    \end{subfigure}
    \hfill
    \begin{subfigure}[t]{0.49\textwidth}
        \phantomcaption{}
        {
        \label{fig:vcnt:shading:performance}
        \centering

        \vspace{0.2cm}
        \hfill\includegraphics[width=0.9\textwidth]{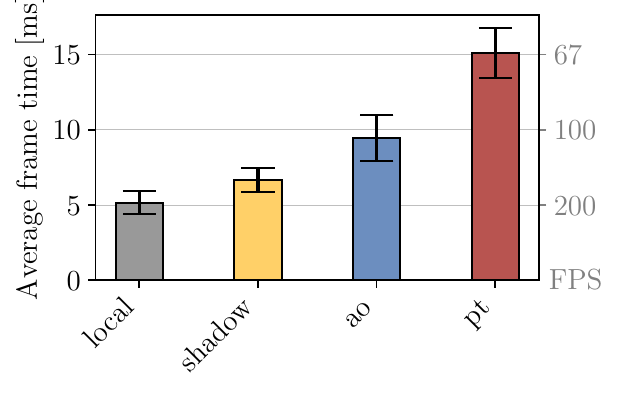}
        }

        \subfiglabel[-4.6cm]{b}
    \end{subfigure}

    \vspace{-0.5cm}

    \begin{subfigure}[t]{\textwidth}
        \caption{}

        \vspace{-0.5em}
        
        \label{fig:vcnt:shading:images}
        \begin{minipage}{0.49\textwidth}
            \centering
            \magnifyTop{
                \includegraphics[width=0.9\columnwidth]{paper/volcanite/results/image/H01-wm_local.jpg}
            }{4}{0.4}{0.6}{0.7}{0.5}{Local\\Shading}{}
        \end{minipage}
        \hfill
        \begin{minipage}{0.49\textwidth}
            \centering
            \magnifyTop{
                \includegraphics[width=0.9\columnwidth]{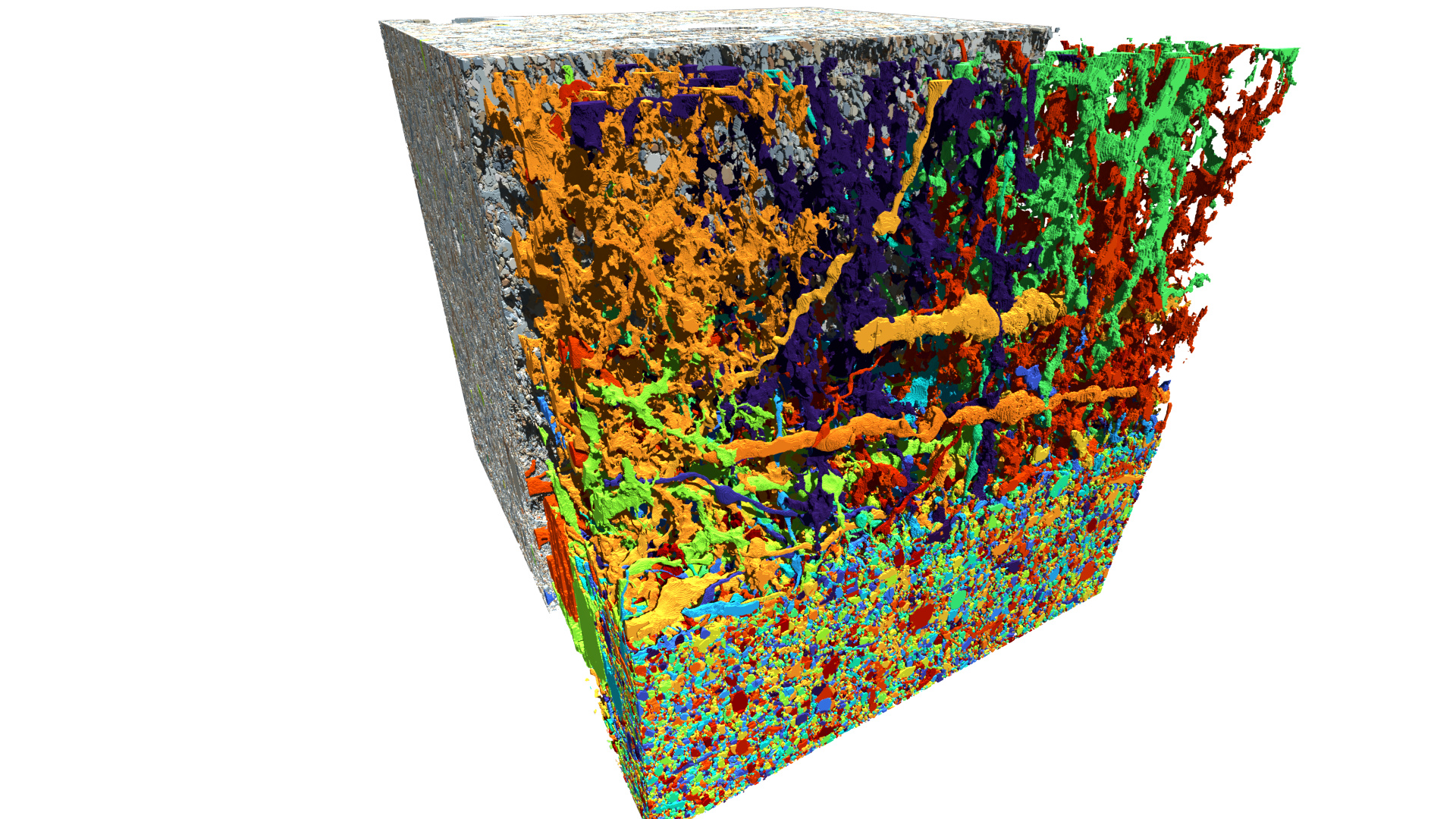}
            }{4}{0.4}{0.6}{0.7}{0.5}{Shadows}{}
        \end{minipage}

        \vspace{-1em}
    
        \begin{minipage}{0.49\textwidth}
            \centering
            \magnifyBottom{
                \includegraphics[width=0.9\columnwidth]{paper/volcanite/results/image/H01-wm_ao.jpg}
            }{4}{0.4}{0.6}{0.7}{0.5}{Ambient\\Occlusion}{}
        \end{minipage}
        \hfill
        \begin{minipage}{0.49\textwidth}
            \centering
            \magnifyBottom{
                \includegraphics[width=0.9\columnwidth]{paper/volcanite/results/image/H01-wm_pt.jpg}
            }{4}{0.4}{0.6}{0.7}{0.5}{Path\\Tracing}{}
        \end{minipage}

        \vspace{-0.5em}
    \end{subfigure}

    \let\spylenstextoptions\oldspylenstextoptions

\end{figure*}

\begin{table*}[pt]
\centering
\caption{Frame times with a static camera over 1024 frames (\textit{Image}) and for a 600 frame long animation (\textit{Video}) where the camera rotates fully around and zooms in to the data. \textit{Frame 1} image timings are high outliers as it decodes all initial bricks once. This cache warm-up quickly vanishes until \textit{Frame 15} while video configurations create highest decoding overheads in later frames when fine resolution bricks are requested.}
\label{tab:vcnt:timings}

\vspace{1em}
{\small Image (static camera)}

\footnotesize


\pgfplotstabletypeset[
    col sep=comma,
    header=true,
    columns={{Data Set},{Shading Mode},{frame min [ms]},{frame avg [ms]},{frame med [ms]},{frame max [ms]},{Frame 00},{Frame 01},{Frame 02},{Frame 03},{Frame 15}},
    columns/{Shading Mode}/.style={column name=Shading, string type, column type={l}},
    columns/{frame min [ms]}/.style={column name=min, column type={r}},
    columns/{frame avg [ms]}/.style={column name=$\varnothing$, column type={r}},
    columns/{frame sdv [ms]}/.style={column name=$\sigma$, column type={r}},
    columns/{frame med [ms]}/.style={column name=med, column type={r}},
    columns/{frame max [ms]}/.style={column name=max, column type={r}},
    columns/{Frame 00}/.style={column name=Frame 0, column type={r}},
    columns/{Frame 01}/.style={column name=Frame 1, column type={r}},
    columns/{Frame 02}/.style={column name=Frame 2, column type={r}},
    columns/{Frame 03}/.style={column name=Frame 3, column type={r}},
    columns/{Frame 15}/.style={column name=Frame 15, column type={r}},
    every head row/.style={after row=\midrule},
    every nth row={4}{before row=\midrule},
    columns/{Data Set}/.style={string type, postproc cell content/.code={
            \pgfmathparse{int(Mod(\pgfplotstablerow,4)}
            \ifnum\pgfmathresult=0
                \pgfkeyssetvalue{/pgfplots/table/@cell content}{%
                   \multirow{4}{*}{\dataNameFromCSV{##1}}%
                }%
            \else
                \pgfkeyssetvalue{/pgfplots/table/@cell content}{}%
            \fi 
    }},
    empty cells with={},
]{paper/volcanite/results/image/image-eval.csv}
\end{table*}

\begin{table*}[pt]
\ContinuedFloat
\centering

\vspace{1em}
{\small Video (moving camera)}

\footnotesize

\pgfplotstabletypeset[
    col sep=comma,
    header=true,
    columns={{Data Set},{Shading Mode},{frame min [ms]},{frame avg [ms]},{frame med [ms]},{frame max [ms]},{Frame 00},{Frame 01},{Frame 02},{Frame 03},{Frame 15}},
    columns/{Shading Mode}/.style={column name=Shading, string type, column type={l}},
    columns/{frame min [ms]}/.style={column name=min, column type={r}},
    columns/{frame avg [ms]}/.style={column name=$\varnothing$, column type={r}},
    columns/{frame sdv [ms]}/.style={column name=$\sigma$, column type={r}},
    columns/{frame med [ms]}/.style={column name=med, column type={r}},
    columns/{frame max [ms]}/.style={column name=max, column type={r}},
    columns/{Frame 00}/.style={column name=Frame 0, column type={r}},
    columns/{Frame 01}/.style={column name=Frame 1, column type={r}},
    columns/{Frame 02}/.style={column name=Frame 2, column type={r}},
    columns/{Frame 03}/.style={column name=Frame 3, column type={r}},
    columns/{Frame 15}/.style={column name=Frame 15, column type={r}},
    every head row/.style={after row=\midrule},
    every nth row={4}{before row=\midrule},
    columns/{Data Set}/.style={string type, postproc cell content/.code={
            \pgfmathparse{int(Mod(\pgfplotstablerow,4)}
            \ifnum\pgfmathresult=0
                \pgfkeyssetvalue{/pgfplots/table/@cell content}{%
                   \multirow{4}{*}{\dataNameFromCSV{##1}}%
                }%
            \else
                \pgfkeyssetvalue{/pgfplots/table/@cell content}{}%
            \fi 
    }},
    empty cells with={},
]{paper/volcanite/results/video/video-eval.csv}
\end{table*}

\begin{table*}[pt]
\centering
\caption{Effect of our cache paletting, where we can store packed palette indices instead of full labels per decoded voxel, on memory and performance. Cache paletting achieves a memory reduction of approximately the \textit{packing factor} depending on the maximum brick palette size.
We render a 1024 frames still image with shadow rays and a cache of 1 GiB. When this cache is not sufficient, some pixels receive fewer valid rendered samples (\textit{SPP}) and frame times increase (worse). Cache paletting allows all required regions to simultaneously reside in the cache and solves these problems. Cache paletting itself has a small overhead making it unsuitable for cases where the cache is sufficient without being packed (same \textit{SPP}).}
\label{tab:vcnt:cache-palette}

\pgfplotstabletypeset[
    col sep=comma,
    header=true,
    column type={r},
    columns={{Data Set},{Cache Palette},{Used Size [MB]},{Packing Factor},{frame avg [ms]},{SPP min},{SPP max}},
    columns/{Shading Mode}/.style={column name=Shading, string type, column type={l}},
    columns/{Used Size [MB]}/.style={column name=Cache Used [MB],column type={>{\raggedleft\arraybackslash}p{1.6cm}},fixed,precision=2},
    columns/{Cache Palette}/.style={column name=Cache Paletting, string type, column type={>{\raggedleft\arraybackslash}p{1.6cm}}},
    columns/{Packing Factor}/.style={column name=Packing Factor, string type, column type={>{\raggedleft\arraybackslash}p{1cm}}},
    columns/{frame avg [ms]}/.style={column name=$\varnothing$ frame [ms], column type={r}},
    columns/{SPP min}/.style={column type={>{\raggedleft\arraybackslash}p{1cm}}},
    columns/{SPP max}/.style={column type={>{\raggedleft\arraybackslash}p{1cm}}},
    row predicate/.code={
            \pgfmathparse{int(Mod(\pgfplotstablerow,8)/2)}
            \ifnum\pgfmathresult=1 %
                \pgfplotstableuserowtrue
            \else
                \pgfplotstableuserowfalse
            \fi
    },
    every head row/.style={after row=\midrule},
    every nth row={2}{before row=\midrule},
    columns/{Data Set}/.style={string type, column type={l}, postproc cell content/.code={
            \pgfmathparse{int(Mod(\pgfplotstablerow,2)}
            \ifnum\pgfmathresult=0
                \pgfkeyssetvalue{/pgfplots/table/@cell content}{%
                   \multirow{2}{*}{\dataNameFromCSV{##1}}%
                }%
            \else
                \pgfkeyssetvalue{/pgfplots/table/@cell content}{}%
            \fi 
    }},
    empty cells with={},
]{paper/volcanite/results/cache-palette/cache-palette-eval.csv}
\end{table*}

\end{appendices}

\FloatBarrier

\bibliographystyle{abbrv-doi-hyperref}
\bibliography{sn-bibliography} %

\end{document}